\documentclass[journal ]{new-aiaa}
\usepackage[utf8]{inputenc}
\usepackage{textcomp}
\usepackage{ulem} 
\usepackage{graphicx}
\usepackage{amsmath}
\usepackage{bm}
\usepackage[version=4]{mhchem}
\usepackage{siunitx}
\usepackage{tablefootnote}
\usepackage{longtable,tabularx}
\usepackage{color}
\usepackage{mathtools}
\usepackage{float}
\usepackage{caption,subcaption}
\usepackage{tikz}
\usepackage[many]{tcolorbox}
\usepackage{color}
\usepackage{epstopdf}
\usepackage[numbers,sort&compress]{natbib}
\def\1{\boldsymbol{1}}

\newtheorem{definition}{Definition}

\newtheorem{remark}{Remark}

\newcounter{example}[section]

\makeatletter
\usetikzlibrary{calc}
   
\makeatother

\title{Learning-Based Stable Optimal Guidance for Spacecraft Close-Proximity Operations}

\author[1]{Kun Wang\footnote{PhD student, School of Aeronautics and Astronautics. E-mail: wongquinn@zju.edu.cn}}
\author[2]{Roberto Armellin\footnote{Professor, Te P\=unaha \=Atea - Space Institute. AIAA Member. E-mail: roberto.armellin@auckland.ac.nz}}
\author[2]{Adam Evans\footnote{PhD student, Te P\=unaha \=Atea - Space Institute. E-mail: aeva686@aucklanduni.ac.nz}}
\author[3]{Harry Holt \footnote{Postdoc, Advanced Concepts Team. E-mail: harry.holt@esa.int}}
\author[1]{Zheng Chen \footnote{Professor, School of Aeronautics and Astronautics. E-mail: z-chen@zju.edu.cn}}

\affil[1]{Zhejiang University, Hangzhou 310027, Zhejiang, China}
\affil[2]{University of Auckland, Auckland 1010, New Zealand}
\affil[3]{European Space Research and Technology Centre,  Noordwijk 2200 AG, Netherlands}

\begin{document}

\maketitle

\begin{abstract}
Machine learning techniques have demonstrated their effectiveness in achieving autonomy and optimality for nonlinear and high-dimensional dynamical systems. However, traditional black-box machine learning methods often lack formal stability guarantees, which are critical for safety-sensitive aerospace applications. This paper proposes a comprehensive framework that combines control Lyapunov functions with supervised learning to provide certifiably stable, time- and fuel-optimal guidance for rendezvous maneuvers governed by Clohessy–Wiltshire dynamics. The framework is easily extensible to nonlinear control-affine systems. A novel neural candidate Lyapunov function is developed to ensure positive definiteness. Subsequently, a control policy is defined, in which the thrust direction vector minimizes the Lyapunov function’s time derivative, and the thrust throttle is determined using minimal required throttle. This approach ensures that all loss terms related to the control Lyapunov function are either naturally satisfied or replaced by the derived control policy. To jointly supervise the Lyapunov function and the control policy, a simple loss function is introduced, leveraging optimal state-control pairs obtained by a polynomial maps based method. Consequently, the trained neural network not only certifies the Lyapunov function but also generates a near-optimal guidance policy, even for the bang-bang fuel-optimal problem. Extensive numerical simulations are presented to validate the proposed method.
\end{abstract}

\section{ Introduction}
\label{intro}
Spacecraft rendezvous generally aims to bring two spacecraft into close proximity, where the chaser spacecraft maneuvers to match the trajectory and velocity of the target spacecraft, thereby achieving controlled relative motion. Following the first successful rendezvous and docking during the Gemini 8 mission in 1966, spacecraft rendezvous has received increasing attention due to its wide applications in various space missions, such as deep space exploration, space debris removal, and on-orbit servicing \cite{flores2014review}. Early rendezvous maneuvers involved a series of human interventions, which constrained the autonomy and robustness of spacecraft operations and hindered the ability to perform complex maneuvers under dynamic or unforeseen conditions. 
Consequently, there is a growing demand for the development of autonomous spacecraft rendezvous guidance \cite{polites1999technology,dittmar2003commercial}, particularly for continuous low-thrust spacecraft, which offer significant advantages in terms of fuel efficiency and prolonged mission durations.

The spacecraft rendezvous problem can be formulated as an optimal control problem (OCP), which is typically solved using numerical methods such as the indirect and direct methods \cite{betts1998survey}. However, these methods are often computationally expensive and not well-suited for on-board implementation. Thanks to recent advances in computing hardware, the concept of computational guidance \& control, as proposed by Lu \cite{lu2017introducing}, has garnered increasing interest in the aerospace community. This approach generates closed-form guidance laws using computationally efficient numerical algorithms onboard. By recursively solving an OCP that is updated at each guidance step with the available state measurements, model predictive control has proven effective in generating closed-loop feedback optimal control solutions for spacecraft rendezvous missions \cite{bodin2011system,di2012model,weiss2015model,hartley2015tutorial,pagone2021pontryagin,bashnick2023fast}. Moreover, model predictive control technology was successfully tested in orbit during the PRISMA technology demonstration mission \cite{bodin2011system}. In addition, by including a proper terminal constraint into the cost function,  a stabilizing implicit controller can be achieved, and the cost function can be interpreted as a
piecewise Control Lyapunov Function (CLF) \cite{bayer2016tube}. However, the resulting problem is generally computationally intractable, making it unsuitable for guidance purposes.  Convex optimization techniques are also popular for spacecraft rendezvous guidance due to their deterministic convergence properties \cite{lu2013autonomous,ortolano2021autonomous,malyuta2023fast}. Nevertheless, aside from problem instances that can be naturally solved using linear, quadratic, or convex programming, tailored sequential optimization methods must be carefully implemented to make problems with general nonconvex constraints computationally tractable.

In recent years, machine learning techniques have proven successful in equipping spacecraft with optimal guidance capabilities. These techniques can be broadly categorized into two classes: supervised learning and reinforcement learning. Supervised learning involves training neural networks on labeled optimal state-control samples, which are obtained offline through traditional optimization methods. In this context, Izzo \textit{et al.} \cite{izzo2020stability} introduced the guidance \& control network, leveraging the low computational times and accuracy in approximating highly nonlinear functions that deep neural networks provide. Once trained, optimal guidance can be generated using simple matrix multiplications and  additions. The guidance \& control network has been successfully applied in various aerospace applications, including powered-descent landing \cite{WANG2024,mulekar2024stable}, orbital transfer \cite{izzo2021real,wang2024real}, rendezvous and docking \cite{federici2021deep}, and attitude control \cite{Yuhan2024}, to name a few. However, within the supervised learning framework, trained neural networks often lack adaptability to new situations unless they are retrained, which can pose limitations in online scenarios where the environment may change over time. In contrast, reinforcement learning emphasizes training agents to make decisions through interaction with their environment and receiving feedback via rewards. Due to its adaptability and exploration capabilities, reinforcement learning has found widespread application in spacecraft rendezvous \cite{broida2019spacecraft,qu2022spacecraft,federici2022meta,tipaldi2022reinforcement,fereoli2024meta}. Nevertheless, the reward function must be carefully designed to balance the fulfillment of terminal conditions and the cost function in the OCP, often leading to learned policies that are suboptimal.
Another drawback of both supervised and reinforcement learning-based guidance methods is their inability to guarantee stability, which is a crucial concern for space autonomy \cite{zhou2017global,shirobokov2021survey,nakamura2022neural}. To address this issue, stabilizing Lyapunov control has been designed for time-optimal transfer between two near-Earth orbits \cite{dalin2015optimal}. The Lyapunov function is formulated as a weighted sum of squared deviations between the current state and the target state, with the weights represented by neural networks that are optimized based on the current state, target state, and spacecraft mass. Holt \textit{et al.} \cite{holt2021optimal} incorporated an actor-critic method into the Lyapunov-based Q-law controller to solve orbit-raising problems. This actor-critic approach made the parameters of the Lyapunov-based Q-law state-dependent, thereby ensuring that the controller could adapt during the transfer. As a result, the issue of user-defined Lyapunov controllers being inherently suboptimal was effectively addressed. Recently, Holt and Armellin \cite{holtISSFD} investigated several different greedy control approaches for proximity operations by integrating CLFs and linear quadratic regulators into reinforcement learning, which improved the optimality compared to reinforcement learning-only approaches.

Unlike traditional machine learning approaches that focus solely on learning or searching for a control policy, certificate learning aims to establish a certificate, such as Lyapunov function for stability or barrier function for safety, that verifies the validity of a known control policy or to jointly learn both a control policy and a certificate \cite{dawson2023safe}. When the control policy is predefined, a neural network is trained to identify a certificate within a space of continuously differentiable functions. This is typically achieved by incorporating the constraints that the certificate function must satisfy into the empirical loss function \cite{richards2018lyapunov,abate2020formal,yin2021stability}. For control-affine dynamical systems with a convex admissible control set,  the policy's optimality can be considered by solving a series of quadratic programming problems during training, as demonstrated in Ref. \cite{dawson2023safe}. Such an approach falls under self-supervised learning, as it does not rely on labeled data for supervision. Recently, Ref. \cite{mulekar2024stable} preconditioned a neural control policy by replicating the open-loop optimal control policy and introduced a stability-constrained imitation learning method. This approach facilitates varying levels of stability certificate verification. However, in existing studies such as \cite{mulekar2024stable,dawson2023safe}, the policy's optimality and certificate functions are generally considered independently, which requires solving separate optimization problems. This separation may result in complex parameter tuning and time-consuming neural network training.

In this paper, we propose a novel learning framework to generate time- and fuel-optimal guidance with certified stability for spacecraft close-proximity operations.  \textcolor{black}{To generate the large datasets required for supervised learning, an efficient method using polynomial maps is employed to rapidly generate optimal trajectories, which are subsequently sampled to obtain optimal state-control pairs.} We then construct a neural candidate Lyapunov function \(V\) that inherently satisfies the positive definiteness property, simplifying the parameter-tuning process in the empirical loss function. The thrust direction vector is designed to minimize the Lyapunov function's time derivative $V_t$.  To enforce the decay condition for the CLF, i.e., \(V_t \leq -\gamma V\) (where \(\gamma\) is the decay rate), we reformulate it using the concept of minimal required throttle \cite{holtISSFD}. In this way, all the loss terms related to the CLF are naturally satisfied or replaced by the derived control policy. Then, a simple loss function is proposed to jointly supervise the Lyapunov function and the corresponding control policy, leveraging optimal state-control pairs.  Unlike existing approaches in Refs. \cite{mulekar2024stable,dawson2023safe}, which address stability and optimality separately, the proposed framework integrates them into a unified process. 
Furthermore, the decay rate \(\gamma\) in the CLF is considered state-dependent, rather than being predefined as a constant in prior works.  The main contributions of this paper are summarized as follows:  
\begin{itemize}
  \item[1)] A neural candidate Lyapunov function is formulated to naturally ensure positive definiteness, significantly simplifying parameter tuning and neural network training; 
  \item[2)] The proposed framework outputs not only a certified Lyapunov function but also a guidance policy that achieves near-optimal performance, even for the bang-bang fuel-optimal problem;
  \item[3)] It is observed that, for the time-optimal problem, considering the decay rate \(\gamma\) as state-dependent not only enhances the training process but also provides slight performance gains compared to using a constant decay rate. Furthermore, the decay rate plays a crucial role in the fuel-optimal problem, where fixing \(\gamma\) as a constant can significantly hinder the convergence of the training process. 
\end{itemize}

The remainder of this paper is structured as follows: Section \ref{SE:problem} outlines the theoretical background on neural Lyapunov function. Section \ref{ProblemStatement} presents the time- and fuel-optimal guidance problems with certified stability.  We detail the proposed framework for learning time- and fuel-optimal guidance with certified stability in Section \ref{SE:method}. Section \ref{SE:Simulations} presents extensive numerical examples to illustrate the efficacy of the proposed approach. Finally, Section \ref{SE:conclusions} summarizes this work.

\section{Background on Neural Lyapunov Function}\label{SE:problem}
Consider a general smooth control-affine dynamical system expressed as
\begin{equation}
  \dot{\boldsymbol{x}} = \boldsymbol{f}(\boldsymbol{x}) + \boldsymbol{g}(\boldsymbol{x})\boldsymbol{u},
  \label{eq:affine}
\end{equation}
where \( \boldsymbol{x} \in \mathcal{\boldsymbol{X}} \subseteq \mathbb{R}^n \) represents the state vector, with $\mathcal{\boldsymbol{X}}$ denoting the admissible state set. The control vector is given by \( \boldsymbol{u} \in \mathcal{\boldsymbol{U}} \subseteq \mathbb{R}^m \), with $\mathcal{\boldsymbol{U}}$ representing the admissible control set. The smooth vector fields \(\boldsymbol{f}: \mathbb{R}^n \to \mathbb{R}^n\) and \(\boldsymbol{g}: \mathbb{R}^n \to \mathbb{R}^{n \times m}\)  are assumed to be locally Lipschitz continuous in both $\boldsymbol{x}$ and $\boldsymbol{u}$.

\begin{definition}(Lyapunov stability \cite{sastry2013nonlinear})
The dynamical system defined in Eq.~(\ref{eq:affine}) with an equilibrium point $\boldsymbol{x}_\text{e} \in \mathcal{\boldsymbol{X}}$ satisfying \( \dot{\boldsymbol{x}}_\text{e} = \boldsymbol{f}(\boldsymbol{x}_\text{e}) + \boldsymbol{g}(\boldsymbol{x}_\text{e})\boldsymbol{u}(\boldsymbol{x}_\text{e}) = \boldsymbol{0} \) is stable under an appropriate controller $\boldsymbol{u}$ in the sense of Lyapunov if, for any given positive number \( \epsilon > 0 \), there exists a positive number \( \delta > 0 \) such that, whenever the initial condition satisfies \( \| \boldsymbol{x}(0) - \boldsymbol{x}_\text{e} \| < \delta \) (where $\| \cdot \|$ denotes the Euclidean vector norm), the resulting system state \( \boldsymbol{x}(t) \) will satisfy \( \| \boldsymbol{x}(t) - \boldsymbol{x}_\text{e}\| < \epsilon \) for all \( t \geq 0 \).  Additionally, if $\lim_{t \rightarrow \infty} \| \boldsymbol{x}(t) \| = \|\boldsymbol{x}_\text{e}\|$, the system is said to be asymptotically stable.
\end{definition}

\begin{definition}(CLF \cite{dawson2023safe}) A continuously differentiable scalar function $V: \mathbb{R}^n \to \mathbb{R}$ is considered a CLF which certifies asymptotic stabilizability of the dynamical system about $\boldsymbol{x}_\text{e}$ if it satisfies the following conditions:
\begin{equation}
    \begin{cases}
    V(\boldsymbol{x}_\text{e}) = 0 \\
    V(\boldsymbol{x}) > 0 \quad \forall \boldsymbol{x} \in \mathcal{\boldsymbol{X}} \setminus \{\boldsymbol{x}_\text{e}\} \\
    \inf\limits_{\boldsymbol{u} \in \mathcal{\boldsymbol{U}}} \left[ L_{\boldsymbol{f}} V(\boldsymbol{x}) + L_{\boldsymbol{g}} V(\boldsymbol{x}) \boldsymbol{u} \right]
    \leq -\gamma V(\boldsymbol{x}) \quad \forall \boldsymbol{x} \in \mathcal{\boldsymbol{X}} 
    \end{cases}
    \label{Eq:Vfunction}
\end{equation}
where  $L_{\boldsymbol{f}} V$  and  $L_{\boldsymbol{g}} V$ denote the Lie derivatives of  $V$ along  $\boldsymbol{f}$  and  $\boldsymbol{g}$, respectively; $\gamma$ is a constant known as the decay rate. 
\end{definition}

Once a CLF \( V \) is identified, it defines the following admissible control set:
\begin{equation}
  \boldsymbol{K}: = \{ \boldsymbol{u} \mid L_{\boldsymbol{f}} V(\boldsymbol{x}) + L_{\boldsymbol{g}} V(\boldsymbol{x}) \boldsymbol{u}  + \gamma  V(\boldsymbol{x}) \leq 0 \},
\end{equation}
such that any Lipschitz feedback controller \( \boldsymbol{u}(\boldsymbol{x}) \) selecting control inputs from the set \( \boldsymbol{K} \) will necessarily ensure the stabilization of the dynamical system.

To learn a CLF \( V(\boldsymbol{x}) \) and derive the feedback policy \( \boldsymbol{u}(\boldsymbol{x}) \) that certifiably stabilizes the dynamical system, two approaches are typically employed. The first approach involves parameterizing both \( V \) and \( \boldsymbol{u} \) as separate neural networks. For notational convenience, we denote the parameter vectors of these neural networks as \( \boldsymbol{\theta}_1 \) and \( \boldsymbol{\theta}_2 \), respectively. The learning process aims to update the parameters \( \boldsymbol{\theta}_1 \) and \( \boldsymbol{\theta}_2 \) to enhance the likelihood of satisfying the conditions outlined in Eq.~(\ref{Eq:Vfunction}). This is accomplished by formulating an empirical loss function known as the Lyapunov risk.

\begin{definition}(Lyapunov risk \cite{chang2019neural})
Consider a neural candidate Lyapunov function $V_{\boldsymbol{\theta}_1}$ and a neural policy $\boldsymbol{u}_{\boldsymbol{\theta}_2}$, the Lyapunov risk $\mathcal{L}_V$ is defined by
\begin{equation}
  \begin{aligned}
    \mathcal{L}_V(\boldsymbol{\theta}_1,\boldsymbol{\theta}_2) =  \frac{1}{N} \sum_{i=1}^{N} &\left[ \lambda_1 V_{\boldsymbol{\theta}_1}^2(\boldsymbol{x}_\text{e}) + \lambda_2 \max(0, -V_{\boldsymbol{\theta}_1}(\boldsymbol{x}_i)) + \right. 
  \left. \lambda_3 \max\left(0, L_{\boldsymbol{f}} V_{\boldsymbol{\theta}_1}(\boldsymbol{x}_i) + L_{\boldsymbol{g}} V_{\boldsymbol{\theta}_1}(\boldsymbol{x}_i) \boldsymbol{u}_{\boldsymbol{\theta}_2}(\boldsymbol{x}_i) + \gamma V_{\boldsymbol{\theta}_1}(\boldsymbol{x}_i) \right) \right],
\label{EQ:existing_method1}
\end{aligned}
\end{equation}
where $\boldsymbol{x}_i$($i = 1, 2, \ldots, N$)  are state vector samples in $\mathcal{\boldsymbol{X}}$, and \( \lambda_j \) (for \( j \in \{1,2,3\} \)) are weighting factors.
\end{definition}

It is important to note that the Lyapunov risk \( \mathcal{L}_V \) does not account for the policy's optimality. To resolve this issue, the second approach uses an “expert” control policy, which can be obtained by solving the OCPs offline, as a basis for the learned control policy. In such case, the Lyapunov risk is built by augmenting Eq.~(\ref{EQ:existing_method1}) with a behavior cloning term that penalizes the difference between the learned control policy $\boldsymbol{u}_{\boldsymbol{\theta}_2}$  and the “expert” control policy $\boldsymbol{u}^{*}$\cite{dawson2023safe}, i.e., 
\begin{equation}
\begin{aligned}
    \mathcal{L}(\boldsymbol{\theta}_1, \boldsymbol{\theta}_2) = 
    \mathcal{L}_V(\boldsymbol{\theta}_1,\boldsymbol{\theta}_2) +  \lambda_4 (\boldsymbol{u}_{\boldsymbol{\theta}_2} - \boldsymbol{u}^{*})^2,
\label{EQ:existing_method2}
\end{aligned}
\end{equation}
where \( \lambda_4 \) is the  weighting factor.

\begin{remark}
The empirical loss function in Eq.(\ref{EQ:existing_method1}) falls under self-supervised learning, as it relies on generating its own labels \cite{dawson2023safe}. In contrast, the “expert” control policy $\boldsymbol{u}^{*}$ provides labels for the empirical loss function in Eq.(\ref{EQ:existing_method2}). Both approaches necessitate the use of separate neural networks to represent the certified Lyapunov function and the control policy, which can complicate training and hinder convergence.
\end{remark}

\section{Problem Statement}\label{ProblemStatement}
\subsection{Relative Motion Dynamics}
\begin{figure}[!htp]
  \begin{center}
  \includegraphics[scale=0.4]{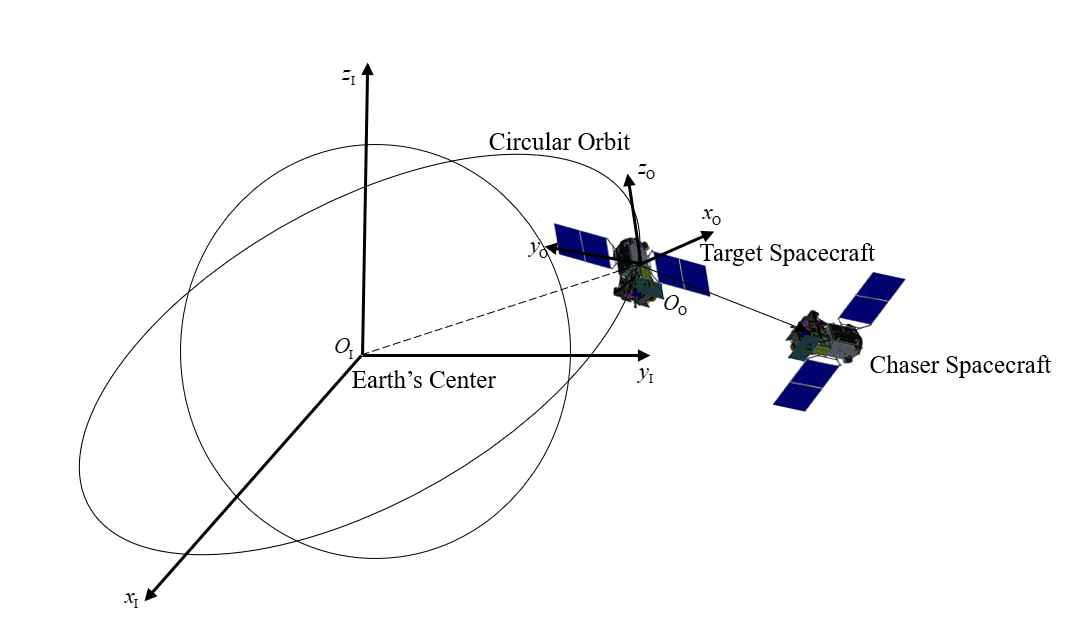}
  \caption{Relative position between the chaser and target spacecraft.}\label{Fig:frame_fig1}
  \end{center}
  \vspace{-0.5cm}
\end{figure}

In this work we consider a two-body point mass dynamical model where the chaser spacecraft is tasked with approaching the target spacecraft that is fixed at the center of the reference frame, as shown in Fig.~\ref{Fig:frame_fig1}. 
The Earth-centered inertial coordinate system is defined by the axes \(x_\text{I}\), \(y_\text{I}\), and \(z_\text{I}\), where the \(z_\text{I}\)-axis points along the Earth's rotational axis, and the \(x_\text{I}\) and \(y_\text{I}\)-axes are in the equatorial plane of the Earth. The origin \(O_\text{I}\) is at the center of the Earth. The circular orbit has an orbital rate of \( n  = \sqrt{\mu/a^3}\) (where $\mu$ is the standard gravitational parameter of the Earth and $a$ is the orbital radius of the target spacecraft’s circular orbit). The circular orbit is traced by the target spacecraft, represented by \(O_\text{O}\) for the target's orbital position. The target spacecraft is described in a Local-Vertical Local-Horizontal (LVLH) coordinate frame, denoted by \(x_\text{O}\), \(y_\text{O}\), and \(z_\text{O}\), with the \(x_\text{O}\)-axis pointing towards the chaser spacecraft, the \(z_\text{O}\)-axis aligned along the normal direction of the circular orbital plane, and the \(y_\text{O}\)-axis completing the right-handed coordinate system. 

We assume that the relative distance between such two spacecraft is far less than their orbital radii. For simplicity, our analysis focuses solely on the in-plane relative motion. Let $x$ and $y$ be the relative position components, $v_x$ and $v_y$ the relative velocity components. Denote by \( \boldsymbol{x} = [x, y, v_x, v_y]^T \) the state vector of the chaser spacecraft, controlled by a throttle \( u \in [0, 1] \) and a unit direction vector \( \boldsymbol{\alpha} = [\alpha_x, \alpha_y]^T \) with $\| \boldsymbol{\alpha} \|=1$. In such case, the relative motion dynamics can be expressed in the LVLH coordinate frame using the linearized Clohessy-Wiltshire equations \cite{clohessy1960terminal}, i.e., 
\begin{align}
  \dot{\boldsymbol{x}} = \mathbf{A} \boldsymbol{x} + \mathbf{B} \boldsymbol{\alpha} u,
  \label{EQ:CW_equation_linear}
\end{align}
with
\[
  \mathbf{A} =
\begin{bmatrix}
0 & 0 & 1 & 0 \\ 
0 & 0 & 0 & 1 \\ 
3 n^2 & 0 & 0 & 2 n \\ 
0 & 0 & -2 n & 0
\end{bmatrix}, \quad
\mathbf{B} =
\begin{bmatrix}
0 & 0 \\ 
0 & 0 \\ 
0 & \frac{T_\text{m}}{m} \\ 
\frac{T_\text{m}}{m} & 0
\end{bmatrix},
\]
where \( T_\text{m} \) is the maximum thrust magnitude, and \( m \) denotes the mass of the chaser spacecraft, which is governed by
\begin{align}
  \dot{m} = -u \frac{T_\text{m}}{I_\text{sp} g_\text{0}},
  \label{EQ:CW_mass_equation}
\end{align}
where \( I_\text{sp} \) is the specific impulse, and \( g_\text{0} \) represents the gravitational acceleration at sea level.

The initial condition for the chaser spacecraft is given by
\begin{align}
  \boldsymbol{x}(0) = [x_0, y_0, v_{x_0}, v_{y_0}]^T,
  \label{EQ:CW_initial_initial}
\end{align}
and the rendezvous condition, i.e., the final condition is expressed as
\begin{align}
  \boldsymbol{x}(t_f) = [0, 0, 0, 0]^T,
  \label{EQ:CW_initial_final}
\end{align}
where the final time \( t_f \) is free in the time-optimal problem and fixed for the fuel-optimal problem.

\begin{remark}
It is important to note that for chaser spacecraft employing electric propulsion systems with a high specific impulse (\( I_\text{sp} > 1000 \, \text{s} \)), mass variation is often considered negligible \cite{pagone2021pontryagin}. However, in this work, we include Eq.~(\ref{EQ:CW_mass_equation}) to ensure the fuel-optimal problem is well-posed when generating the training dataset. When training the neural network to replicate the open-loop optimal control policy, we ignore the mass variation. This approach aims to highlight the proposed method's ability to replicate bang-bang control for the fuel-optimal problem, even though it lacks physical significance when the mass variation is completely negligible.
\end{remark}

The time- and fuel-optimal guidance problems with certified stability are now defined as follows.
\subsection{Time-Optimal Guidance with Certified Stability}
Given the initial condition $\boldsymbol{x}(0)$ in Eq.~(\ref{EQ:CW_initial_initial}), find a CLF $V(\boldsymbol{x})$ and a guidance policy $\boldsymbol{u}(\boldsymbol{x})$ (consisting of throttle $u \in [0,1]$ and direction vector $\boldsymbol{\alpha}$) that drives the chaser spacecraft governed by Eq.~(\ref{EQ:CW_equation_linear}) to a final condition that is 
arbitrarily close to the equilibrium point $\boldsymbol{x}_\text{e} = \boldsymbol{x}(t_f)$ defined in Eq.~(\ref{EQ:CW_initial_final}), such that the cost function
$
J = \int_0^{t_f} 1 \, dt
$
is minimized.
\subsection{Fuel-Optimal Guidance with Certified Stability}
Given the initial condition $\boldsymbol{x}(0)$ in Eq.~(\ref{EQ:CW_initial_initial}), the  fixed final time $t_f$, it is obvious that the optimal control policy is dependent on the state and the 
time-to-go $t_g$ of \cite{wang2022nonlinear}
\begin{align}
t_g = t_f - t.
\end{align}
In such case, the guidance problem aims to  find a CLF $V(t_g,\boldsymbol{x})$ and a  policy $\boldsymbol{u}(t_g,\boldsymbol{x})$ (consisting of throttle $u \in [0,1]$ and direction vector $\boldsymbol{\alpha}$) that drives the chaser spacecraft governed by Eq.~(\ref{EQ:CW_equation_linear}) to a final condition at $t_f$ that is arbitrarily close to the equilibrium point $\boldsymbol{x}_\text{e} = \boldsymbol{x}(t_f)$ defined in Eq.~(\ref{EQ:CW_initial_final}), such that the cost function
$
J = \int_0^{t_f} u \, dt
$
is minimized.

\section{Methodology}\label{SE:method}
This section first describes the necessary conditions for the time- and fuel-optimal problems, which are used to build the training dataset consisting of optimal state-control pairs. Then, we will present the strategy to learn the optimal control policy with certified stability.
\subsection{Training Dataset Generation}
In this section, we derive the necessary conditions based on Pontryagin's Minimum Principle. Then, the training dataset is collected by solving a sufficient number of Two-Point Boundary-Value Problems (TPBVPs). Before proceeding, we shall present the constant parameters for the time- and fuel-optimal problems, as shown in Table~\ref{constants}.
\begin{table}[h!]
  \centering
  \caption{Constant parameters}
  \begin{tabular}{ccc}
  \hline
  \textbf{Name} & \textbf{Variable} & \textbf{Value} \\
  \hline
  Gravitational acceleration at sea level & $g_\text{0}$ & 9.80665 m/s\textsuperscript{2} \\
  Standard gravitational parameter & $\mu$ & $3.986 \times 10^5$ km\textsuperscript{3}/s\textsuperscript{2} \\
  Earth's radius & $R_\text{e}$ & 6371 km \\
  Radius of the target spacecraft’s circular orbit & $a$ & $R_\text{e} + 500$ km \\
  Specific impulse & $I_{\text{sp}}$ & 3300 s \\
  Maximum thrust magnitude & $T_\text{m}$ & 2.5 mN \\
  Chaser spacecraft's initial mass & $m$ & 30 kg \\
  \hline
  \end{tabular}
  \label{constants}
  \end{table}
\subsubsection{Time-optimal problem}
Notice that the mass variation is ignored for the time-optimal problem.
Denote by $\boldsymbol{\lambda} = [\lambda_x,\lambda_y,\lambda_{v_x},\lambda_{v_y}]^T$ the co-state vector. The Hamiltonian is formulated as 
\begin{align}
    \mathscr{H} = 1  + \lambda_x v_x + \lambda_y v_y + \lambda_{v_x} (3n^2 x + 2n v_y + u\frac{T_{\text{m}}}{m}\alpha_x) + \lambda_{v_y} (-2n v_x + u\frac{T_{\text{m}}}{m}\alpha_y).
\label{EQ:Ham_time_optimal}
\end{align}

According to Pontryagin's Minimum Principle \cite{Pontryagin}, the co-state equations are
\begin{align}
\begin{cases}
\dot{\lambda}_x =  -3n^2 \lambda_{v_x},\\
\dot{\lambda}_y =   0, \\
\dot{\lambda}_{v_x} =  -\lambda_x + 2 \lambda_{v_y}, \\
\dot{\lambda}_{v_y} = -{\lambda}_y - 2 n {\lambda}_{v_x}.
\end{cases}
\label{EQ:dot_p_time_optimal}
\end{align}
The optimal unit direction vector $\boldsymbol{\alpha}^* = [\alpha^*_x,\alpha^*_y]^T$ should minimise the Hamiltonian, leading to
\begin{align}
\left[ \begin{array}{c}
  \alpha^*_x \\
  \alpha^*_y
\end{array}
\right]  = - \frac{1}{\sqrt{\lambda^2_{v_x} + \lambda^2_{v_y}}}\left[ \begin{array}{c}
    \lambda_{v_x} \\
    \lambda_{v_y}
\end{array}
\right].
\label{EQ:optimal_H1_time_optimal}
\end{align}
In addition, the optimal throttle should be kept at its maximum \cite{evans2024high}, i.e.,
\begin{align}
  u^* = 1.
\label{EQ:optimal_u_time_optimal}
\end{align}

Since the final time $t_f$ is free, we have the stationary condition 
\begin{equation}
  \mathscr{H}(t_f) = 0.
  \label{EQ:H_law_time_optimal}
\end{equation}

Equations~(\ref{EQ:dot_p_time_optimal}),(\ref{EQ:optimal_H1_time_optimal}), (\ref{EQ:optimal_u_time_optimal}), and (\ref{EQ:H_law_time_optimal}) can be used to build the following shooting function:
\begin{equation}
\boldsymbol{\Psi} (\boldsymbol{\lambda}(0); t_f) = [x(t_f),y(t_f),v_x(t_f), v_y(t_f), \mathscr{H}(t_f)]^T=\boldsymbol{0},
\label{EQ:TPBVP_law_time_optimal}
\end{equation}
where $\boldsymbol{\lambda}(0)$ and $t_f$ are the initial guesses of the initial co-state vector and the final time, respectively.

\textcolor{black}{In order to generate the large datasets necessary for supervised learning, numerous TPBVPs need to be solved. Typical methods such as indirect shooting often suffer from convergence issues, making such a method slow and unreliable for generating extensive databases. Instead, a technique using polynomial maps which was recently developed in the literature \cite{EVANS202517} is utilized in this work. To summarize, solving a single TPBVP and expanding the result with respect to the initial state using differential algebraic techniques produces the high-order Taylor expansion of the flow. Manipulating the resulting polynomials to enforce the same optimality conditions from the original TPBVP then provides a \textit{polynomial map}. The evaluation of such a map with a randomized initial state returns a new optimal trajectory, which may be sampled accordingly. Moreover, in order to ensure accuracy of the map, a technique known as automatic domain splitting is utilized to generate sets of polynomials which span a defined initial domain which guarantees the polynomials are to a desired accuracy. The generation of trajectories is then reduced to the lightweight and rapid evaluation of polynomials, providing an efficient method for database creation. For full details of the methodology the reader is referred to the referenced work.}

\textcolor{black}{ Polynomial maps are leveraged in this work, which enable an even distribution of trajectories to be obtained which span an extensive state-space domain, attributes which are highly beneficial for the training process. The initial domain is centered on $\bm{x}(0)= [500~\text{m}, -500~\text{m}, 1~\text{m/s}, -1~\text{m/s}]^T$ and consists of positional uncertainties of $\pm 75$ m and $\pm 150$ m in the $x$- and $y$-directions respectively, and velocity uncertainties of $\pm 0.05$ m/s in both $x$- and $y$-directions. The trajectories are each split into $k$ equal segments over their corresponding time of flights. A uniform distribution is then used to sample a random state from within each segment. Three datasets are created which will be used in the training process: a training set $\mathcal{D}_{\textrm{train}}$ with $k=1000$ consisting of 10,000,000 samples; a validation set $\mathcal{D}_{\textrm{val}}$ with $k=100$ consisting of 100,000 samples; and a test set $\mathcal{D}_{\textrm{test}}$ with $k=1$ consisting of 1000 samples. Each sample consists of a state and corresponding optimal control direction.
}

\subsubsection{Fuel-optimal problem}
Unlike the time-optimal problem, the mass variation, even being negligible, has to be included here. Denote by $\lambda_m$ the mass co-state.
The Hamiltonian is constructed as 
\begin{align}
    \mathscr{H} = u  + \lambda_x v_x + \lambda_y v_y + \lambda_{v_x} (3n^2 x + 2n v_y + u\frac{T_{\text{m}}}{m}\alpha_x) + \lambda_{v_y} (-2n v_x + u\frac{T_{\text{m}}}{m}\alpha_y) - \lambda_m u\frac{T_\text{m}}{I_\text{sp}g_\text{0}}.
\label{EQ:Ham_fual_optimal}
\end{align}

The co-state equations in Eq.~(\ref{EQ:dot_p_time_optimal}) still hold. The mass co-state equation is
\begin{align}
  \dot{\lambda}_m = -\frac{\partial \mathscr{H}}{\partial m} = u T_{\text{m}}\frac{\lambda_{v_x} \alpha_x + \lambda_{v_y} \alpha_y}{m^2}.
\label{EQ:lambda_mass_optimal}
\end{align}

The optimality condition for the thrust direction vector in Eq.~(\ref{EQ:optimal_H1_time_optimal})  also holds. We assume that the optimal throttle $u^*$ is bang-bang, i.e.,
\begin{align}
  u^* = 
  \left\{ 
      \begin{array}{lc}
          1, & S \leq 0 \\
          0, & S > 0 \\
      \end{array}
  \right.
  \label{EQ:magnitudeUcha6}
\end{align}
where $S$ is the switching function defined by 
\begin{align}
  S = \frac{\partial \mathscr{H}}{\partial u} = 1 + \frac{T_{\text{m}}}{m} (\lambda_{v_x} \alpha_x + \lambda_{v_y} \alpha_y) - \lambda_m \frac{T_\text{m}}{I_\text{sp}g_\text{0}}.
\label{EQ:SF}
\end{align}

To avoid the numerical difficulties from the discontinuous bang-bang control in Eq.~(\ref{EQ:magnitudeUcha6}), we use the smoothing technique from Ref. \cite{di2014high} to approximate Eq.~(\ref{EQ:magnitudeUcha6}) by 
\begin{align}
  u^* \approx u(S,\rho) = \frac{1}{1 + \exp(\rho S)},
  \label{EQ:magnitudeUcha6_smooth}
\end{align}
where $\rho$ is the smoothing constant set to $600$.

Since the final mass is free, it leads to
\begin{align}
  \lambda_m(t_f)= 0.
\label{EQ:lambda_m_tf}
\end{align}

Equations~(\ref{EQ:dot_p_time_optimal}),(\ref{EQ:optimal_H1_time_optimal}), (\ref{EQ:lambda_mass_optimal}), (\ref{EQ:magnitudeUcha6_smooth}), and (\ref{EQ:lambda_m_tf}) can be used to build the following shooting function:
\begin{equation}
\boldsymbol{\Psi} (\boldsymbol{\lambda}(0)) = [x(t_f),y(t_f),v_x(t_f), v_y(t_f), \lambda_m(t_f)]^T=\boldsymbol{0},
\label{EQ:TPBVP_law_fuel_optimal}
\end{equation}
where $\boldsymbol{\lambda}(0)$ is the initial guesses of the initial co-state vector.

\textcolor{black}{Polynomial maps are once again leveraged for database generation. The initial domain is identical to that of the previous time-optimal case. Three datasets are created for use in the training process: a training set, a validation set, and a test set. The number of samples and number of segments the individual trajectories are split into for each of the datasets are the same as those previously used for the time-optimal database. Each sample consists of a state, time-to-go and corresponding optimal control direction and thrust throttle.}

\subsection{Learning Optimal Control Policy with Certified Stability}
\subsubsection{Time-optimal problem}
We define a neural network \( \boldsymbol{\phi} \) parameterized by \( \boldsymbol{\theta} \) as follows:
\begin{equation}
\boldsymbol{\phi}_{\boldsymbol{\theta}}(\boldsymbol{x}): \mathbb{R}^4 \rightarrow \mathbb{R}^2,
\label{EQ:neural_law_cha6}
\end{equation}
where the input is the state vector \( \boldsymbol{x} \) and the output is given by 
\(
\boldsymbol{\phi} = [\phi, \gamma]^T \quad \text{with}  \quad \gamma > 0.
\) This formulation allows the decay rate to be state-dependent.

Considering Eq.~(\ref{Eq:Vfunction}) and using the output \( \phi_{\boldsymbol{\theta}} \) from Eq.~(\ref{EQ:neural_law_cha6}), we construct a scalar-valued network \( V_{\boldsymbol{\theta}} \) as follows:
\begin{align}
V_{\boldsymbol{\theta}}(\boldsymbol{x}) := \left[\phi_{\boldsymbol{\theta}}(\boldsymbol{x}) - \phi_{\boldsymbol{\theta}}(\boldsymbol{x}_\text{e})\right]^2.
\label{EQ:vfunction_cha6}
\end{align}
It is straightforward to verify that this automatically enforces $V_{\boldsymbol{\theta}}(\boldsymbol{x})$ to be positive semi-definite and equal to zero at $\boldsymbol{x}_\text{e}$.

Driving the chaser spacecraft to the final condition is equivalent to reducing the value of \( V_{\boldsymbol{\theta}}(\boldsymbol{x}) \) to zero. We aim to derive a control policy that minimizes the time derivative of \( V_{\boldsymbol{\theta}}(\boldsymbol{x}) \), expressed as follows:
\begin{align}
\min_{\|{\boldsymbol{\alpha}}\|=1 \& u \in [0,1]} \left\{ \dot{V}_{\boldsymbol{\theta}}(\boldsymbol{x}) = \frac{\partial V_{\boldsymbol{\theta}}}{\partial \boldsymbol{x}} (\mathbf{A} \boldsymbol{x} + \mathbf{B} {\boldsymbol{\alpha}} u) \right\}.
\label{Eq:control_mini}
\end{align}

To minimize \( \dot{V}_{\boldsymbol{\theta}}(\boldsymbol{x}) \), the greedy unit direction vector \( \boldsymbol{\alpha} \) can be determined by \cite{holtISSFD}
\begin{align}
\boldsymbol{\alpha} = -\frac{\left(\frac{\partial V_{\boldsymbol{\theta}}}{\partial \boldsymbol{x}} \mathbf{B}\right)^T}{\left\|\frac{\partial V_{\boldsymbol{\theta}}}{\partial \boldsymbol{x}} \mathbf{B}\right\|}.
\label{Eq:control_mini_direction_1}
\end{align}

By combining Eqs.~(\ref{Eq:control_mini}) with ~(\ref{Eq:control_mini_direction_1}), 
 the decay condition in Eq.~(\ref{Eq:Vfunction}) becomes
\begin{align}
\frac{\partial V_{\boldsymbol{\theta}}}{\partial \boldsymbol{x}} \mathbf{A} \boldsymbol{x} - \|\frac{\partial V_{\boldsymbol{\theta}}}{\partial \boldsymbol{x}} \mathbf{B}\| u \leq -\gamma_{\boldsymbol{\theta}} V_{\boldsymbol{\theta}}.
\label{Eq:control_mini_thrust}
\end{align}

It is evident that the minimal required throttle, denoted by $ \underline{u} $, to satisfy Eq.~(\ref{Eq:control_mini_thrust}) is 
\begin{align}
\underline{u} = \frac{\frac{\partial V_{\boldsymbol{\theta}}}{\partial \boldsymbol{x}} \mathbf{A} \boldsymbol{x} + \gamma_{\boldsymbol{\theta}} V_{\boldsymbol{\theta}}}{\|\frac{\partial V_{\boldsymbol{\theta}}}{\partial \boldsymbol{x}} \mathbf{B}\|}.
\label{Eq:control_mini_thrust_equal}
\end{align}

Since \(\|\frac{\partial V_{\boldsymbol{\theta}}}{\partial \boldsymbol{x}} \mathbf{B}\| \geq 0\), it is evident that any throttle selected from the following set will not only guarantee the stabilization of the dynamical system but also satisfy the constraints of the admissible throttle set:
\begin{equation}
\{ u \mid u \geq \underline{u}, \quad u \in [0,1] \}.
\label{Time_optimalset}
\end{equation}

In the case of the minimal required throttle \( \underline{u} \leq 1 \), 
the optimal throttle \( {u}^* = 1 \) lies within the set defined in Eq.~(\ref{Time_optimalset}).  Conversely, if the minimal required throttle \( \underline{u} > 1 \), it means that the decay condition will be violated. In such case, we replace the decay condition using the minimal required throttle. Recall that the positive definiteness of the neural Lyapunov function is already enforced by Eq.~(\ref{EQ:vfunction_cha6}). To guarantee both stability and optimality, we construct 
the following empirical loss function:
\begin{equation}
\begin{aligned}
\mathcal{L}(\boldsymbol{\theta}) = 
\frac{1}{N} \sum_{i=1}^{N} \left[ \lambda_1 \max(0, \underline{u}_i - 1) + \lambda_2 \left(1 - \boldsymbol{\alpha}_i \cdot \boldsymbol{\alpha}^*_i\right) + \lambda_3 (V_{\boldsymbol{\theta}}(\boldsymbol{x}_{\text{nom}})-1)^2\right],
\label{EQ:existing_method_paper}
\end{aligned}
\end{equation}
where \( \boldsymbol{\alpha} \) is the unit direction vector from Eq.~(\ref{Eq:control_mini_direction_1}), and \( \boldsymbol{\alpha}^* \) is the open-loop optimal unit direction vector. The parameters \( \lambda_1 \),  \( \lambda_2 \) and \( \lambda_3 \) are weighting factors. $V_{\boldsymbol{\theta}}(\boldsymbol{x}_{\text{nom}})$ represents the Lyapunov function for a nominal state input $\boldsymbol{x}_{\text{nom}}$.
The first term encourages the minimal required throttle in Eq.~(\ref{Eq:control_mini_thrust_equal}) to comply with the admissible control set \( u \in [0,1] \), while the second term aims to replicate the optimal unit direction vector. The third term poses a normalization constraint on $V$, and it does not introduce conservativeness since a valid Lyapunov function can always be scaled \cite{majumdar2013control}. 
It is important to note that there is no need to replicate the throttle, as it remains constant, i.e., $u^*=1$ \cite{evans2024high}.

\begin{remark}
To guarantee the positive definiteness of the neural Lyapunov function, some prior works (e.g., \cite{dawson2023safe}) propose learning a function \( \boldsymbol{\omega}(\boldsymbol{x}): \mathbb{R}^4 \rightarrow \mathbb{R}^h \), and constructing the Lyapunov function as \( V(\boldsymbol{x}) = \boldsymbol{\omega}(\boldsymbol{x})^T\boldsymbol{\omega}(\boldsymbol{x}) \). However, this approach still requires incorporating the minimum condition \( V(\boldsymbol{x}_\text{e}) = 0 \) into the empirical loss function. In contrast, our neural Lyapunov function, as defined in Eq.~(\ref{EQ:vfunction_cha6}), eliminates this requirement entirely. Furthermore, unlike the empirical loss function in Eq.~(\ref{EQ:existing_method2}), which relies on four weighting factors, our empirical loss function in Eq.~(\ref{EQ:existing_method_paper}) requires only three weighting factors to supervise both the certificate \( V(\boldsymbol{x}) \) and the control policy  \( \boldsymbol{\alpha}(\boldsymbol{x}) \). This significantly simplifies parameter tuning.
\end{remark}
\subsubsection{Fuel-optimal problem}
Unlike the time-optimal problem, in which the input of the neural Lyapunov function is the state vector, we should include the time-to-go $t_g$ as well for the fuel-optimal problem. Therefore, we define a neural network $\boldsymbol{\phi}$ parameterized by ${\boldsymbol{\theta}}$
\begin{equation}
\boldsymbol{\phi}_{\boldsymbol{\theta}}(t_g, \boldsymbol{x}): \mathbb{R}^5 \rightarrow \mathbb{R}^2,
\label{EQ:neural_law_fuel_optimal}
\end{equation}
where the output is still $\boldsymbol{\phi}  = [\phi,\gamma]^T$ with $\gamma>0$. 

Then a scalar-valued network $V_{\boldsymbol{\theta}}$ is built as
\begin{align}
V_{\boldsymbol{\theta}}(t_g, \boldsymbol{x}) := \left[\phi_{\boldsymbol{\theta}}(t_g, \boldsymbol{x}) - \phi_{\boldsymbol{\theta}}(0, \boldsymbol{x}_\text{e})\right]^2.
\label{EQ:vfunction_cha6fuel_optimal}
\end{align}

In such case, the time derivative of $V_{\boldsymbol{\theta}}(t_g, \boldsymbol{x})$ becomes
\begin{align}
  \frac{\text{d}}{\text{d}t}V_{\boldsymbol{\theta}}(t_g, \boldsymbol{x}) = 
 \frac{\partial V_{\boldsymbol{\theta}}}{\partial t_g} \frac{\text{d} t_g}{\text{d} t} + \frac{\partial V_{\boldsymbol{\theta}}}{\partial \boldsymbol{x}} \frac{\text{d} \boldsymbol{x}}{\text{d} t} = - \frac{\partial V_{\boldsymbol{\theta}}}{\partial t_g} + \frac{\partial V_{\boldsymbol{\theta}}}{\partial \boldsymbol{x}} (\mathbf{A} \boldsymbol{x} + \mathbf{B} {\boldsymbol{\alpha}} u).
    \label{Eq:control_mini_fuel_optimal}
\end{align}

The unit direction vector $\boldsymbol{\alpha}$ that minimizes $\dot{V}_{\boldsymbol{\theta}}(t_g, \boldsymbol{x})$ must be
\begin{align}
     {\boldsymbol{\alpha}}
     = - \frac{(\frac{\partial V_{\boldsymbol{\theta}}}{\partial \boldsymbol{x}} \mathbf{B})^T}{
     \|\frac{\partial V_{\boldsymbol{\theta}}}{\partial \boldsymbol{x}} \mathbf{B}\|}.
     \label{Eq:control_mini_direction}
\end{align}

The minimal required throttle $\underline{u}$ to satisfy the third condition in Eq.~(\ref{Eq:Vfunction})  is
\begin{align}
  \underline{u} = 
     (\frac{\partial V_{\boldsymbol{\theta}}}{\partial \boldsymbol{x}} \mathbf{A} \boldsymbol{x}
     +\gamma_{\boldsymbol{\theta}} V_{\boldsymbol{\theta}} - \frac{\partial V_{\boldsymbol{\theta}}}{\partial t_g}) /\|\frac{\partial V_{\boldsymbol{\theta}}}{\partial \boldsymbol{x}} \mathbf{B}\|.
     \label{Eq:control_mini_thrust_equal}
 \end{align}

Then, the empirical loss function is constructed as
 \begin{equation}
  \begin{aligned}
    \mathcal{L}(\boldsymbol{\theta}) = 
    \frac{1}{N} \sum_{i=1}^{N} \left[ \lambda_1 (\underline{u}_i - u^*_i)^2
     + \right. 
  \left. \lambda_2 (1-\boldsymbol{\alpha}_i \cdot \boldsymbol{\alpha}^*_i) + \lambda_3 (V_{\boldsymbol{\theta}}(t_{g_\text{nom}}, \boldsymbol{x}_{\text{nom}})-1)^2 \right],
\label{EQ:existing_method_paper_fuel}
\end{aligned}
\end{equation}
where $\boldsymbol{\alpha}$ is the unit direction vector in Eq.~(\ref{Eq:control_mini_direction}) and $\boldsymbol{\alpha}^*$ is the open-loop optimal unit direction vector, and $\lambda_1$, $\lambda_2$, and $\lambda_3$ are the weighting factors.  $V_{\boldsymbol{\theta}}(\boldsymbol{x}_{\text{nom}})$ represents the Lyapunov function for a nominal state input $(t_{g_\text{nom}}, \boldsymbol{x}_{\text{nom}})$. The first and second terms aim to replicate the open-loop optimal throttle and unit direction vector, respectively. The third term is used to normalize the neural Lyapunov function.

\subsection{Neural Network Training and Implementation}
The hyperparameter configuration for neural network training is as follows: a learning rate of 0.0001, a batch size of 2000, 100 epochs for the time-optimal problem, and 200 epochs for the fuel-optimal problem. The Adam optimizer is employed to minimize their respective empirical loss functions. The network architecture consists of 3 fully connected hidden layers for the time-optimal problem and 4 fully connected hidden layers for the fuel-optimal problem, with 64 neurons in each layer. The {\it tanh} activation function is used in the hidden layers, while the output layer uses a linear activation function for \( \phi \) and an exponential activation function for \( \gamma \). In addition, the parameters in Eqs.~(\ref{EQ:existing_method_paper}) and (\ref{EQ:existing_method_paper_fuel}) are chosen after a number of preliminary trials. Specifically, the weighting factors are set as $\lambda_1 = 1$, $\lambda_2 = 1$, and $\lambda_3 = 0.1$ for the time-optimal problem, while they are set as $\lambda_1 = 1.5$, $\lambda_2 = 1$, and $\lambda_3 = 0.1$ for the fuel-optimal problem. 
The nominal state for the time-optimal problem is set to $\boldsymbol{x}_{\text{nom}} = [500~\text{m}, -500~\text{m}, 1~\text{m/s}, -1~\text{m/s}]^T$. The nominal state for the fuel-optimal problem is chosen as $(t_{g_\text{nom}}, \boldsymbol{x}_{\text{nom}}) = [14,400~\text{s}, 500~\text{m}, -500~\text{m}, 1~\text{m/s}, -1~\text{m/s}]^T$. To demonstrate the effects of the decay rate $\gamma$ on the training process, we also consider some scenarios with constant decay rates. In such cases, the neural network in Eq.~(\ref{EQ:neural_law_cha6}) becomes ${\phi}_{\boldsymbol{\theta}}(\boldsymbol{x}): \mathbb{R}^4 \rightarrow \mathbb{R}^1$; and the neural network in Eq.~(\ref{EQ:neural_law_fuel_optimal})  becomes ${\phi}_{\boldsymbol{\theta}}(t_g, \boldsymbol{x}): \mathbb{R}^5 \rightarrow \mathbb{R}^1$.

\begin{figure}[!htp]
  \begin{center}
  \includegraphics[scale=0.2]{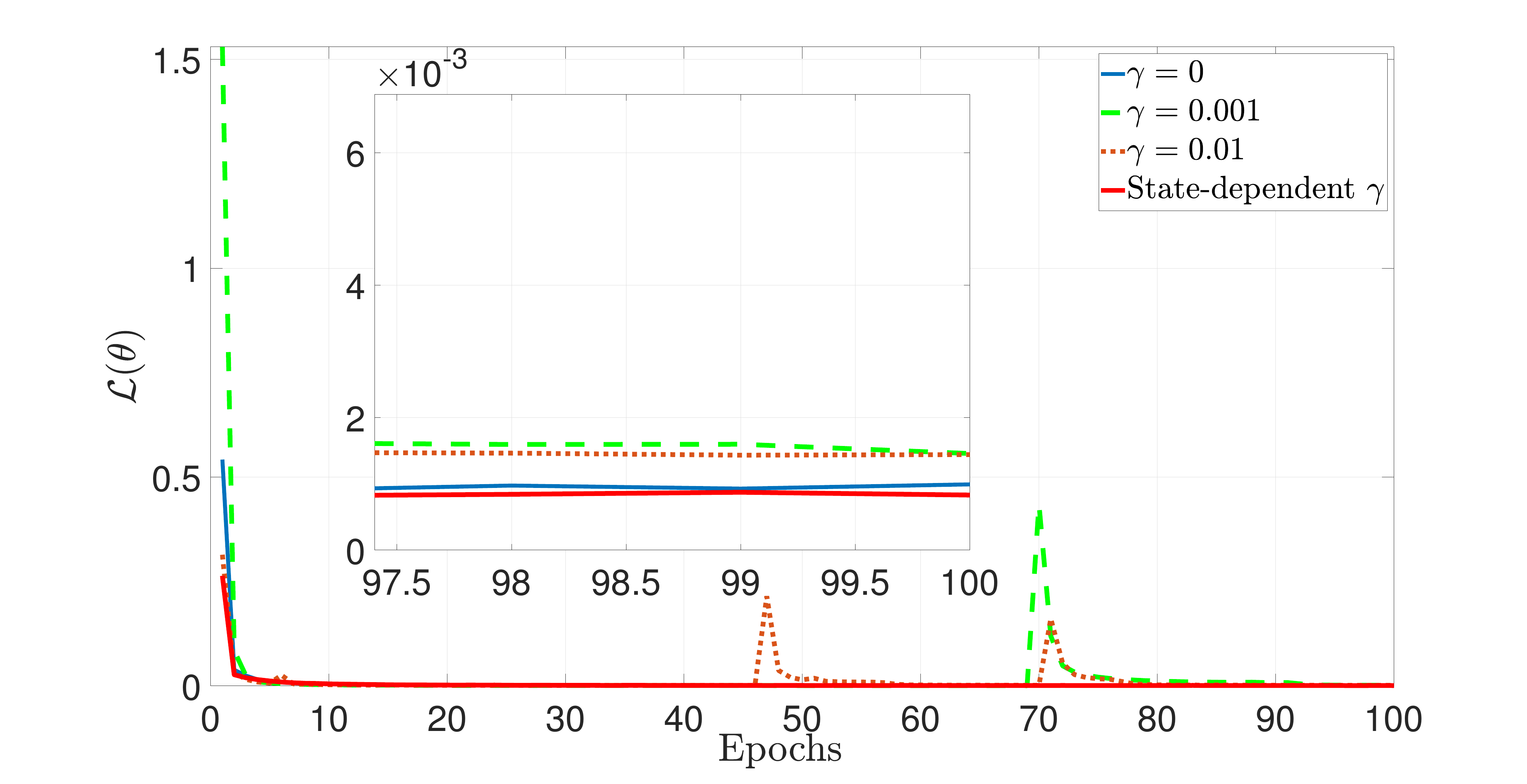}
  \caption{Empirical loss function profiles for the time-optimal problem with different decay rates.}\label{Fig:loss_time_optimal}
  \end{center}
  \vspace{-0.5cm}
\end{figure}

The training process for the time-optimal problem takes approximately 4.3 hours on a laptop.
Figure~\ref{Fig:loss_time_optimal} illustrates the empirical loss function profiles under different decay rate configurations. When the decay rate \( \gamma \) is treated as a state-dependent parameter, the loss exhibits the smoothest and most rapid reduction compared to configurations with constant decay rates. In scenarios with constant decay rates (\( \gamma = 0, 0.001, \text{and } 0.01 \)), the empirical loss functions converge to approximately \( 9.9 \times 10^{-4} \), \( 1.4 \times 10^{-3} \), and \( 1.4 \times 10^{-3} \), respectively. Notably, the state-dependent decay rate scenario achieves a loss of \( 8.3 \times 10^{-4} \), which is the lowest, as highlighted in the enlarged view in Fig.~\ref{Fig:loss_time_optimal}. The state-dependent decay rate not only accelerates convergence but also ensures smoother optimization dynamics, which is evident in the absence of oscillations compared to the constant decay rate settings. This underscores the effectiveness of adopting a state-dependent decay rate for improving the training process.

\begin{figure}[!htp]
  \begin{center}
  \includegraphics[scale=0.2]{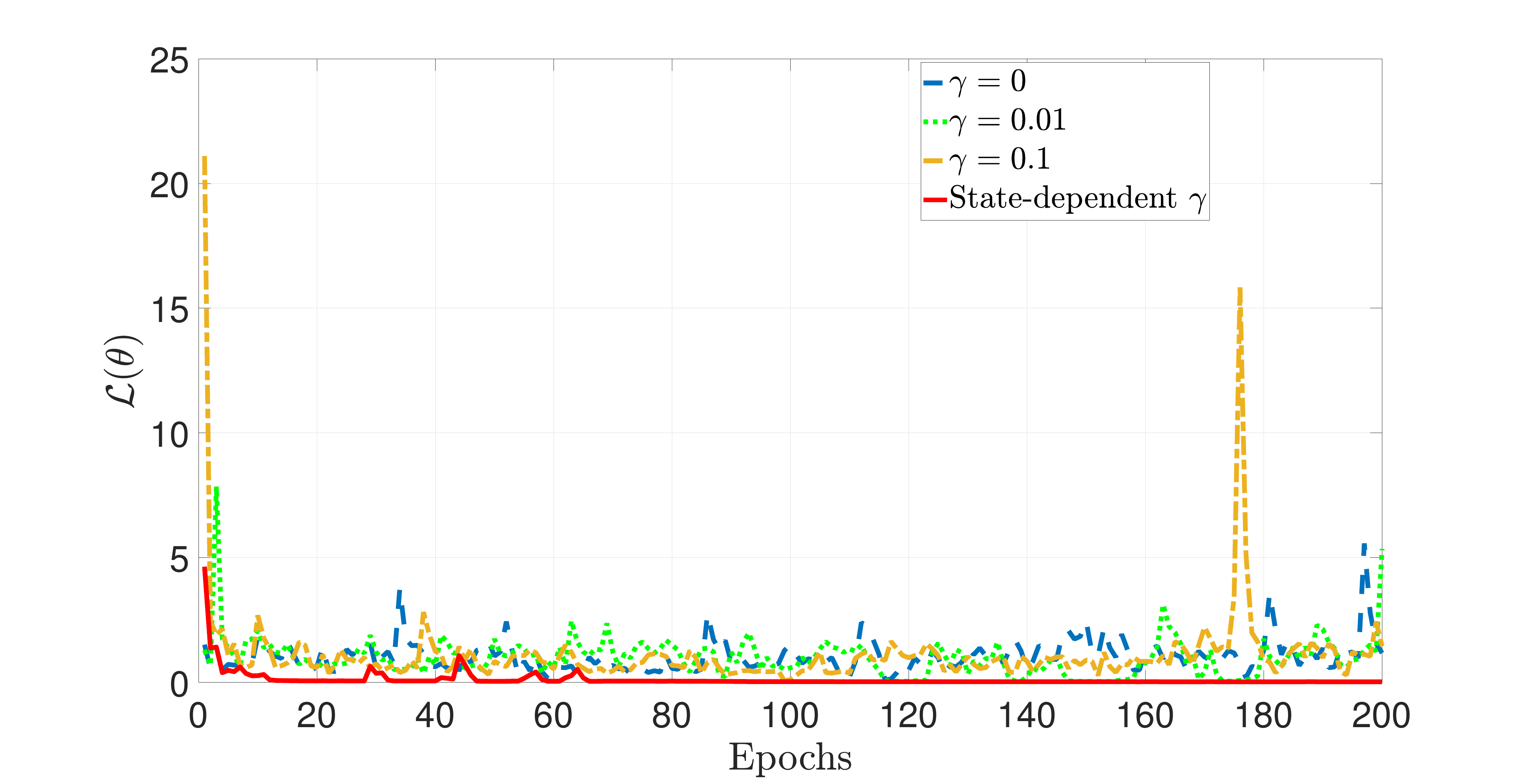}
  \caption{Empirical loss function profiles for the fuel-optimal problem with different decay rates.}\label{Fig:loss_fuel_optimal}
  \end{center}
  \vspace{-0.5cm}
\end{figure}

The training process for the fuel-optimal problem takes approximately 17.5 hours.
Figure~\ref{Fig:loss_fuel_optimal} illustrates the empirical loss function profiles under different decay rate configurations. The scenario with a state-dependent decay rate \( \gamma \) demonstrates the smoothest and most rapid reduction in loss, converging to approximately \( 2.2 \times 10^{-2} \) within the first 80 epochs. In contrast, the scenarios with constant decay rates (\( \gamma = 0 \), \( \gamma = 0.01 \), and \( \gamma = 0.1 \)) exhibit slower and less stable convergence behaviors. Among these, the configuration with \( \gamma = 0.1 \) shows significant oscillations and fails to achieve meaningful convergence, while the other constant decay rate settings struggle to approach a steady-state value. This comparison highlights the effectiveness of employing a state-dependent decay rate in achieving faster and more stable loss minimization for the fuel-optimal problem.  

\begin{figure}[!htp]
  \begin{center}
  \includegraphics[scale=0.3]{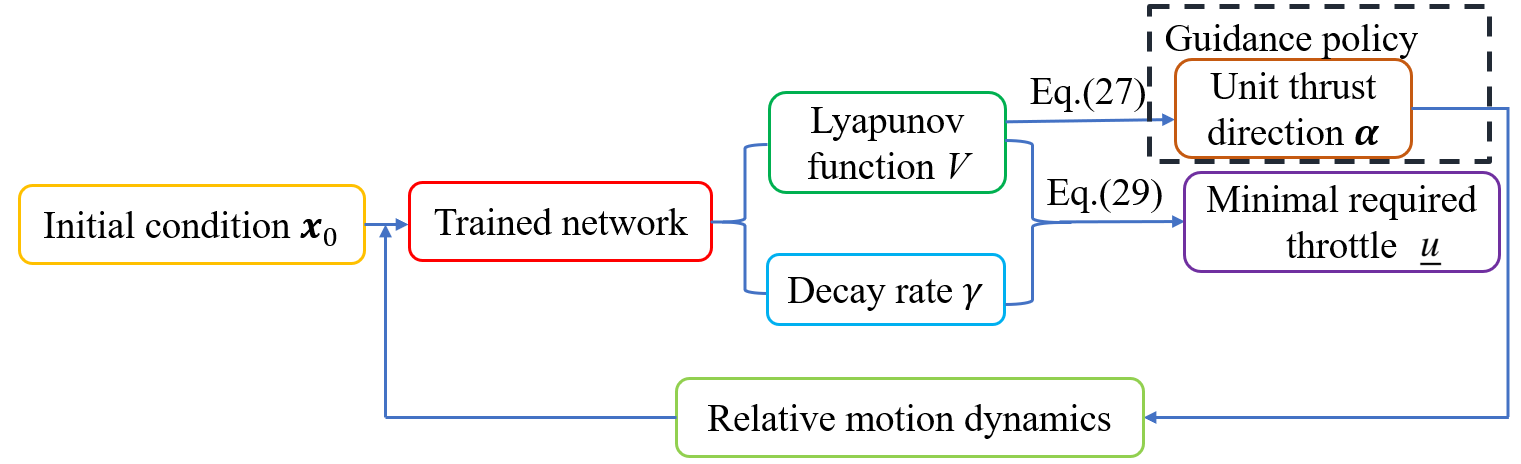}
  \caption{Closed-loop guidance diagram for the time-optimal problem.}\label{Fig:Closed_looptimeoptimal}
  \end{center}
  \vspace{-0.5cm}
\end{figure}

\begin{figure}[!htp]
  \begin{center}
  \includegraphics[scale=0.3]{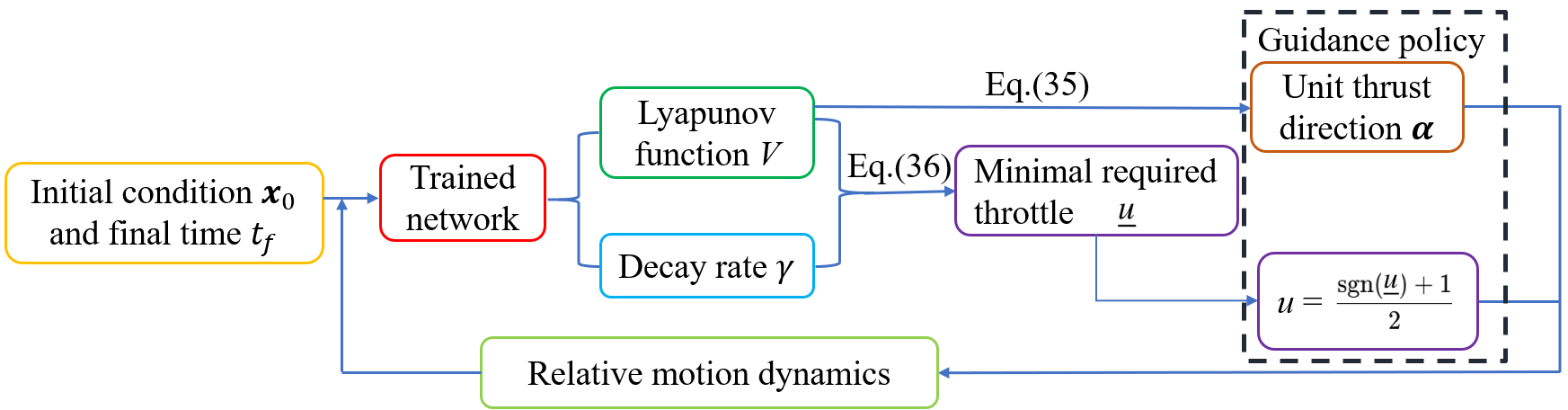}
  \caption{Closed-loop guidance diagram for the fuel-optimal problem.}\label{Fig:Closed_loopfueloptimal}
  \end{center}
  \vspace{-0.5cm}
\end{figure}

Based on the strategy outlined in the previous subsection, the procedures for applying the proposed method to the time-optimal and fuel-optimal problems are illustrated in Figs.~\ref{Fig:Closed_looptimeoptimal} and \ref{Fig:Closed_loopfueloptimal}, respectively. For the time-optimal problem, it is noteworthy that the throttle remains fixed at \( u = 1 \). In contrast, for the fuel-optimal problem, the throttle is set to \( u = 1 \) only when the minimal required throttle \( \underline{u} > 0 \); otherwise, it is set to \( u = 0 \). This is governed by the expression \( u = \frac{\text{sgn}(\underline{u})+1}{2} \).
\section{Numerical Simulations}\label{SE:Simulations}
In this section, we evaluate the efficacy of the proposed method on both the time-optimal and fuel-optimal problems by examining its performance in terms of stability, optimality, and robustness. Additionally, the computational cost of the proposed approach is analyzed. The learned guidance policy is updated every \(3.6\) seconds, and all simulations are conducted on a laptop equipped with an AMD Ryzen 7-5800H CPU running at 3.2 GHz.
\subsection{Time-Optimal Problem}
\subsubsection{Stability and Optimality Performance}
\begin{figure}[!htp]
    \centering
    \begin{subfigure}[t]{0.48\textwidth}
      \centering
      \includegraphics[width=\linewidth]{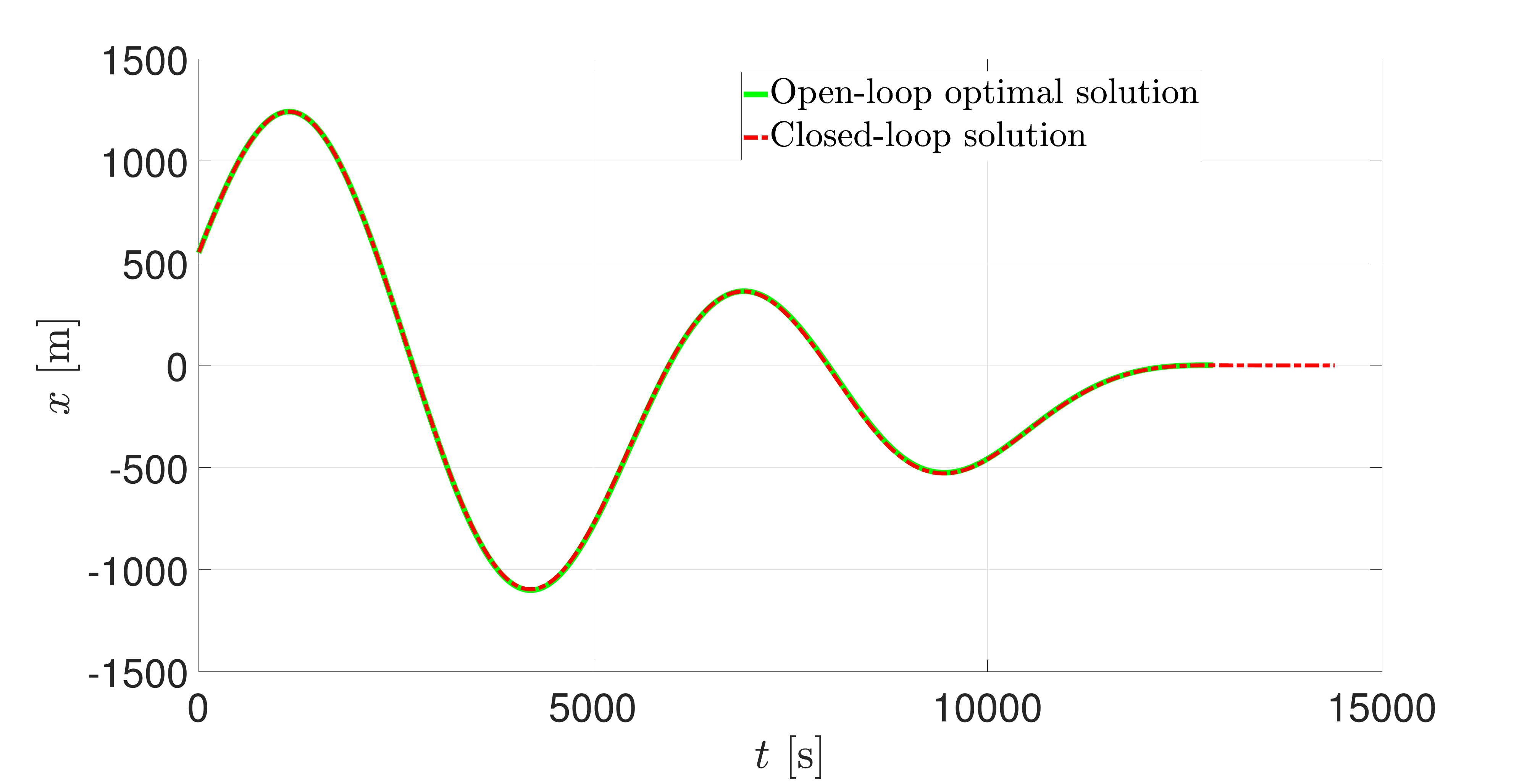}
      \caption{Relative position $x$ profile}
      \label{Fig:CaseA_x}
    \end{subfigure}
    \hfill
    \begin{subfigure}[t]{0.48\textwidth}
      \centering
      \includegraphics[width=\linewidth]{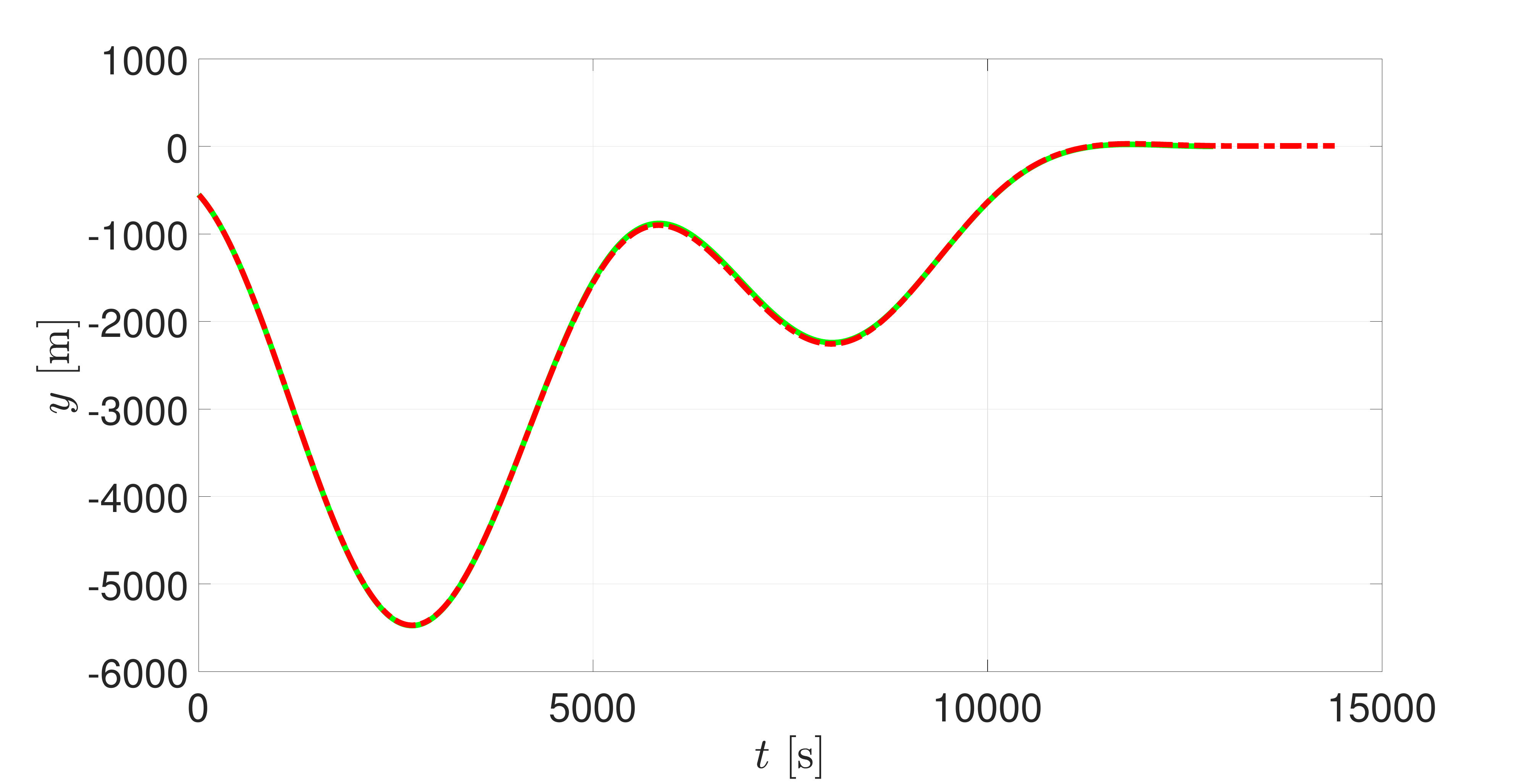}
      \caption{Relative position $y$ profile}
      \label{Fig:CaseA_y}
    \end{subfigure}\\[1em]
    
    \begin{subfigure}[t]{0.48\textwidth}
      \centering
      \includegraphics[width=\linewidth]{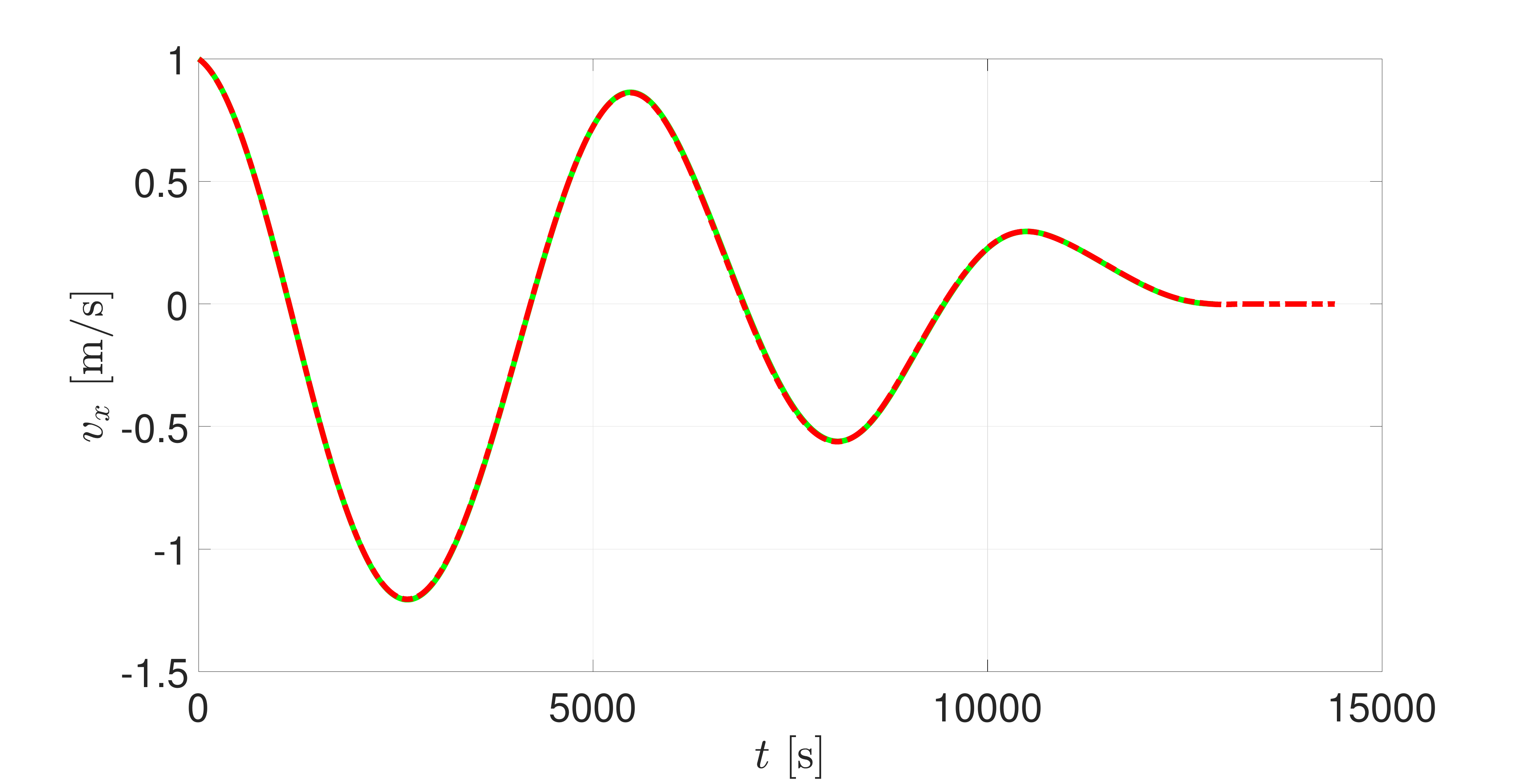}
      \caption{Relative velocity $v_x$ profile}
      \label{Fig:CaseA_vx}
    \end{subfigure}
    \hfill
    \begin{subfigure}[t]{0.48\textwidth}
      \centering
      \includegraphics[width=\linewidth]{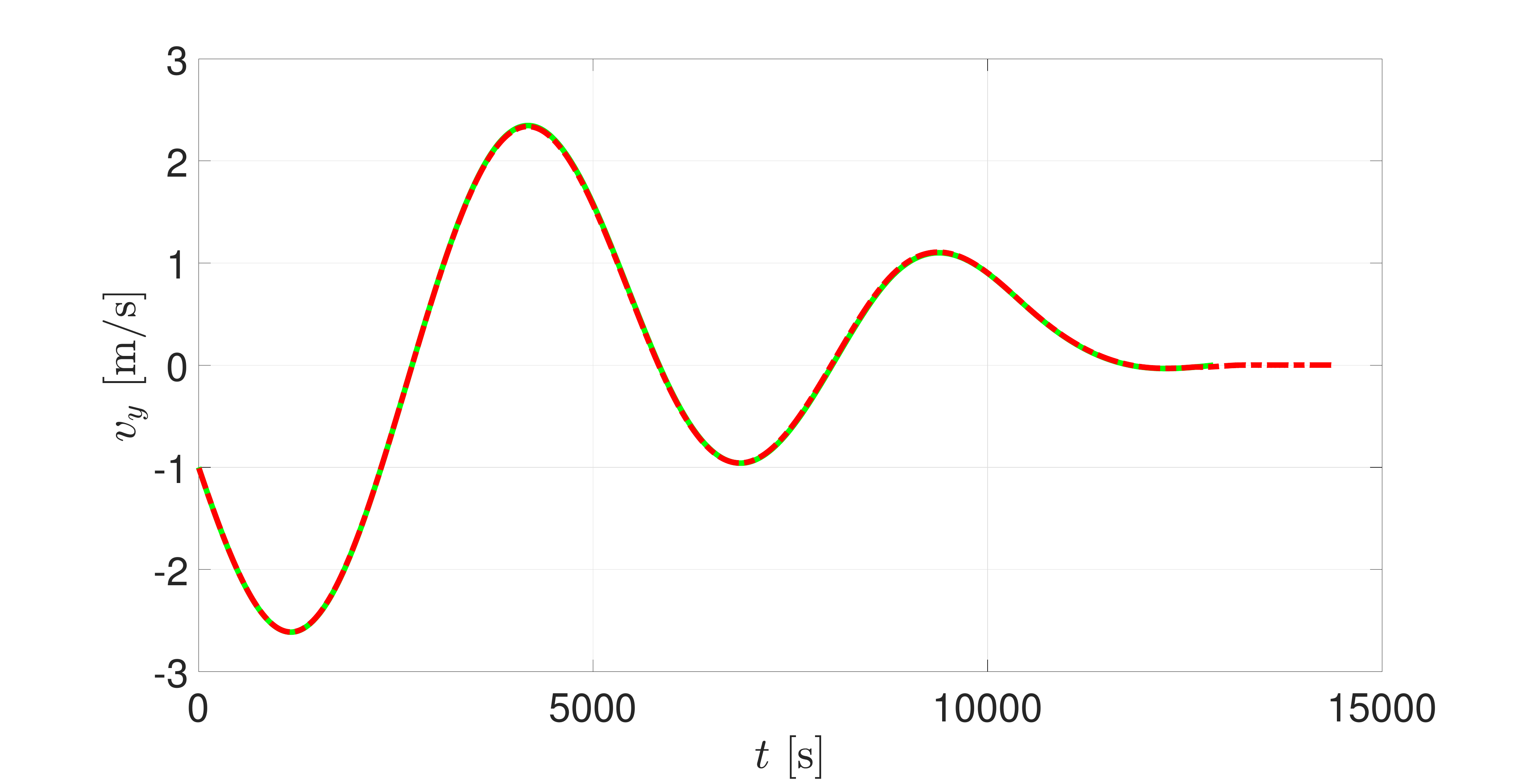}
      \caption{Relative velocity $v_y$ profile}
      \label{Fig:CaseA_vy}
    \end{subfigure}
    \caption{Comparison of individual state profiles for the time-optimal problem.}
    \label{Fig:Simulation_results_caseA}
    \vspace{-0.5cm}
  \end{figure}

\begin{figure}[!htp]
    \begin{center}
    \includegraphics[scale=0.2]{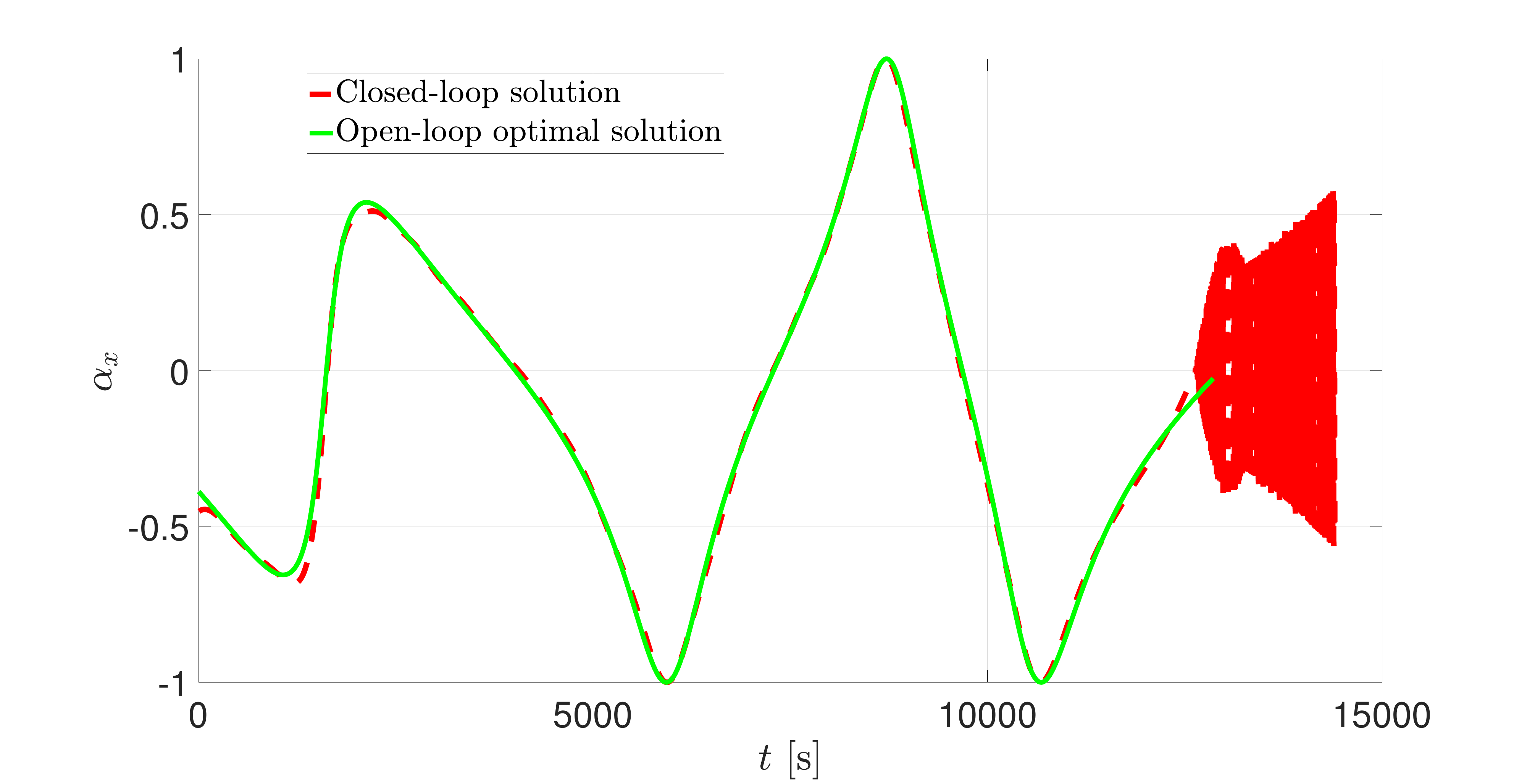}
    \caption{Thrust direction component $\alpha_x$ profile.}\label{Fig:CaseA_direction}
    \end{center}
    \vspace{-0.5cm}
\end{figure}

We consider a nonnominal initial condition of \(\boldsymbol{x}_0 = [550~\text{m}, -550~\text{m}, 1~\text{m/s}, -1~\text{m/s}]^T\). The proposed method is applied to guide the chaser spacecraft, with the solution to the shooting function defined in Eq.~(\ref{EQ:TPBVP_law_time_optimal}) used as a benchmark for comparison. Figure~\ref{Fig:Simulation_results_caseA} depicts the comparison of relative position and velocity components obtained by the proposed method and the indirect method.
The closed-loop solutions are shown to align perfectly with the open-loop optimal solutions. 
The offline-computed optimal final time is found to be \(t^*_f = 1, 2860 \) s. Using the proposed method, the state of the chaser spacecraft  at \(t^*_f \)  is \([-0.4~\text{m}, 6.3~\text{m}, -0.0009~\text{m/s}, -0.0144~\text{m/s}]^T\), which is very close to the target state. Instead of terminating the simulation at \( t^*_f \), the dynamics are further propagated using the proposed method. Notably, the chaser spacecraft, guided by the proposed approach, remains in the vicinity of the target state for \( t > t^*_f \).
A comparison of the thrust direction \(\alpha_x\) profiles is shown in Fig.~\ref{Fig:CaseA_direction}, demonstrating that the proposed method can generate near-optimal solutions.

\begin{remark} 
The learned guidance policy keeps the chaser spacecraft close to the exact rendezvous condition by causing fluctuations, as shown in Fig.~\ref{Fig:CaseA_direction}. As the chaser spacecraft approaches the target spacecraft, the small divisor in Eq.~(\ref{Eq:control_mini_direction}) contributes to the chattering phenomenon observed in Fig.~\ref{Fig:CaseA_direction}. To achieve exact rendezvous, a strategy that divides the rendezvous process into two stages based on the value of the Lyapunov function \( V \) can be employed. In this approach, the learned guidance policy is applied in the first stage, while the open-loop rendezvous control law is used in the second stage. For further details about the two-stage strategy, refer to Ref. \cite{gurfil2023spacecraft}. 
\end{remark}

\begin{figure}[!htp]
    \centering
    \begin{subfigure}[t]{0.45\textwidth}
    \centering
    \includegraphics[width = \linewidth]{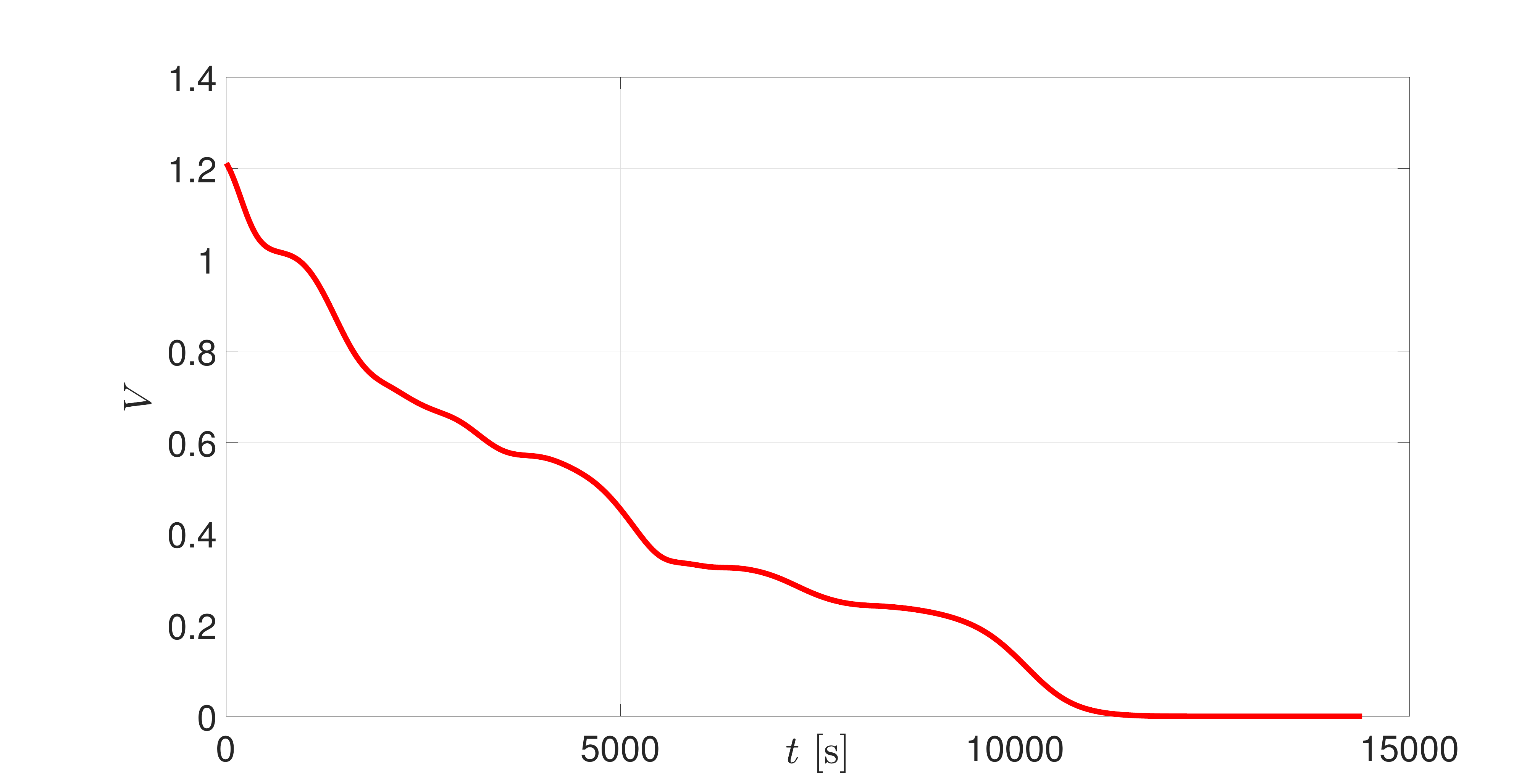}
    \caption{Lyapunov function}
    \label{Fig:CaseA_u}
    \end{subfigure}
    \hspace{0.05\textwidth}
    \begin{subfigure}[t]{0.45\textwidth}
    \centering
    \includegraphics[width = \linewidth]{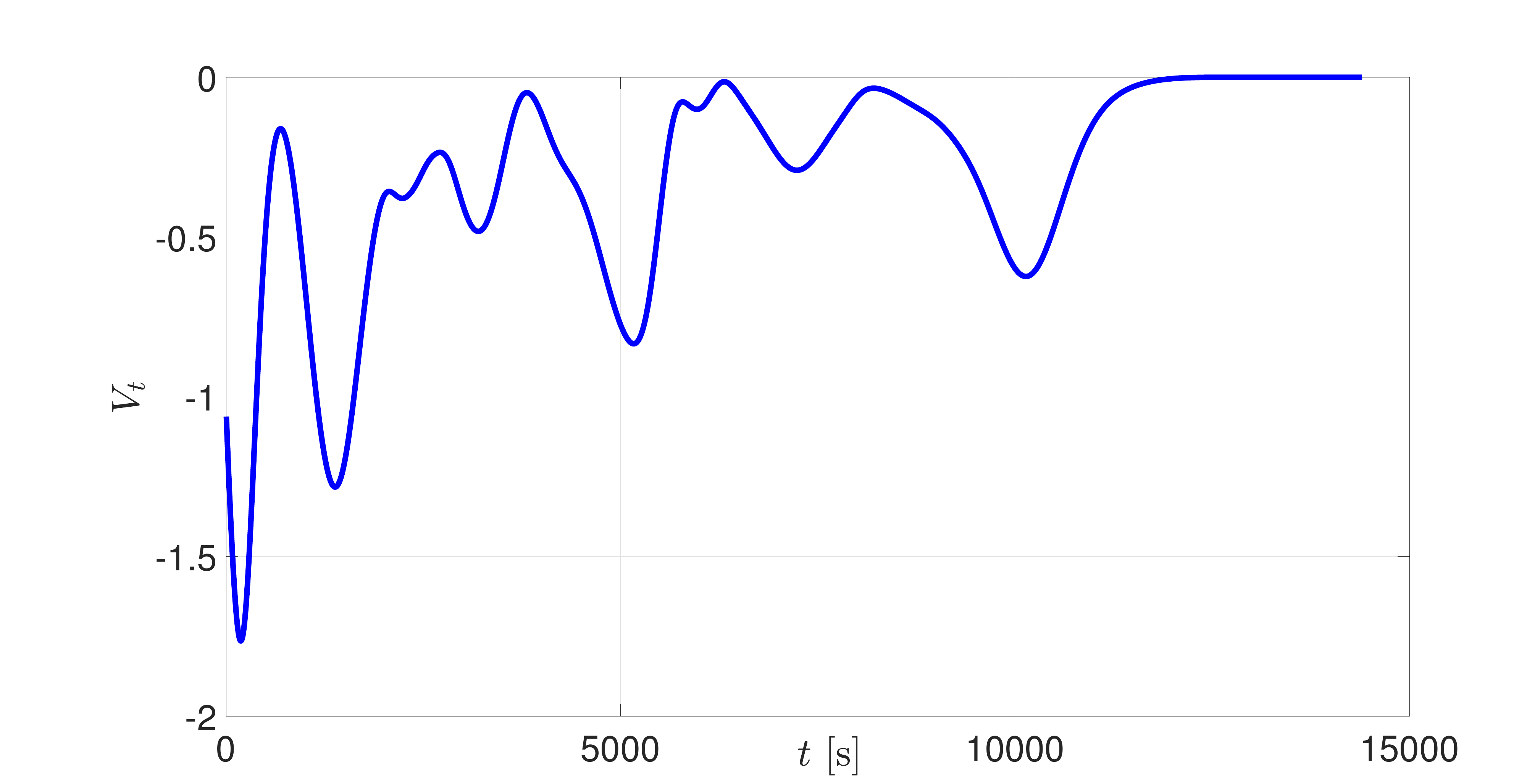}
    \caption{Lyapunov function's time derivative}
    \label{Fig:CaseA_J}
    \end{subfigure}
    \caption{Lyapunov function and its time derivative profiles for the time-optimal problem.}
  \label{Fig:CaseA_control_profile}
  \vspace{-0.5cm}
    \end{figure}

\begin{figure}[!htp]
      \begin{center}
      \includegraphics[scale=0.2]{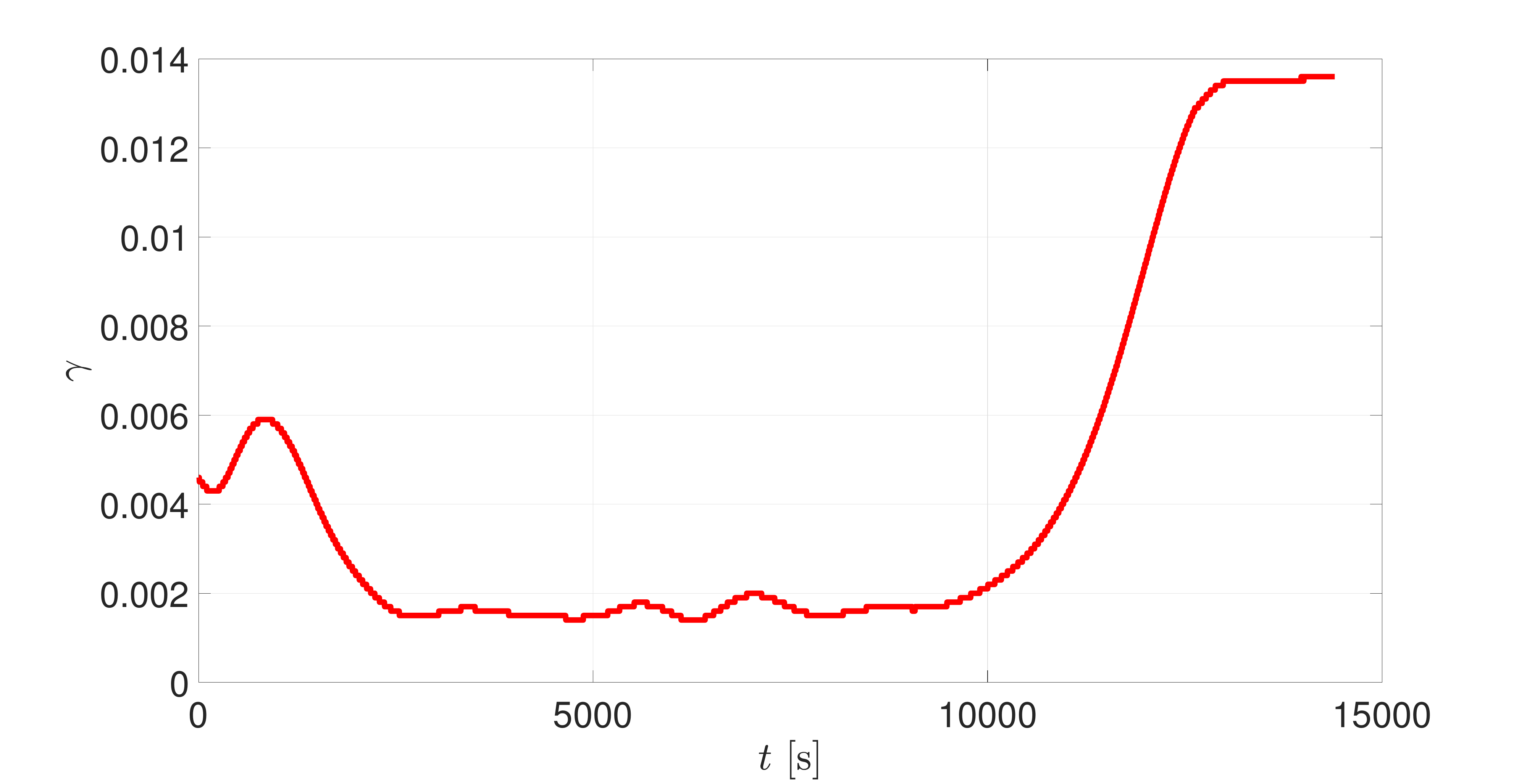}
      \caption{State-dependent decay rate profile for the time-optimal problem.}\label{Fig:Optimal_decayrate}
      \end{center}
      \vspace{-0.5cm}
\end{figure}

\begin{figure}[!htp]
      \begin{center}
      \includegraphics[scale=0.2]{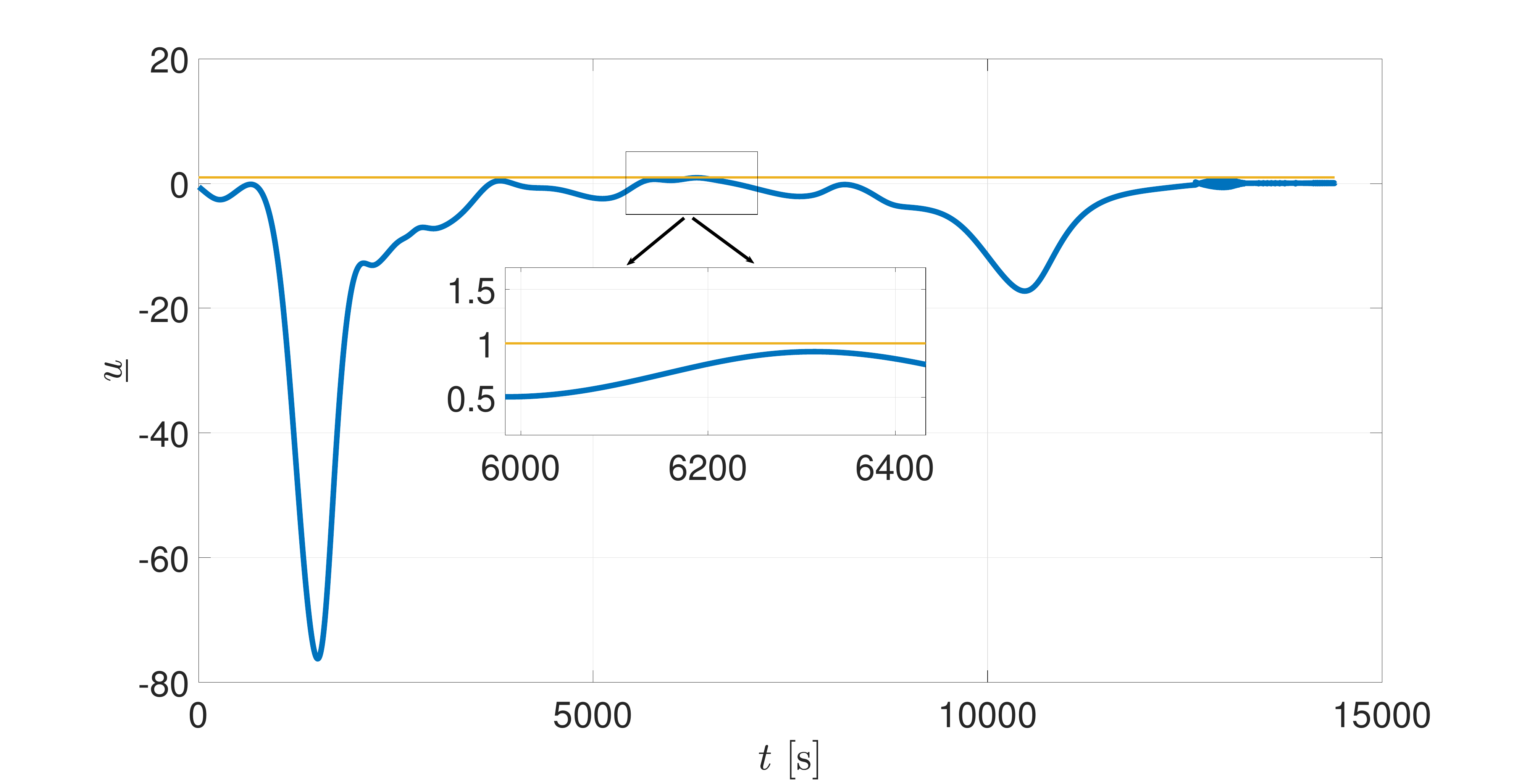}
      \caption{Minimal required throttle profile for the time-optimal problem.}\label{Fig:CaseA_minimal_required_control}
      \end{center}
      \vspace{-0.5cm}
\end{figure}

Figure~\ref{Fig:CaseA_control_profile} shows the Lyapunov function $V$ and its time derivative $V_t$ throughout the entire rendezvous. It is evident that the Lyapunov function maintains positive definiteness and continuously decreases over time. Additionally, the decay rate defined in Eq.~(\ref{Eq:Vfunction}) is presented in Fig.~\ref{Fig:Optimal_decayrate}. It can be observed that the decay rate fluctuates during most of the propagation, except near the end. Once the chaser spacecraft is guided close to the target state, the changes in the decay rate become negligible.
Moreover, Fig.~\ref{Fig:CaseA_minimal_required_control} illustrates the minimal required throttle profile throughout the entire rendezvous. The maximum value of the minimal required throttle is found to be 0.9242, as highlighted in the enlarged view within the figure. This indicates that applying a constant-magnitude throttle of \(u = 1\) ensures that Eq.~(\ref{Eq:control_mini_thrust}) always holds. Additionally, once the chaser spacecraft approaches the target state, the minimal required throttle stays very close to zero.
\subsubsection{Impact of the Decay Rate on Guidance Performance}
From Fig.~\ref{Fig:Optimal_decayrate}, it can be observed that the state-dependent decay rate remains very small throughout. Furthermore, the learned guidance policy, represented by $\boldsymbol{\alpha}$ in Eq.~(\ref{Eq:control_mini_direction_1}), depends solely on the neural Lyapunov function $V$ and is independent of the decay rate $\gamma$. It has already been demonstrated that treating the decay rate as a state-dependent parameter can enhance the training process, as shown in Fig.~\ref{Fig:loss_time_optimal}. In this subsection, we further investigate the impact of the decay rate on guidance performance.

\begin{figure}[!htp]
    \centering
    \begin{subfigure}[t]{0.48\textwidth}
      \centering
      \includegraphics[width=\linewidth]{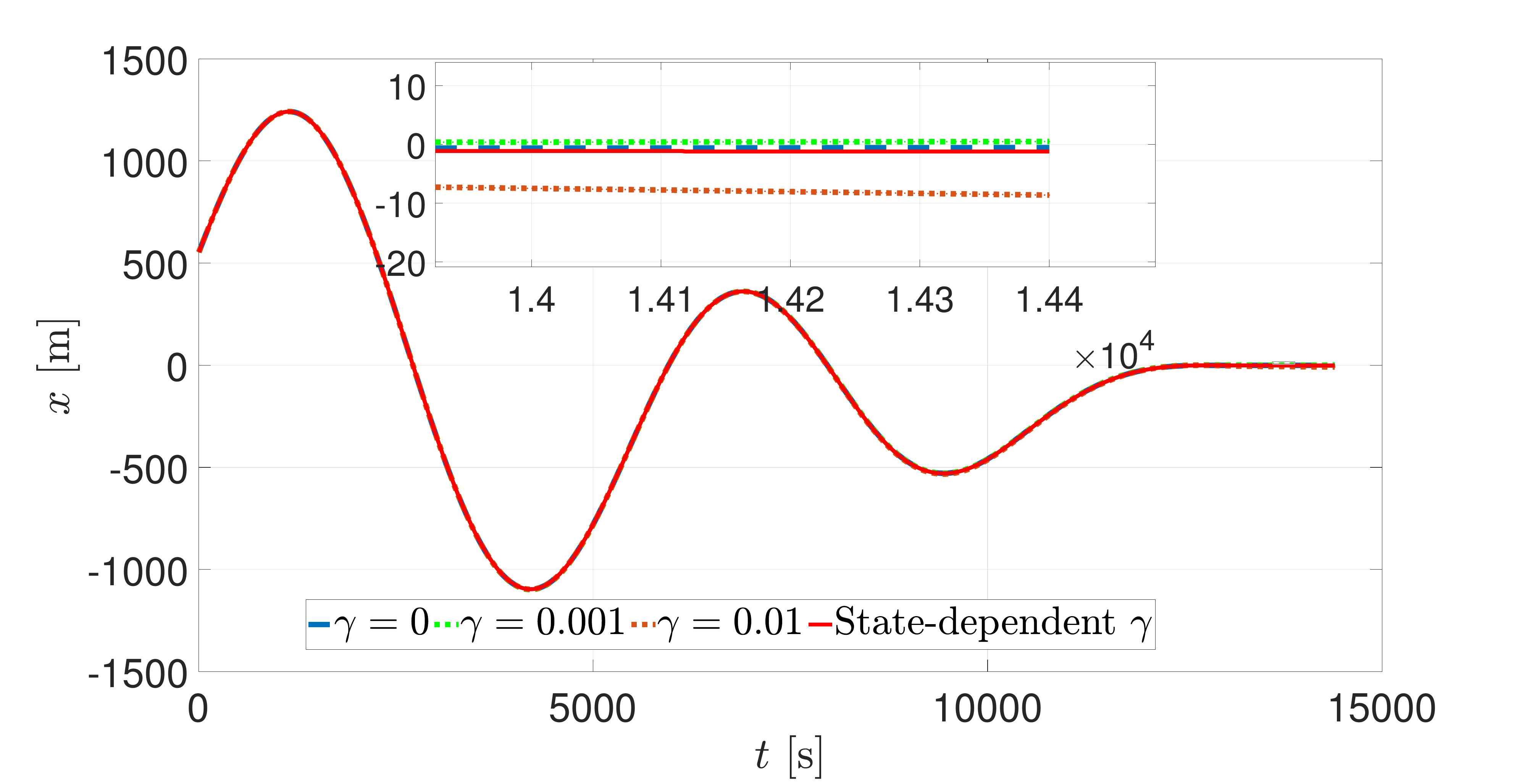}
      \caption{Relative position $x$ profile}
      \label{Fig:CaseA_x_different}
    \end{subfigure}
    \hfill
    \begin{subfigure}[t]{0.48\textwidth}
      \centering
      \includegraphics[width=\linewidth]{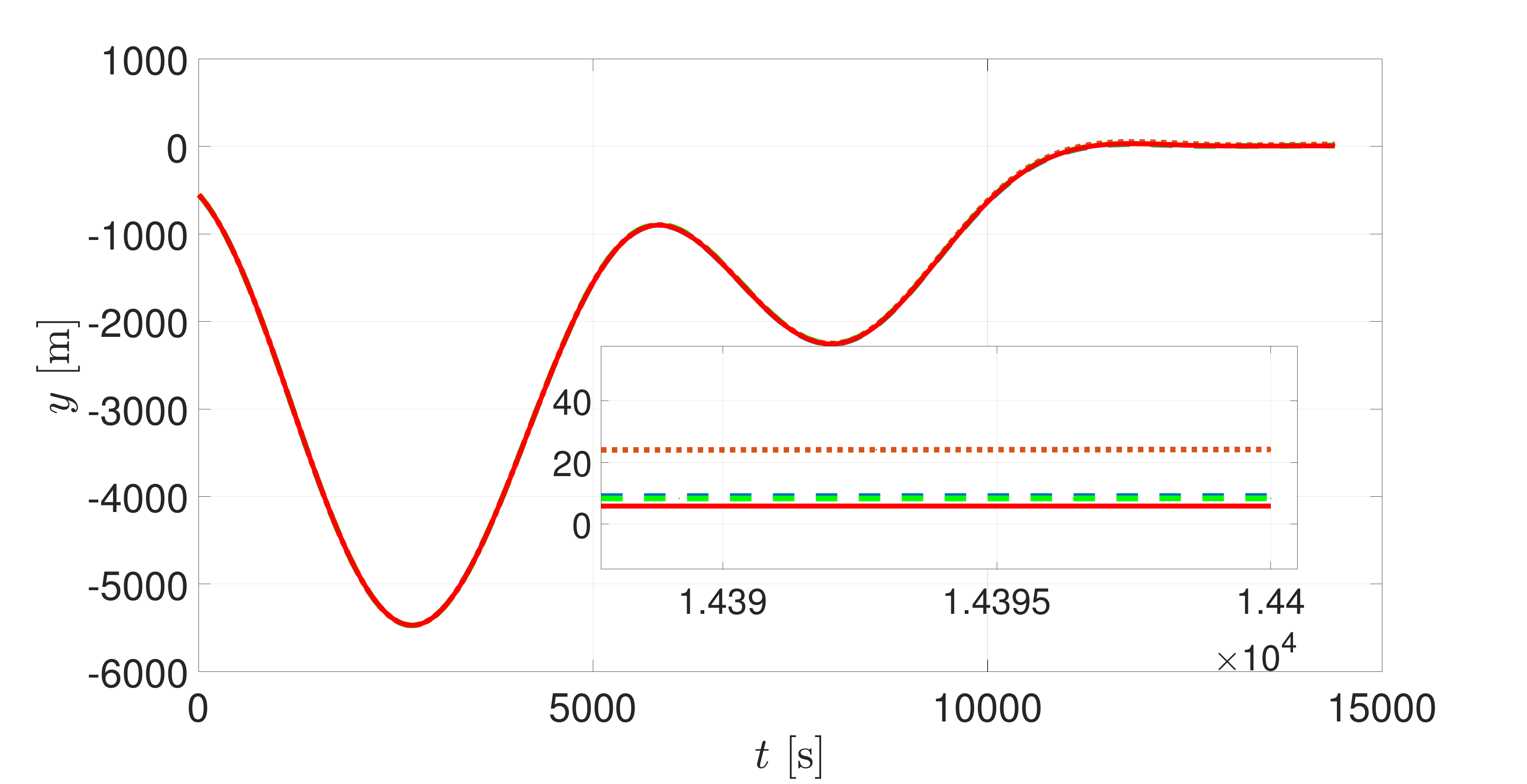}
      \caption{Relative position $y$ profile}
      \label{Fig:CaseA_y_different}
    \end{subfigure}\\[1em]
    
    \begin{subfigure}[t]{0.48\textwidth}
      \centering
      \includegraphics[width=\linewidth]{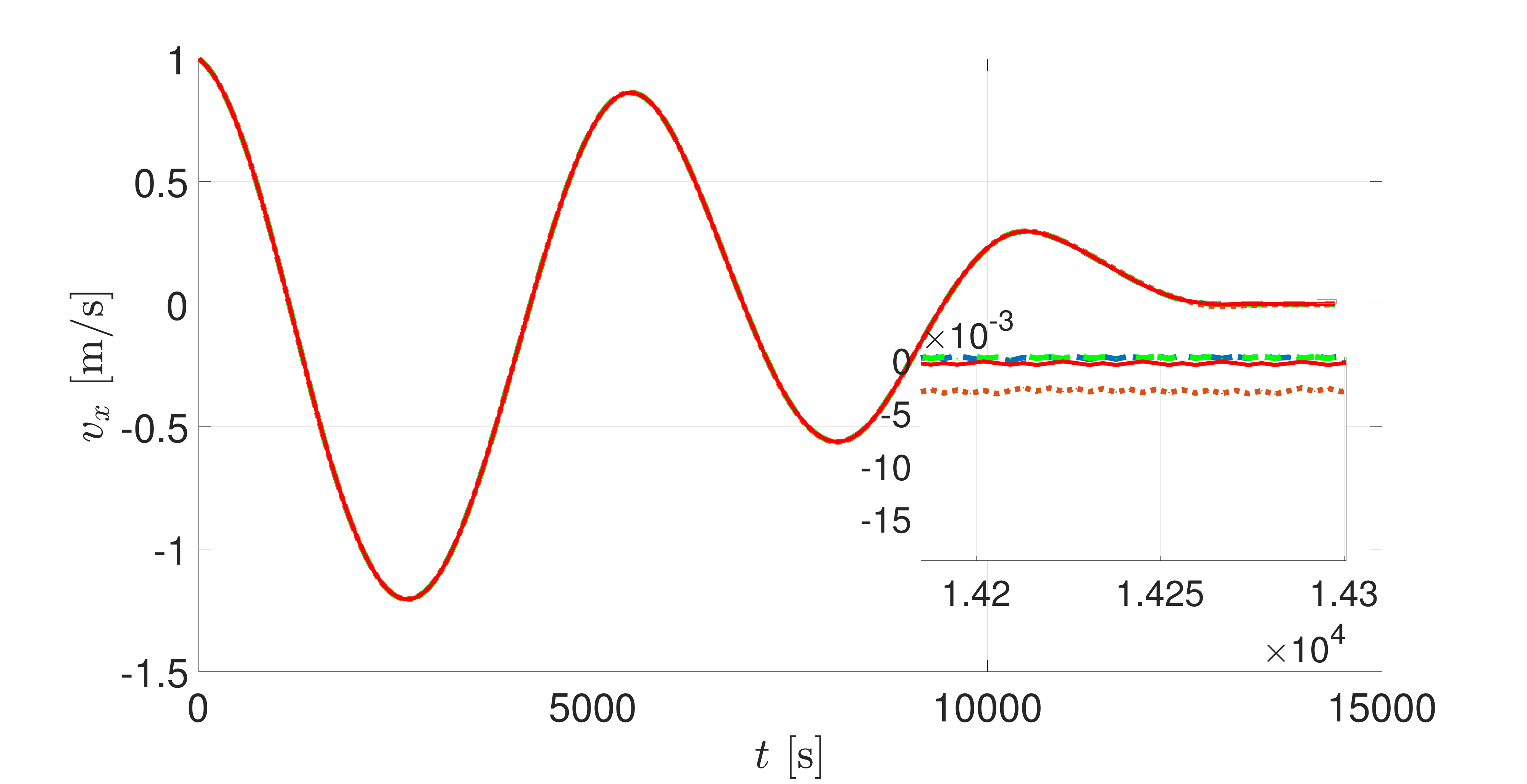}
      \caption{Relative velocity $v_x$ profile}
      \label{Fig:CaseA_vx_different}
    \end{subfigure}
    \hfill
    \begin{subfigure}[t]{0.48\textwidth}
      \centering
      \includegraphics[width=\linewidth]{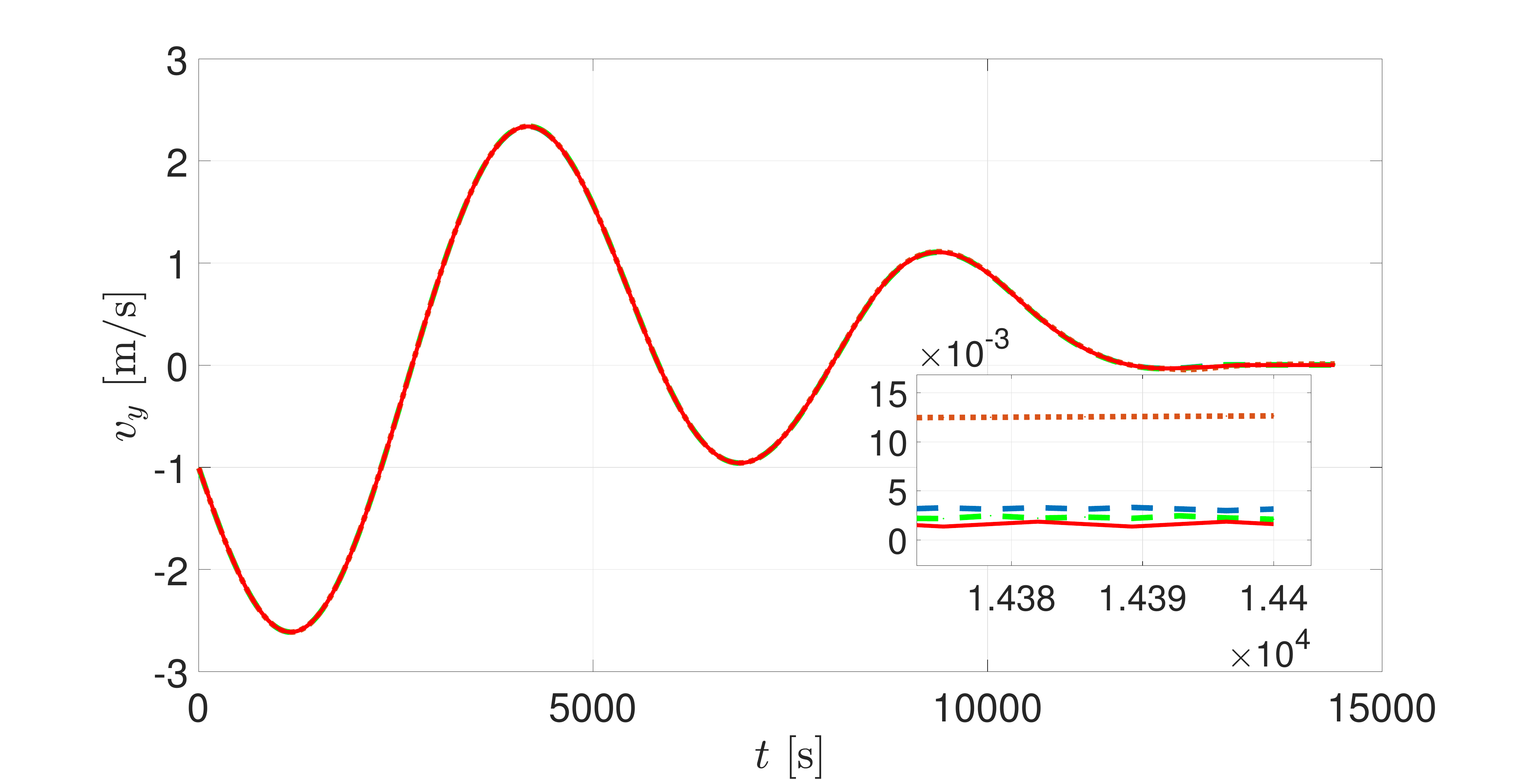}
      \caption{Relative velocity $v_y$ profile}
      \label{Fig:CaseA_vy_different}
    \end{subfigure}
    \caption{Comparison of individual state profiles under different decay rate configurations.}
    \label{Fig:Simulation_results_caseA_different}
    \vspace{-0.5cm}
\end{figure}

We consider the same initial condition as in the previous subsection and apply four guidance strategies using different decay rate configurations: $\gamma = 0$, $\gamma = 0.001$, $\gamma = 0.01$, and a state-dependent $\gamma$. Figure~\ref{Fig:Simulation_results_caseA_different} presents the individual state profiles for these four guidance strategies. The results are nearly identical across all four, but they lead to different state errors, as illustrated in the enlarged views. Table~\ref{diffent_gamma} summarizes the final state errors for the different guidance strategies. It is evident that the final state errors in terms of position $y$ and velocity components $v_x$ and $v_y$ are smaller for the state-dependent decay rate scenario than for the constant decay rate scenarios. Specifically, the final position errors for the four strategies are $9.1230$ m, $8.3283$ m, $25.6999$ m, and $5.9583$ m, respectively. Additionally, the final velocity errors are $0.0031$ m/s, $0.0022$ m/s, $0.0129$ m/s, and $0.0016$ m/s. These results indicate that the strategy with a state-dependent decay rate achieves the best guidance performance.
\begin{table}[h!]
  \centering
  \caption{Final state errors resulted from different guidance strategies}
  \setlength{\tabcolsep}{3.5pt} 
  \renewcommand{\arraystretch}{1.1} 
  \begin{tabular}{ccccc}
  \hline
  & \text{relative position} $x$ (m) & \text{relative position} $y$ (m) & \text{relative velocity} $v_x$ (m/s) & \text{relative velocity} $v_y$ (m/s) \\ 
  \hline
  $\gamma = 0$ & -0.523  & 9.108 & 0.00026 & 0.0031 \\
  $\gamma = 0.001$ & 0.505  & 8.313 & 0.00052 & 0.0021 \\ 
  $\gamma = 0.01$  & -8.612  & 24.214 & -0.0028 & 0.0126 \\
  State-dependent $\gamma$ & -1.230 & 5.831  & -0.0003 & 0.0016 \\ \hline
  \end{tabular}
  \label{diffent_gamma}
\end{table}

\subsubsection{Robustness Performance}
To further assess the guidance performance, we apply the following perturbations to the same initial conditions as described in the previous subsection, i.e.,
  \begin{align}
    \left\{ (\delta{x_0}, \delta{y_0}, \delta{{v_x}_0}, \delta{{v_y}_0}) \mid 
    \delta{x_0} \in [-18,+18]~\text{m}, \quad \delta{y_0} \in [-26,+26]~\text{m}, \right. \nonumber \\
    \left.
    \delta{{v_x}_0} \in [-0.015,+0.015]~\text{m/s}, \quad \delta{{v_y}_0} \in [-0.015,+0.015]~\text{m/s} \right\}.
\label{initial_condition_definition}
\end{align}
We consider the rendezvous successful if the guided final state, using the proposed method, converges within a ball defined by
\begin{equation}
    \mathcal{B} = \left\{ (x, y, v_x, v_y) \mid \sqrt{x^2 + y^2} < 10~\text{m} \quad \text{and} \quad \sqrt{v_x^2 + v_y^2} < 0.02~\text{m/s}\right\}.
    \label{ball_definition}
\end{equation}
    
\begin{figure}[!htp]
  \centering
  \begin{subfigure}[t]{0.45\textwidth}
  \centering
  \includegraphics[width = \linewidth]{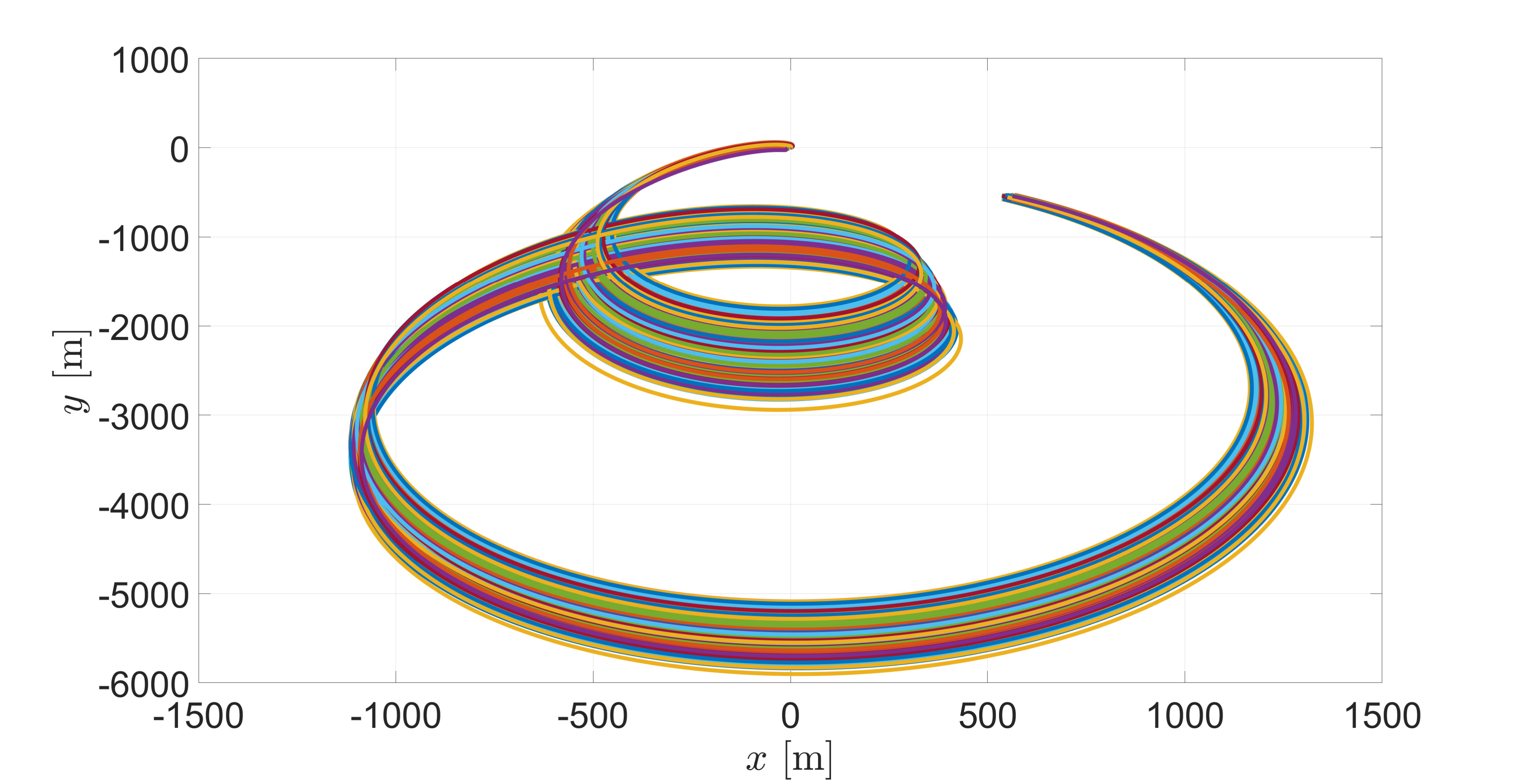}
  \caption{Relative position components}
  \label{Fig:CaseB_many_traj}
  \end{subfigure}
  \hspace{0.05\textwidth}
  \begin{subfigure}[t]{0.45\textwidth}
  \centering
  \includegraphics[width = \linewidth]{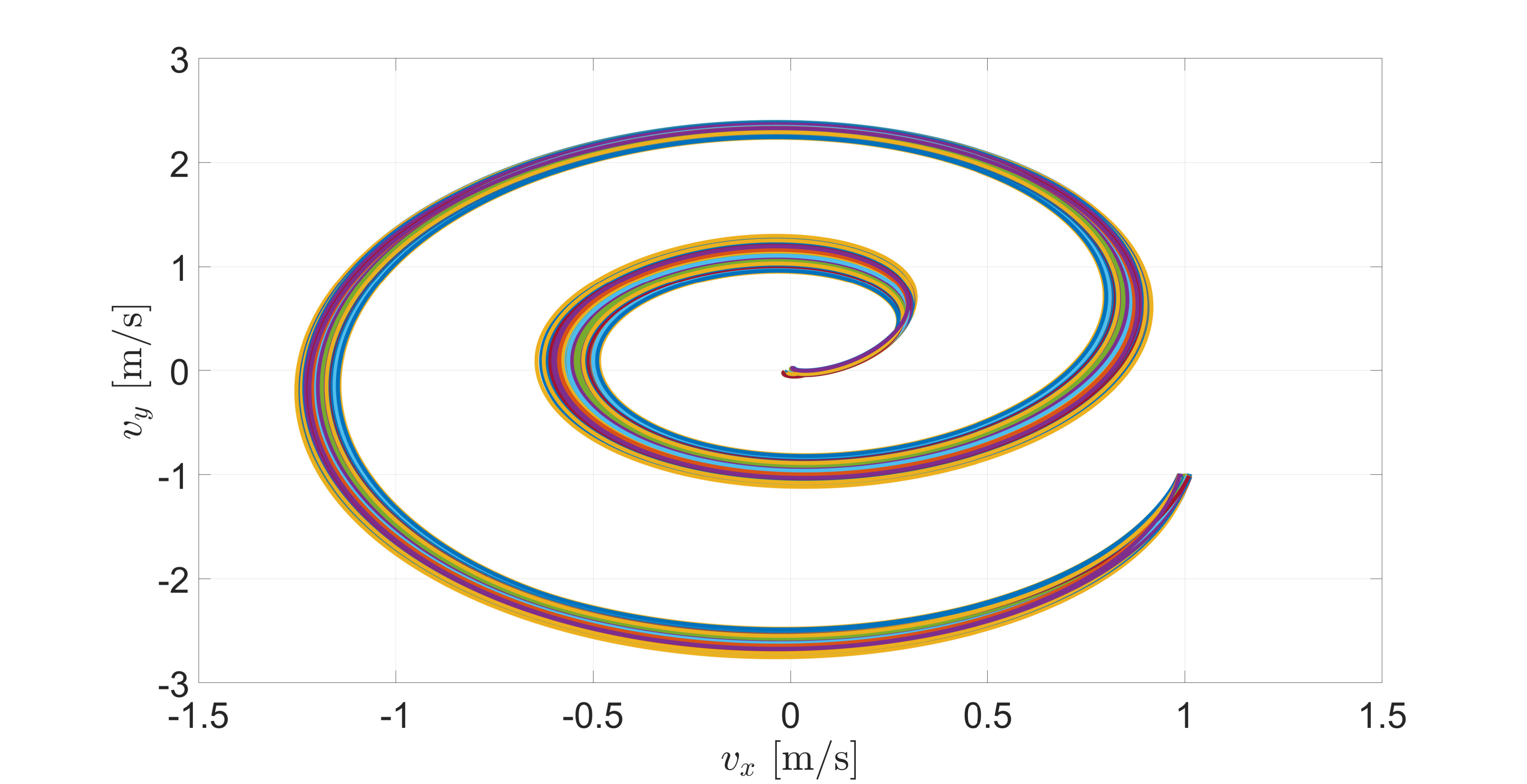}
  \caption{Relative velocity components}\label{Fig:CaseB_many_vel}
  \end{subfigure}
  \caption{Relative position and velocity components under initial condition perturbations.}
\label{Fig:CaseA_optimal_many}
\vspace{-0.5cm}
  \end{figure}

\begin{figure}[!htp]
  \begin{center}
  \includegraphics[scale=0.15]{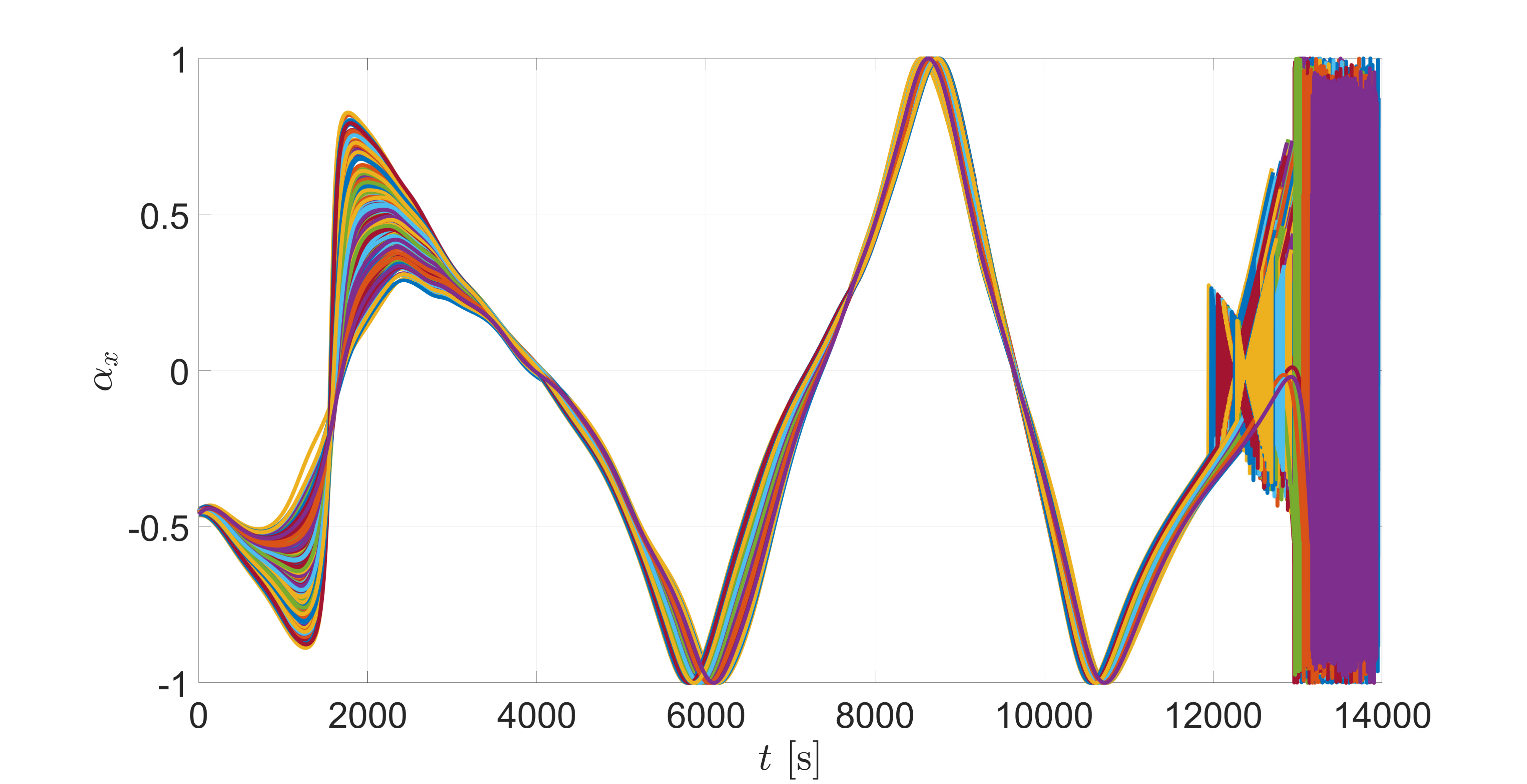}
  \caption{Thrust direction component $\alpha_x$ profiles under initial condition perturbations.}\label{Fig:many_time_directions}
  \end{center}
  \vspace{-0.5cm}
\end{figure}

\begin{figure}[!htp]
    \begin{center}
    \includegraphics[scale=0.2]{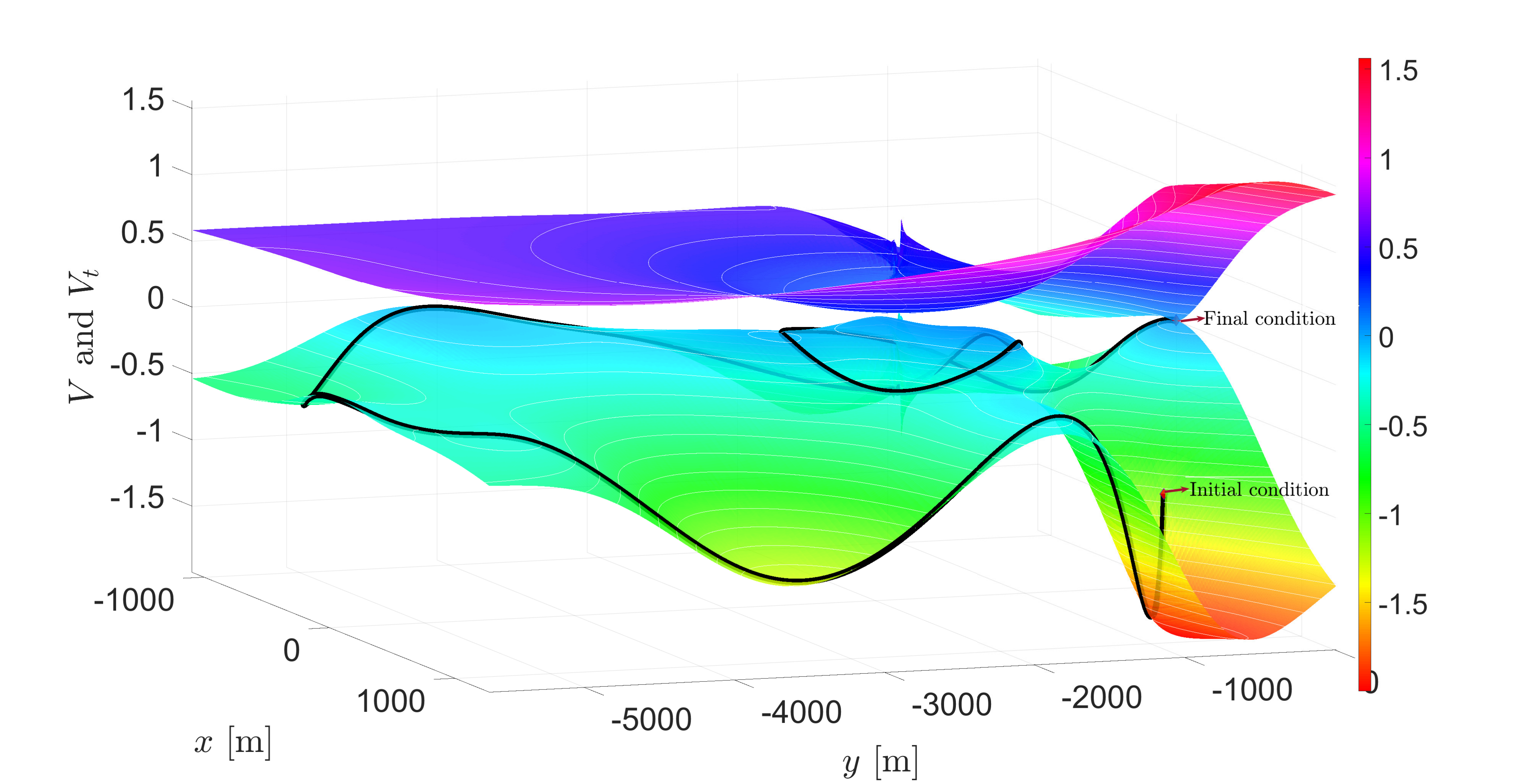}
    \caption{Lyapunov function and its time derivative w.r.t. the relative position components.}\label{Fig:CaseB_many_traj_lya}
    \end{center}
    \vspace{-0.5cm}
\end{figure}

\begin{figure}[!htp]
  \begin{center}
  \includegraphics[scale=0.2]{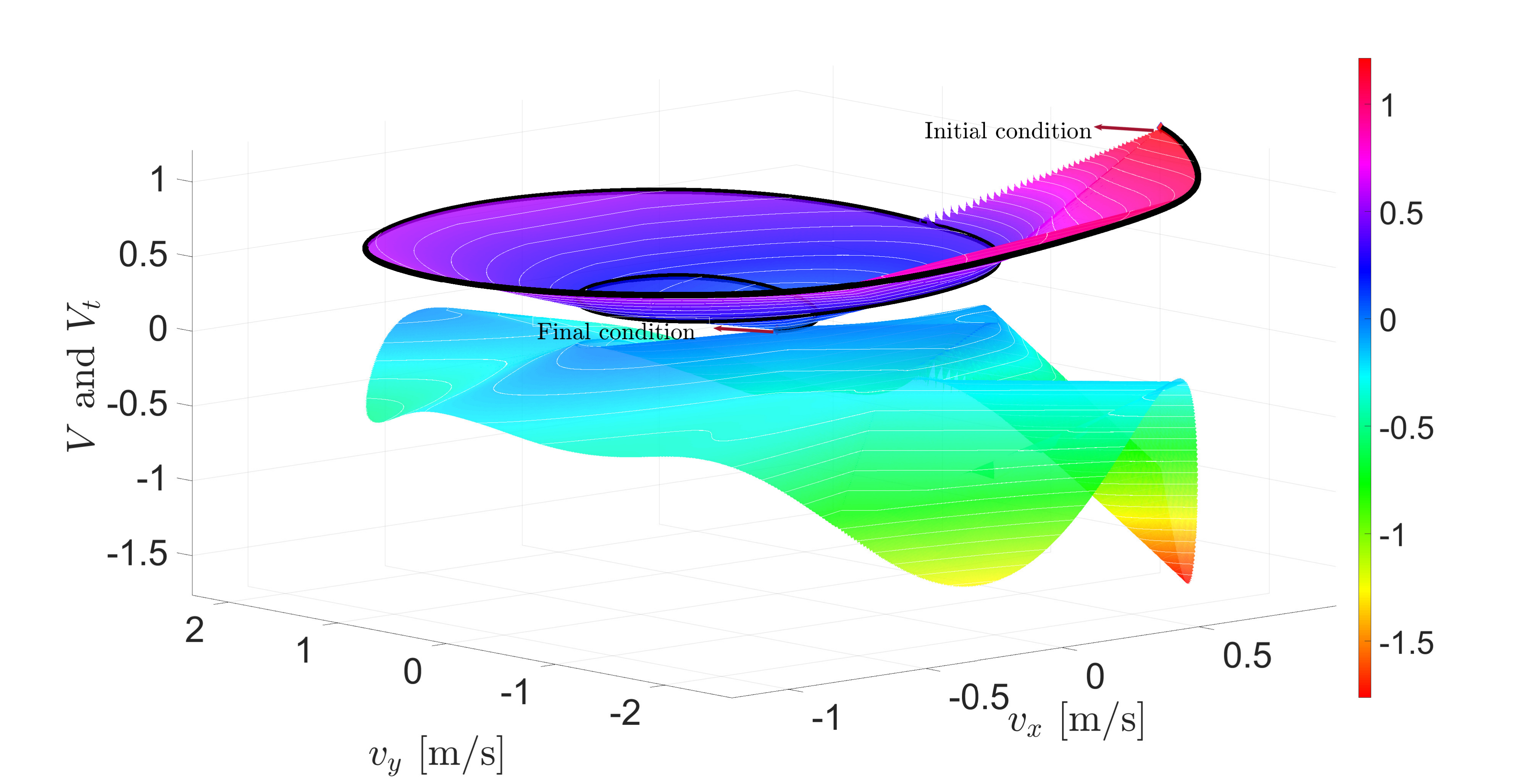}
  \caption{Lyapunov function and its time derivative w.r.t. the relative velocity components.}\label{Fig:CaseB_many_vel_lya}
  \end{center}
  \vspace{-0.5cm}
\end{figure}

A total of 200 random perturbated initial conditions are tested. All 200 cases successfully achieve rendezvous using the proposed method, as demonstrated by Fig.~\ref{Fig:CaseA_optimal_many}. 
Figure~\ref{Fig:many_time_directions} displays the thrust direction profiles generated by the proposed method for perturbated initial conditions. Additionally, Figs.~\ref{Fig:CaseB_many_traj_lya} and \ref{Fig:CaseB_many_vel_lya} show the Lyapunov function (upper surface) and its time derivative (lower surface) as functions of the relative position and velocity components, respectively. These smooth surfaces are generated through quadratic or cubic interpolation, with the relative position and velocity components serving as the grids for interpolating both the Lyapunov function and its time derivative. A representative trajectory is emphasized with a solid black line to highlight the results. The Lyapunov function remains positive throughout the trajectory, except at the final condition. Despite the linear nature of the system dynamics, both the Lyapunov function and its time derivative exhibit significant nonlinear complexity, as they must guarantee convergence to the equilibrium point within minimal time.
\subsection{Fuel-Optimal Problem}
\subsubsection{Stability and Optimality Performance}
\begin{figure}[!htp]
  \centering
  \begin{subfigure}[t]{0.48\textwidth}
  \centering
  \includegraphics[width = \textwidth]{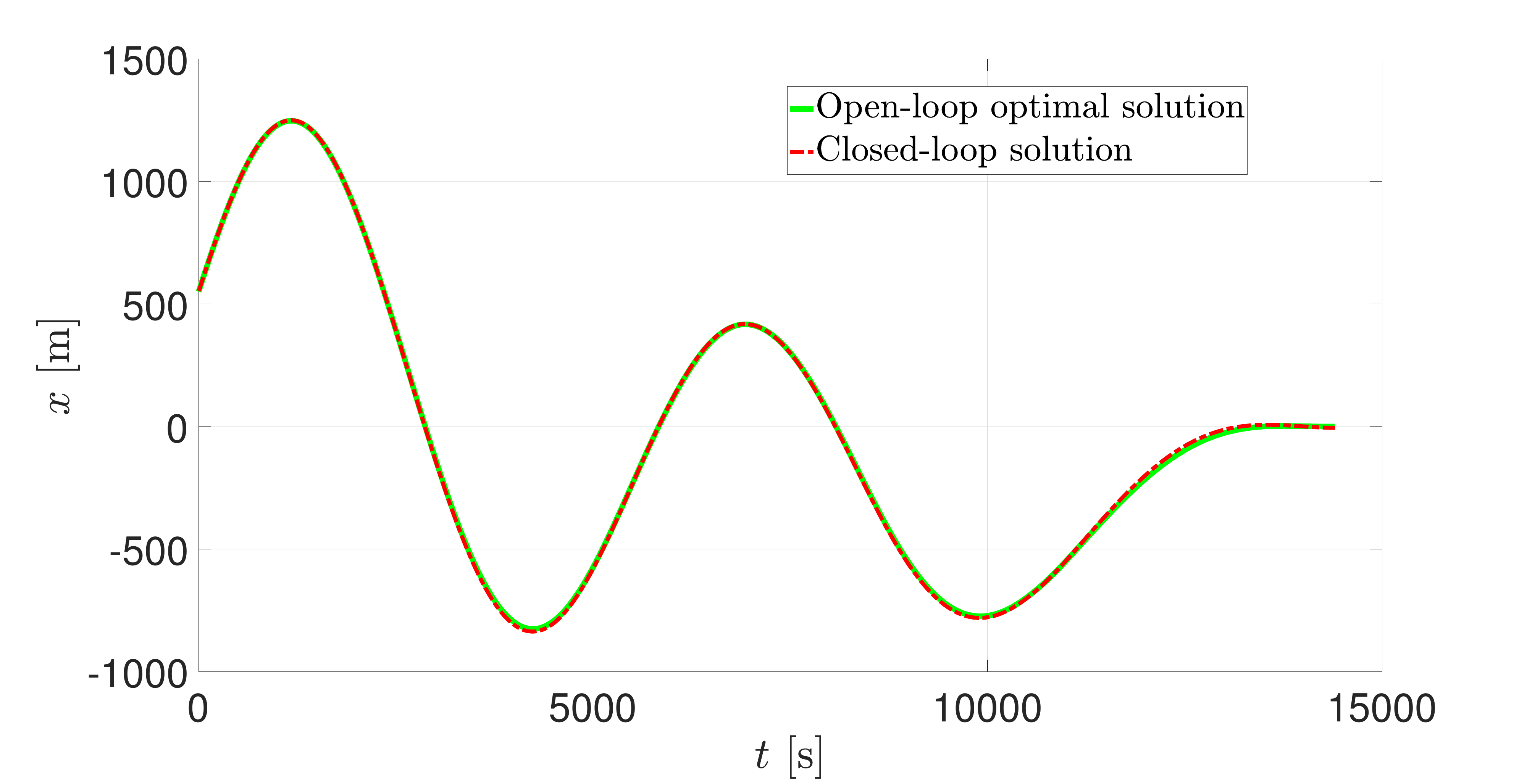}
  \caption{Relative position $x$ profile}
  \label{Fig:CaseA_x_fuel}
  \end{subfigure}
  \hfill
  \begin{subfigure}[t]{0.48\textwidth}
  \centering
  \includegraphics[width = \textwidth]{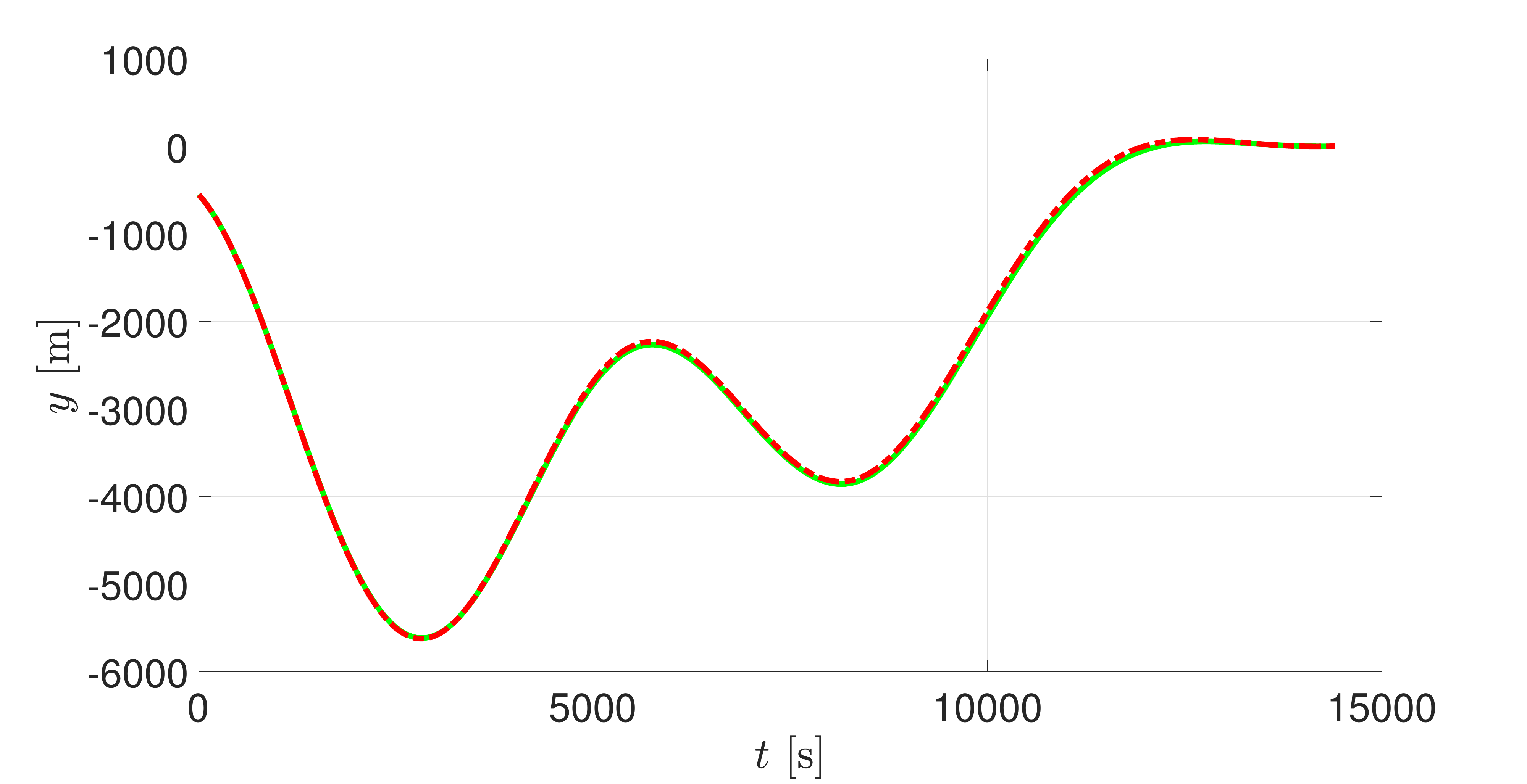}
  \caption{Relative position $y$ profile}
  \label{Fig:CaseA_y_fuel}
  \end{subfigure}\\[1em]

  \begin{subfigure}[t]{0.48\textwidth}
  \centering
  \includegraphics[width = \textwidth]{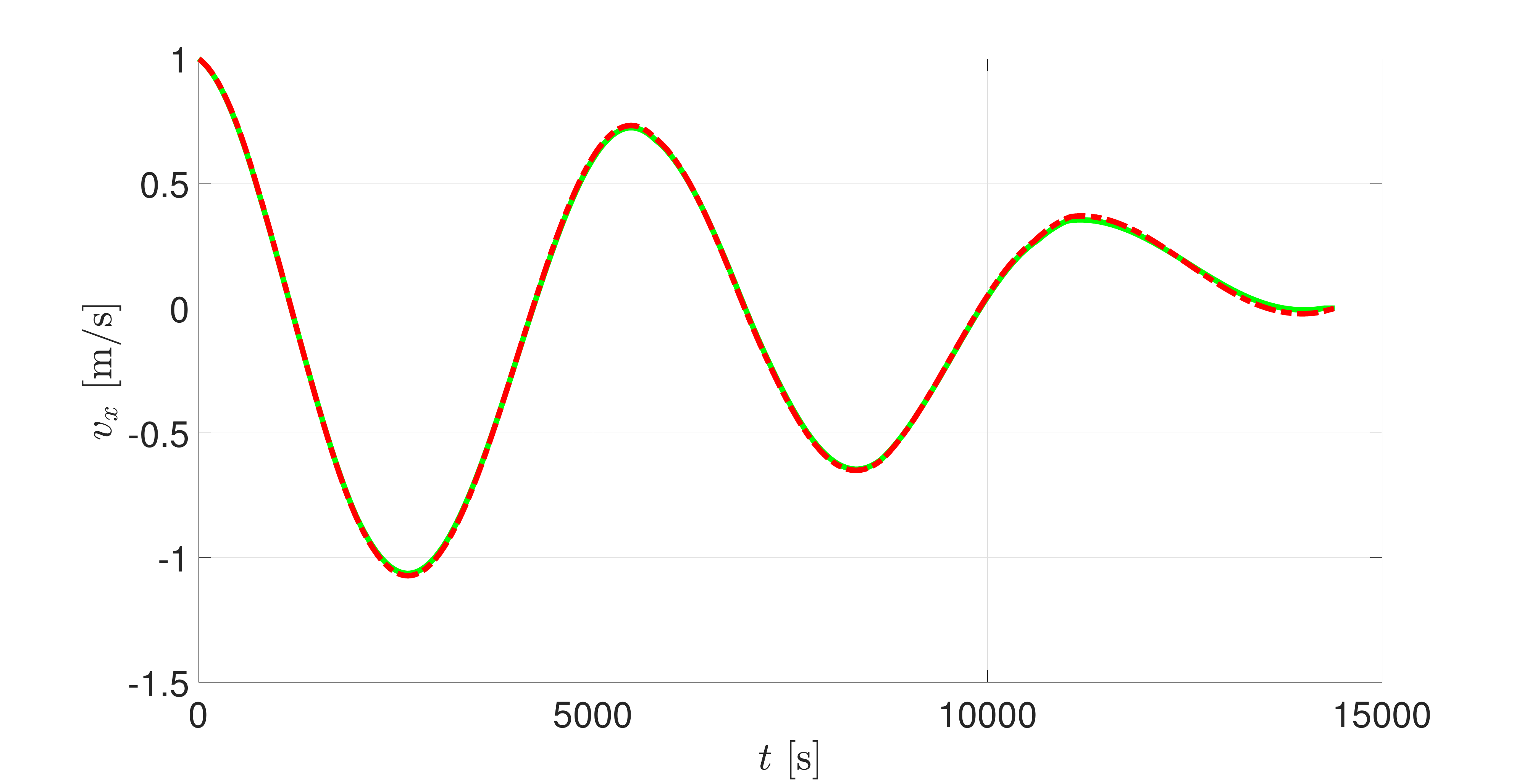}
  \caption{Relative velocity $v_x$ profile}
  \label{Fig:CaseA_vx_fuel}
  \end{subfigure}
  \hfill
  \begin{subfigure}[t]{0.48\textwidth}
  \centering
  \includegraphics[width = \textwidth]{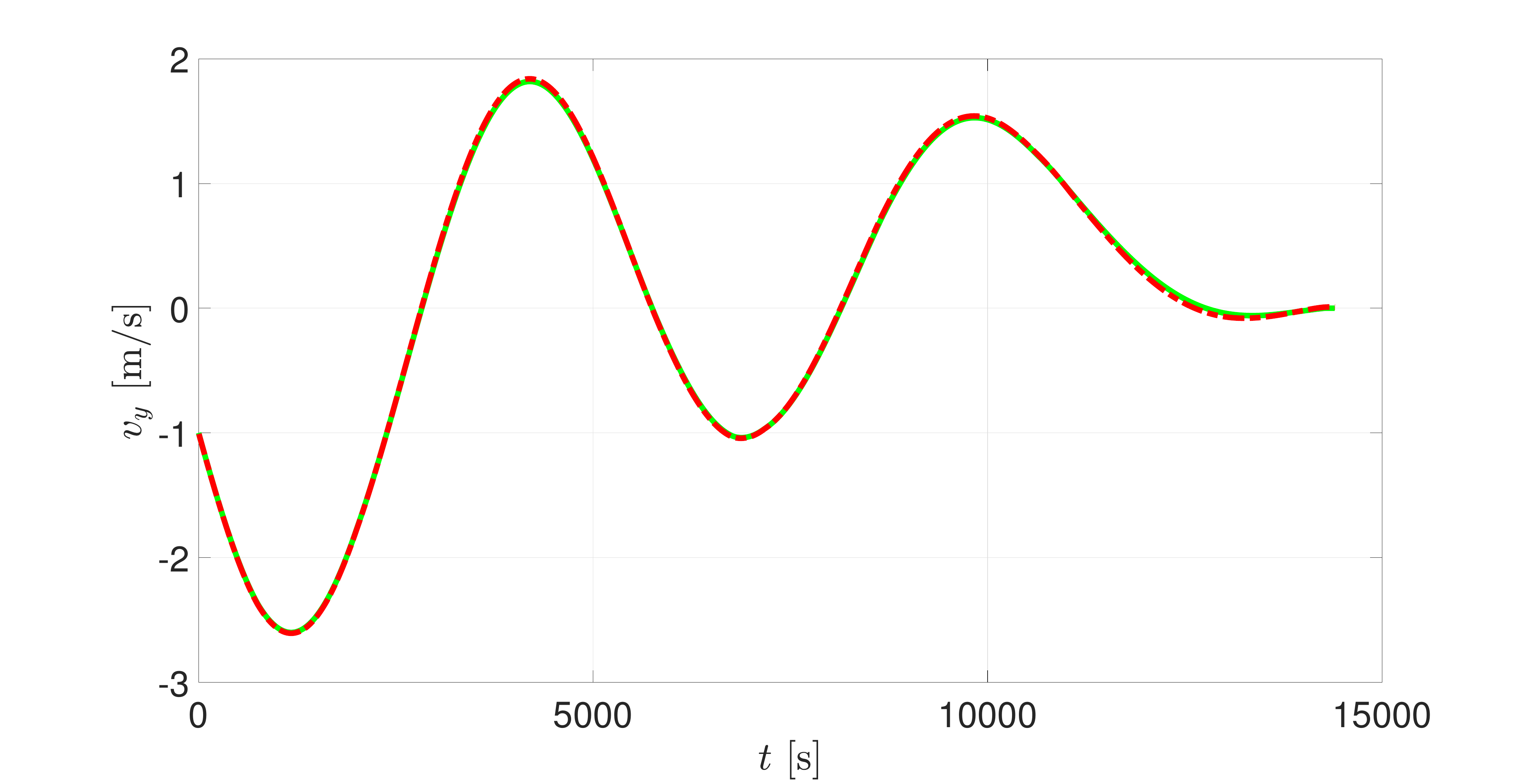}
  \caption{Relative velocity $v_y$ profile}
  \label{Fig:CaseA_vy_fuel}
  \end{subfigure}
  \caption{Comparison of individual state profiles for the fuel-optimal problem.}
  \label{Fig:Simulation_results_fuel_optimal}
\end{figure}

We consider the same initial condition as the time-optimal problem, but the rendezvous time is fixed to be $1 4,400$ s. When using the proposed guidance strategy in Fig.~\ref{Fig:Closed_loopfueloptimal}, the guidance process terminates when the time-to-go $t_g$ reaches zero. Figure~\ref{Fig:Simulation_results_fuel_optimal} compares the results obtained from the proposed and indirect shooting methods. One can observe that both the relative position and velocity components have an almost perfect match. The final state of the chaser spacecraft guided by the proposed method is $[-5.4~\text{m}, 1~\text{m}, 0~\text{m/s}, 0.0092~\text{m/s}]^T$, which is very close to the target state.

\begin{figure}[!htp]
    \centering
    \begin{subfigure}[t]{0.45\textwidth}
      \centering
      \includegraphics[width = \linewidth]{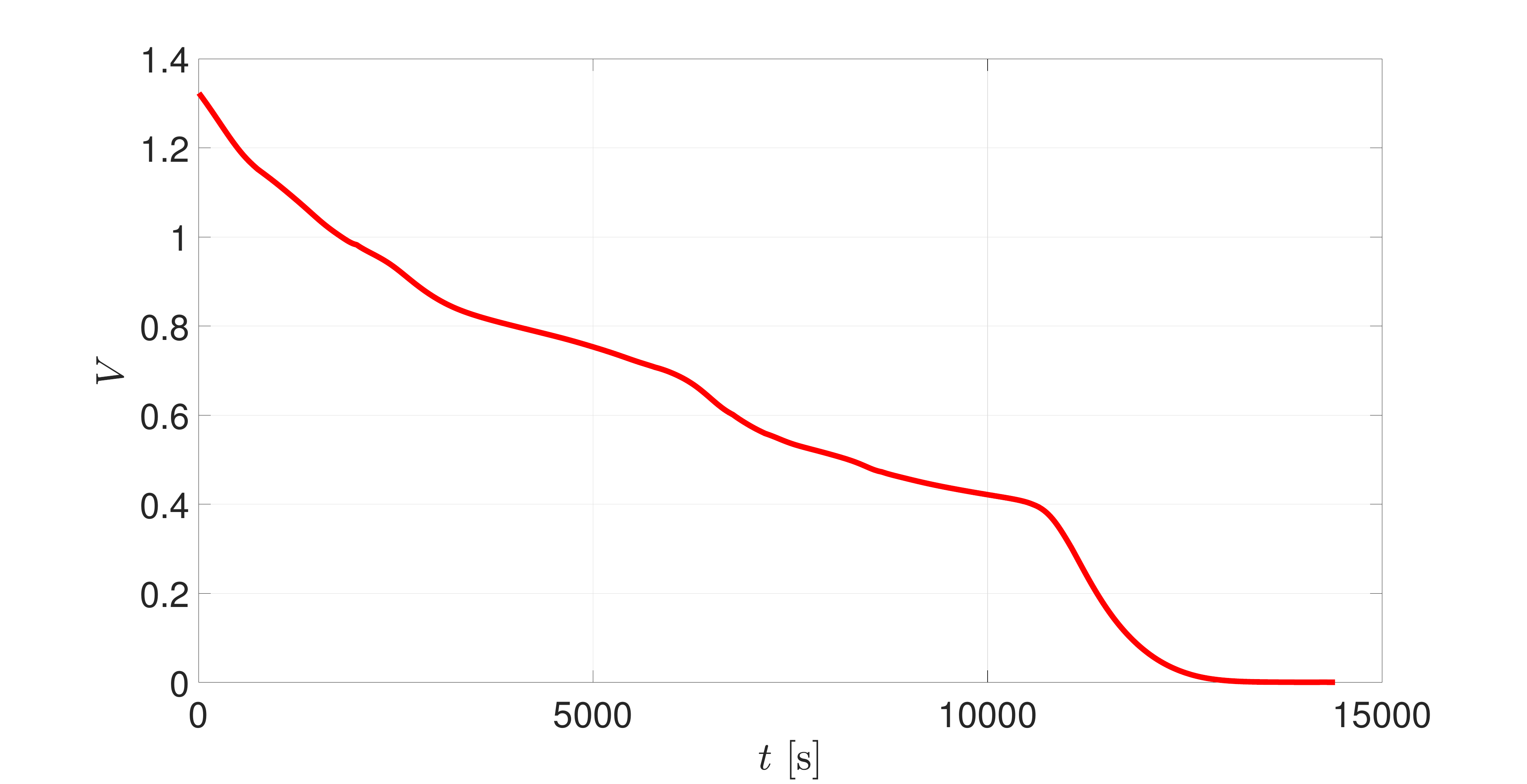}
      \caption{Lyapunov function}
      \label{Fig:CaseA_u_fuel}
    \end{subfigure}
    \hspace{0.05\textwidth}  
    \begin{subfigure}[t]{0.45\textwidth}
      \centering
      \includegraphics[width = \linewidth]{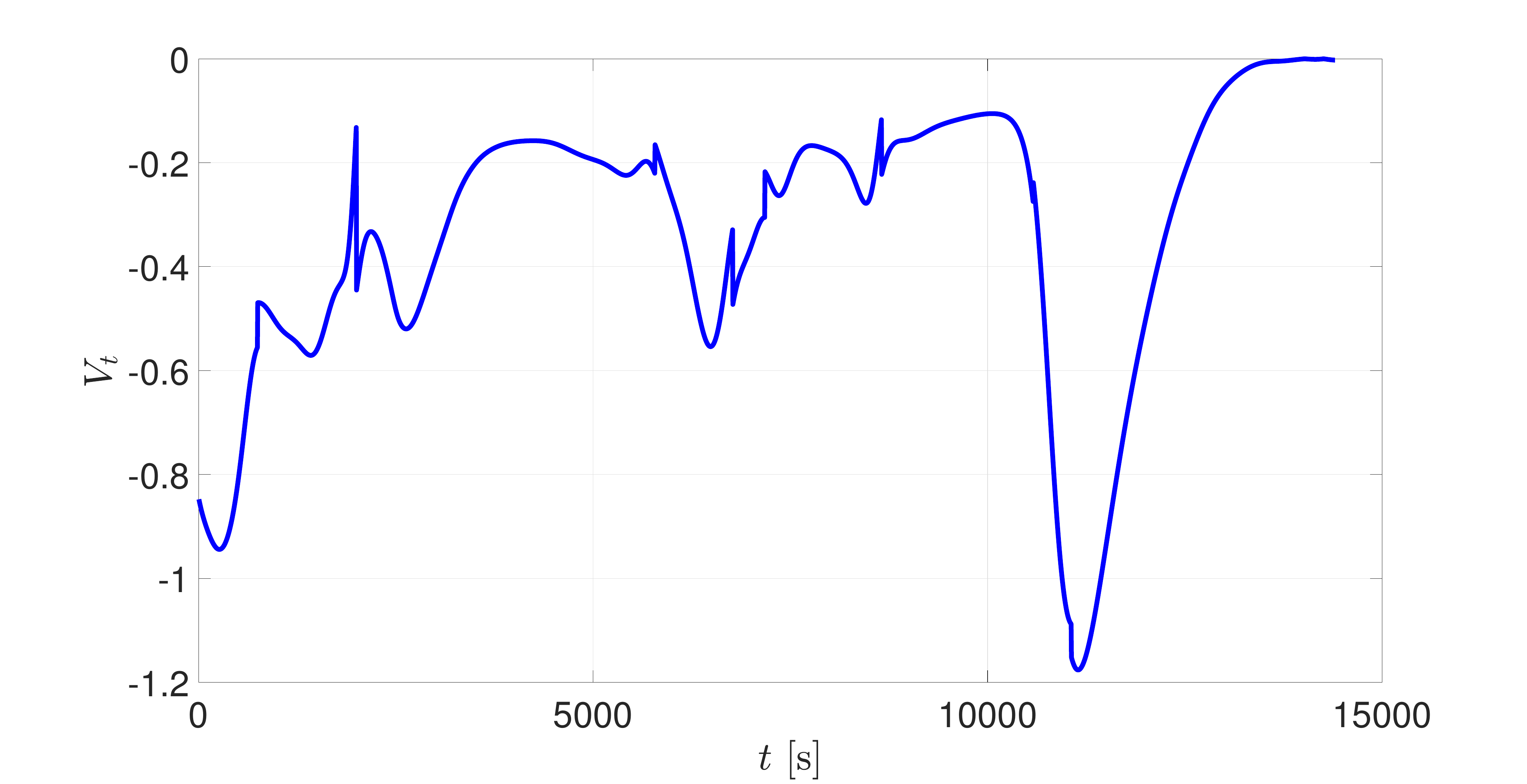}
      \caption{Lyapunov function's time derivative}
      \label{Fig:CaseA_J_fuel}
    \end{subfigure}
    \caption{Lyapunov function and its time derivative profiles for the fuel-optimal problem.}
    \label{Fig:CaseA_control_profile_fuel}
    \vspace{-0.5cm}
  \end{figure}

\begin{figure}[!htp]
  \begin{center}
  \includegraphics[scale=0.18]{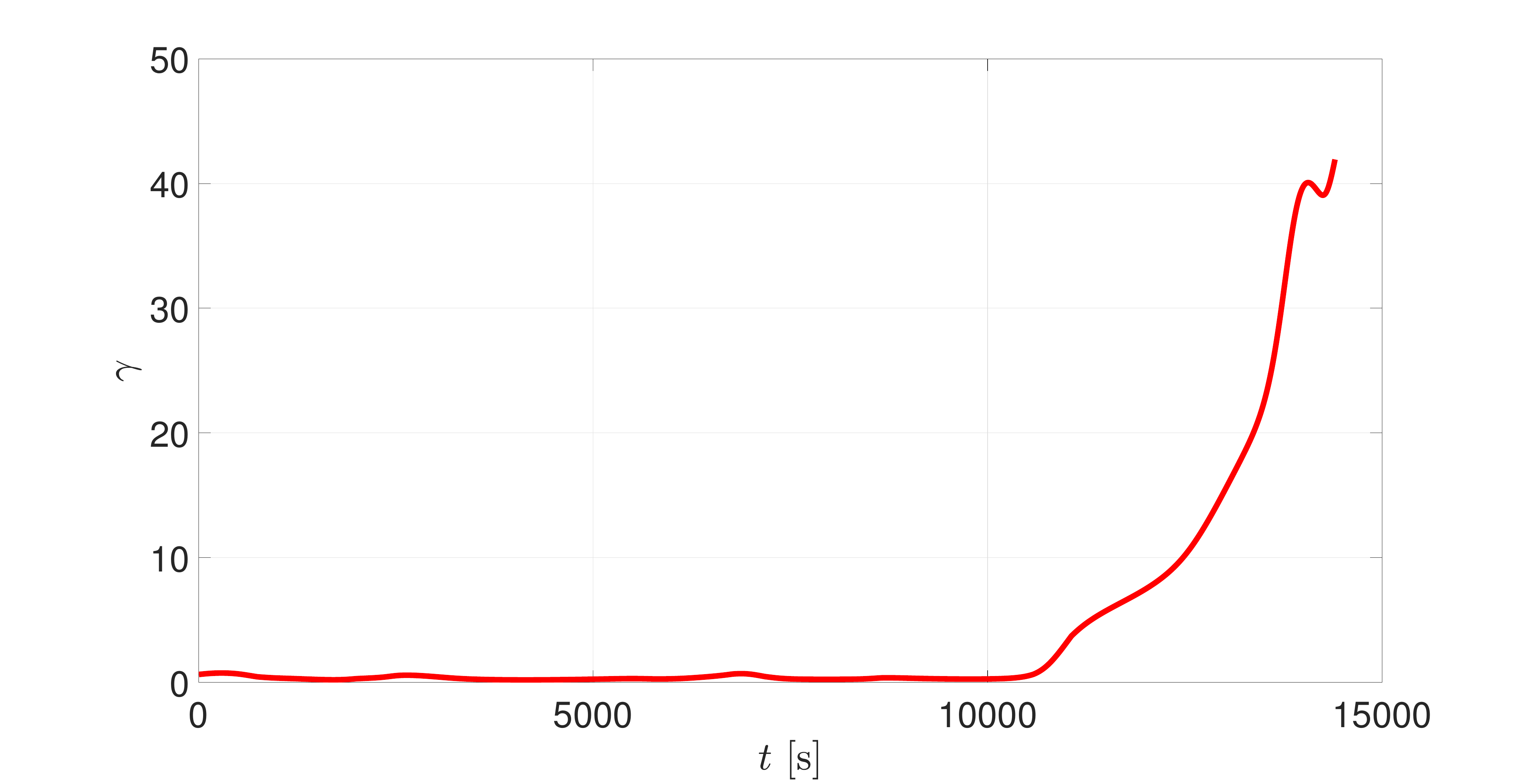}
  \caption{State-dependent decay rate profile for the fuel-optimal problem.}\label{Fig:Optimal_decayrate_fuel}
  \end{center}
  \vspace{-0.5cm}
\end{figure}

\begin{figure}[!htp]
  \begin{center}
  \includegraphics[scale=0.18]{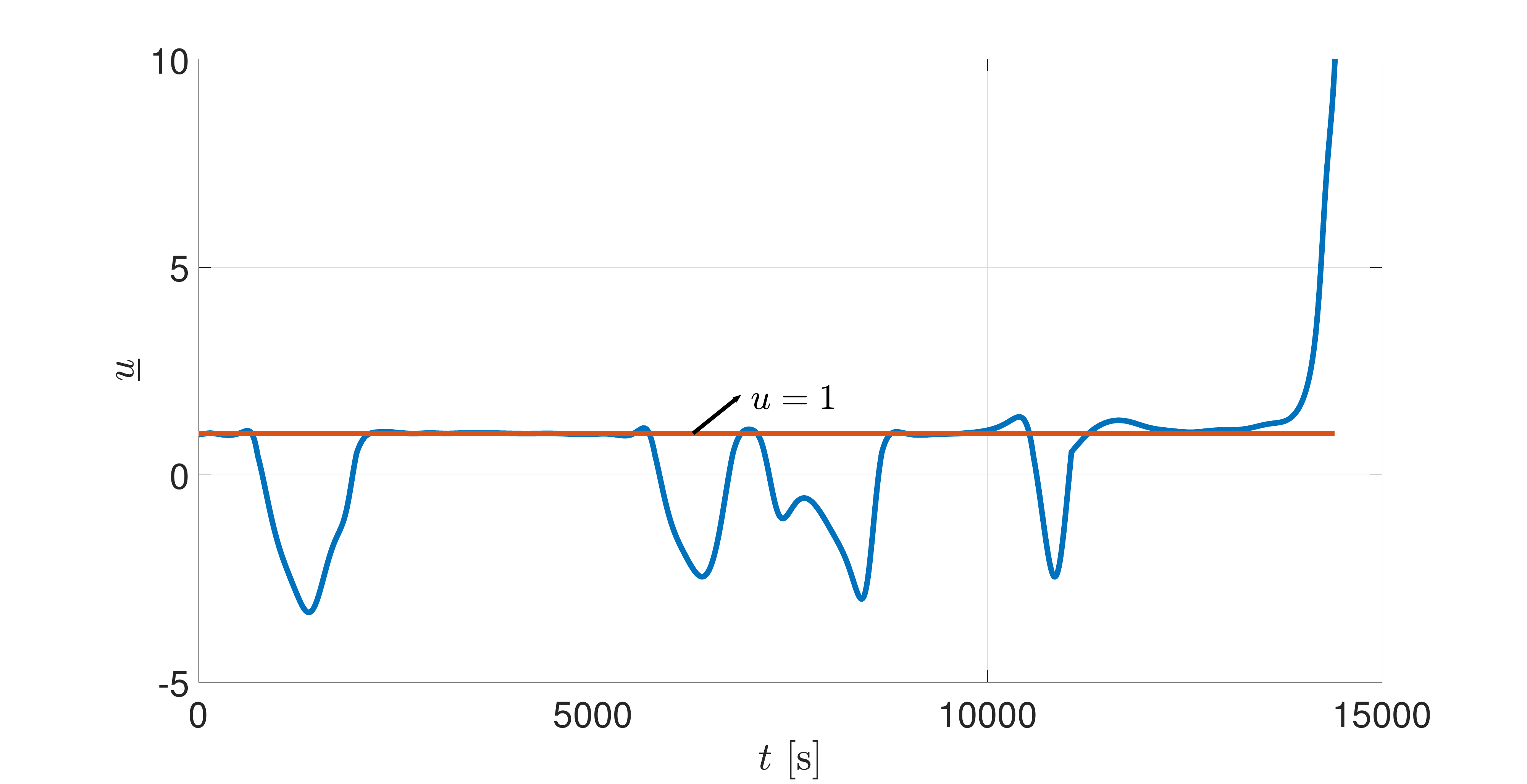}
  \caption{Minimal required throttle profile for the fuel-optimal problem.}\label{Fig:minimal_required_control_fuel}
  \end{center}
  \vspace{-0.5cm}
\end{figure}

Figure~\ref{Fig:CaseA_control_profile_fuel} shows the Lyapunov function and its time derivative along the rendezvous trajectory. Compared to the time-optimal problem in Fig.~\ref{Fig:CaseA_control_profile}, the  Lyapunov function's time derivative here is generally less smooth. The decay rate, as illustrated in Fig.~\ref{Fig:Optimal_decayrate_fuel}, exhibits a marked difference from the time-optimal scenario shown in Fig.~\ref{Fig:Optimal_decayrate}. Unlike the consistently small decay rate observed in the time-optimal problem, the decay rate significantly increases as the chaser spacecraft approaches the target spacecraft. A similar trend is observed in the minimal required throttle profile, as depicted in Fig.~\ref{Fig:minimal_required_control_fuel}. It is worth noting that while the maximal admissible constraint of \( u = 1 \) is occasionally exceeded, particularly as the chaser spacecraft nears the target, the Lyapunov function maintains a consistent decrease, as confirmed by Fig.~\ref{Fig:CaseA_J_fuel}.

\begin{figure}[!htp]
  \centering
  \begin{subfigure}[t]{0.7\textwidth}
  \centering
  \includegraphics[width = \textwidth]{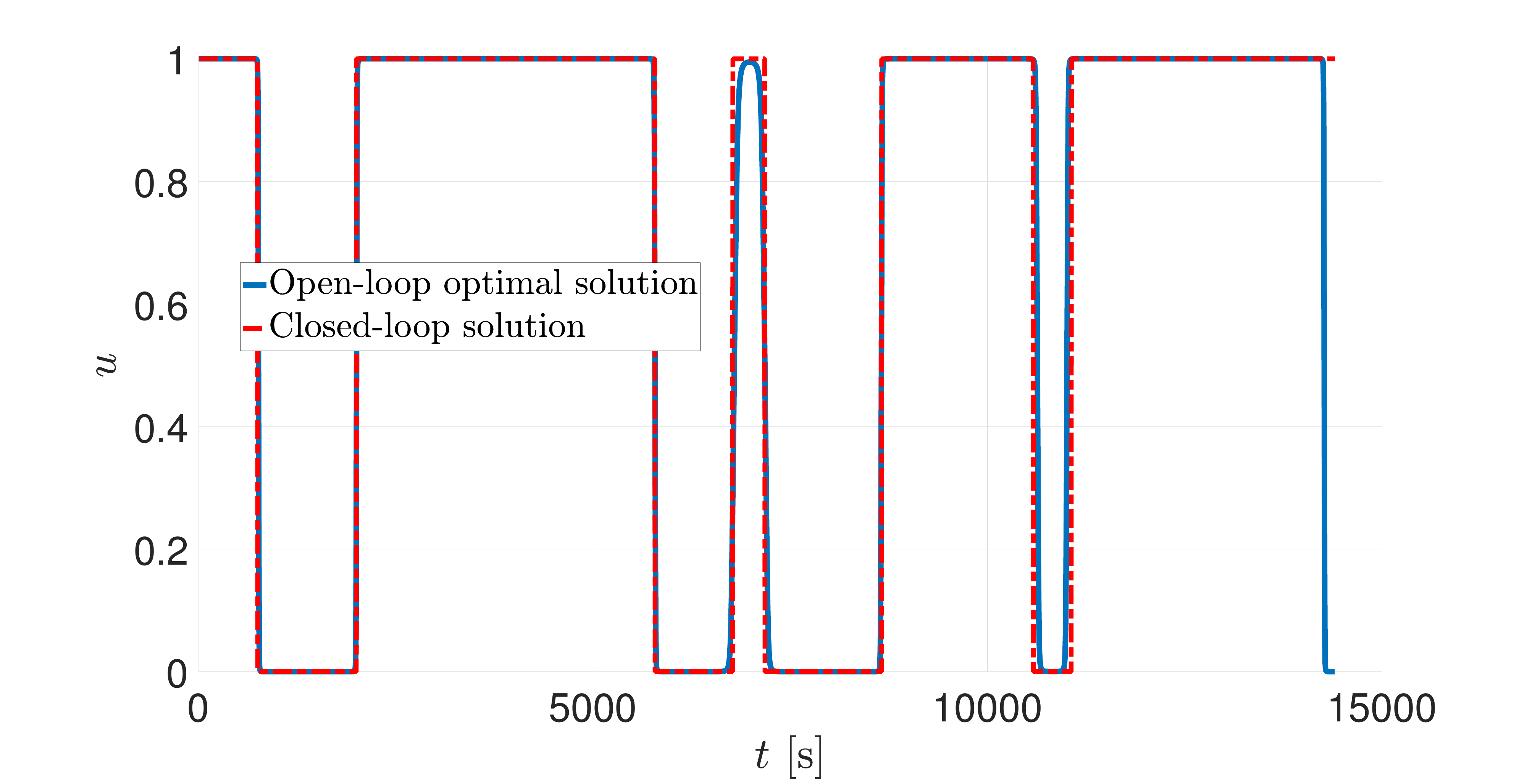}
  \caption{Throttle}
  \label{Fig:Fuel_optimal_u}
  \end{subfigure}
  \hspace{0.05\textwidth}
  \begin{subfigure}[t]{0.7\textwidth}
  \centering
  \includegraphics[width = \textwidth]{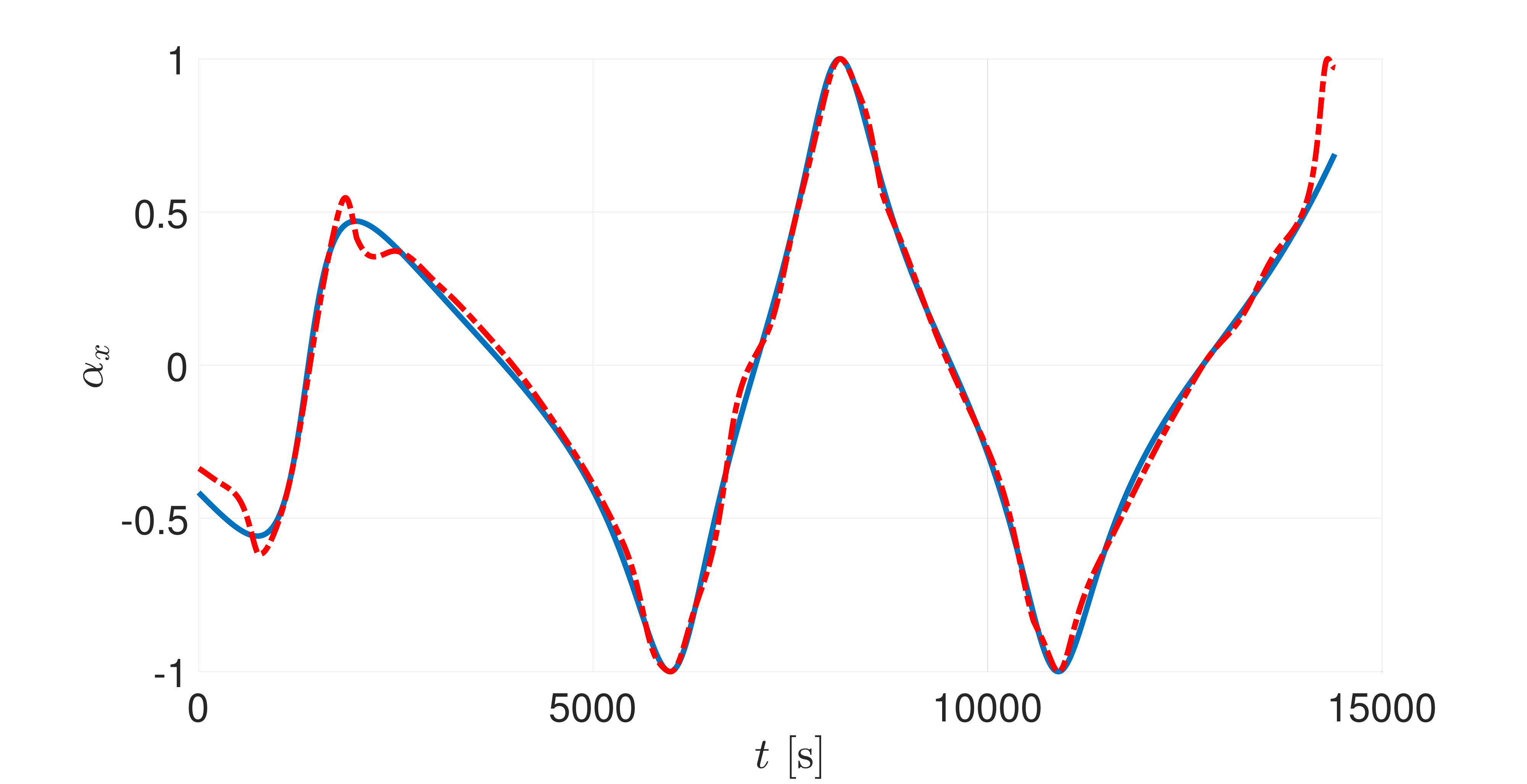}
  \caption{Thrust direction component $\alpha_x$}
  \label{Fuel_optimal_alpha}
  \end{subfigure}
  \caption{Comparison between the learned guidance policy and the open-loop optimal control policy.}
\label{Fig:CaseA_control_profile_fuel_comparison}
\vspace{-0.5cm}
  \end{figure}

By leveraging the learned Lyapunov function and decay rate, the guidance policy—comprising the throttle and thrust direction—is compared with the open-loop optimal solutions illustrated in Fig.~\ref{Fig:CaseA_control_profile_fuel_comparison}. The open-loop optimal throttle exhibits nine switches, and the closed-loop strategy closely matches the first eight switches, deviating only at the final one. However, the closed-loop thrust direction is less accurate than the open-loop optimal solution. This discrepancy can partly be attributed to the training process prioritizing the throttle over the thrust direction in the loss function, as defined in Eq.~(\ref{EQ:existing_method_paper_fuel}). It is worth noting that the discontinuous bang-bang throttle is challenging to replicate using conventional supervised learning methods (see Refs. \cite{WANG2024,origer2023guidance}). However, with the proposed strategy, the replicated throttle not only maintains a bang-bang nature but also closely approximates the open-loop bang-bang solution. Furthermore, the abrupt changes in throttle result in less smooth variations in the Lyapunov function's time derivative, as illustrated in Fig.~\ref{Fig:CaseA_J_fuel}. 

We further compare the fuel consumption \( \Delta V \), which  is computed by  
\[
\Delta V = \int_0^{t_f} u \frac{T_\text{m}}{m} \, dt.
\] 
The optimal value of \( \Delta V \) obtained using the indirect shooting method is \( 0.8467~\text{m/s} \), while the proposed method yields \( 0.8499~\text{m/s} \). This corresponds to a negligible penalty of just \( 0.3779\% \), highlighting the effectiveness of the proposed approach.

\subsubsection{Robustness Performance}
\begin{figure}[!htp]
  \centering
  \begin{subfigure}[t]{0.45\textwidth}
  \centering
  \includegraphics[width = \textwidth]{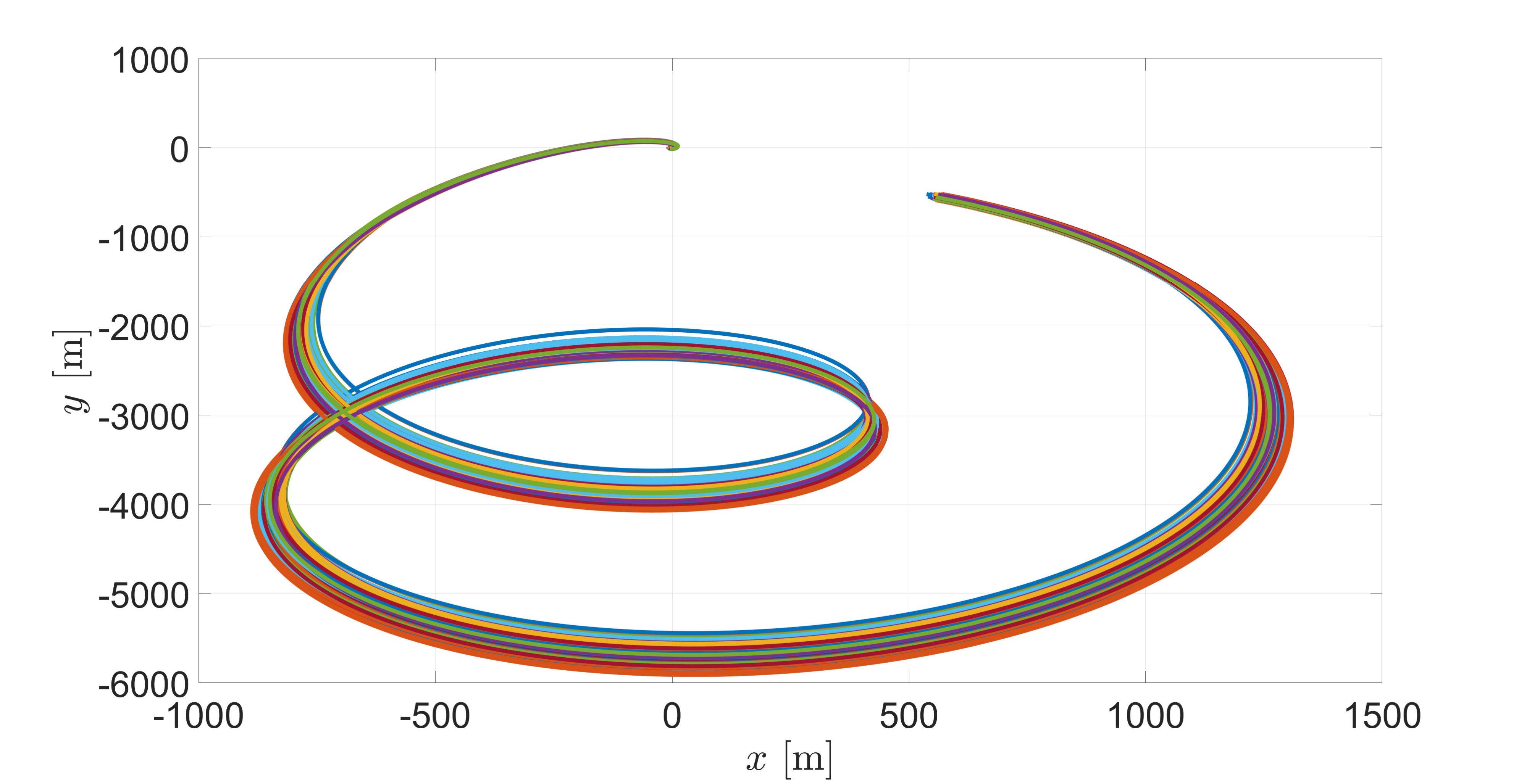}
  \caption{Relative position components}
  \label{Fig:CaseB_many_traj_fuel}
  \end{subfigure}
  \hspace{0.05\textwidth}
  \begin{subfigure}[t]{0.45\textwidth}
  \centering
  \includegraphics[width = \textwidth]{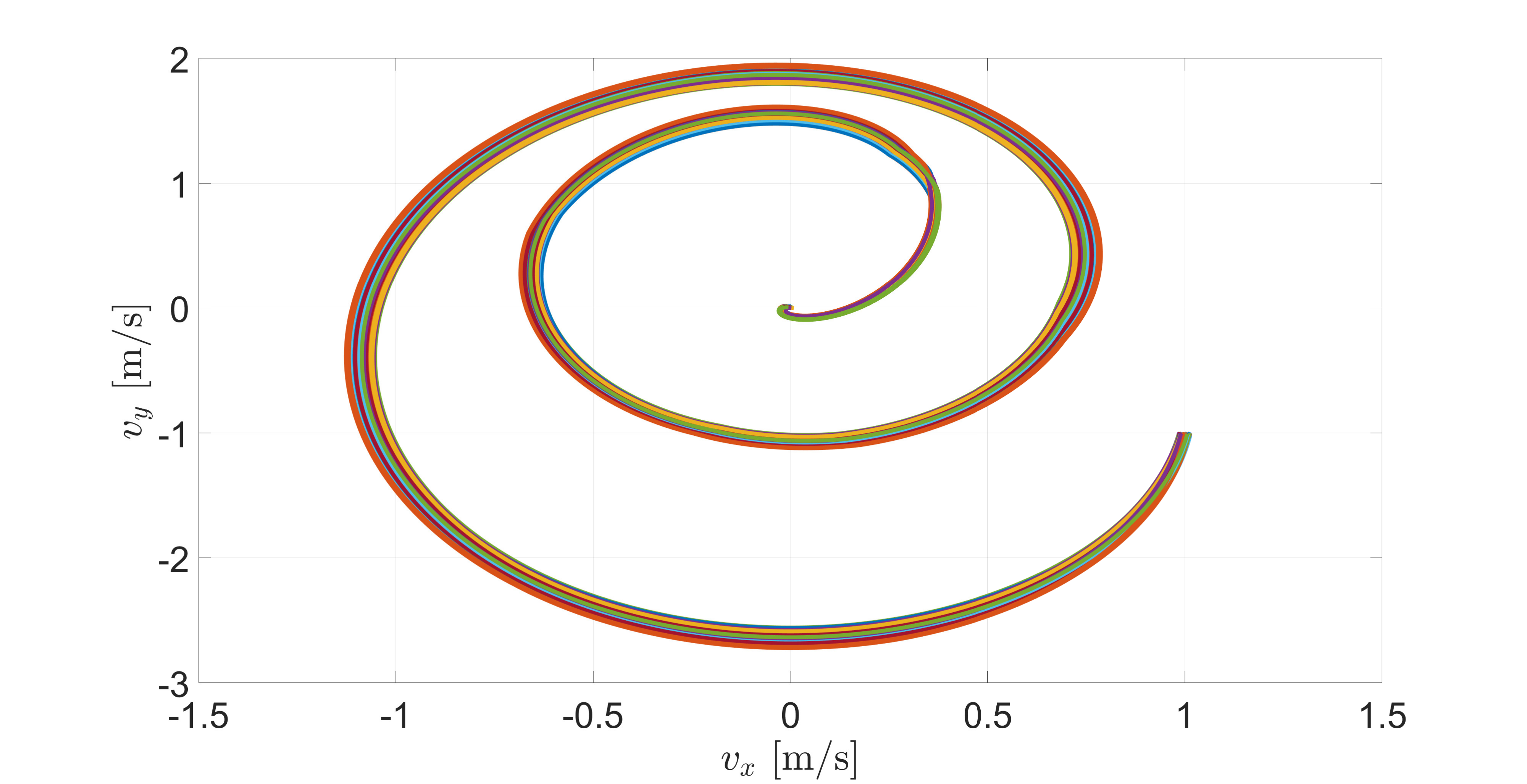}
  \caption{Relative velocity components}\label{Fig:CaseB_many_vel_fuel}
  \end{subfigure}
  \caption{Relative position and velocity components under initial condition perturbations.}
\label{Fig:CaseA_optimal_many_fuel}
\vspace{-0.5cm}
  \end{figure}

\begin{figure}[!htp]
    \centering
    \begin{subfigure}[t]{0.45\textwidth}
    \includegraphics[width = \textwidth]{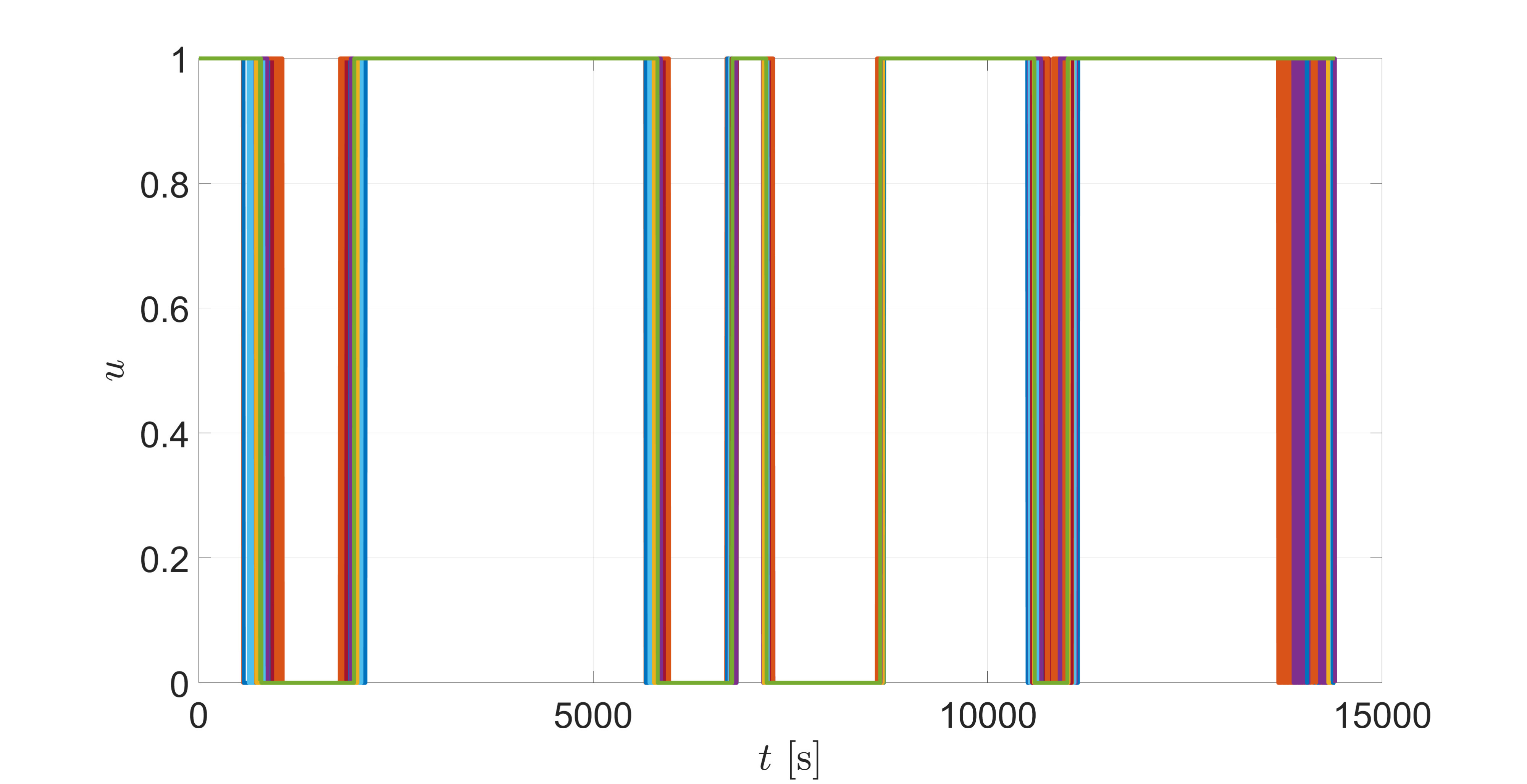}
    \caption{Throttle}
    \label{Fig:CaseB_many__fuel_u}
    \end{subfigure}
    \hspace{0.05\textwidth}
    \begin{subfigure}[t]{0.45\textwidth}
    \centering
    \includegraphics[width = \textwidth]{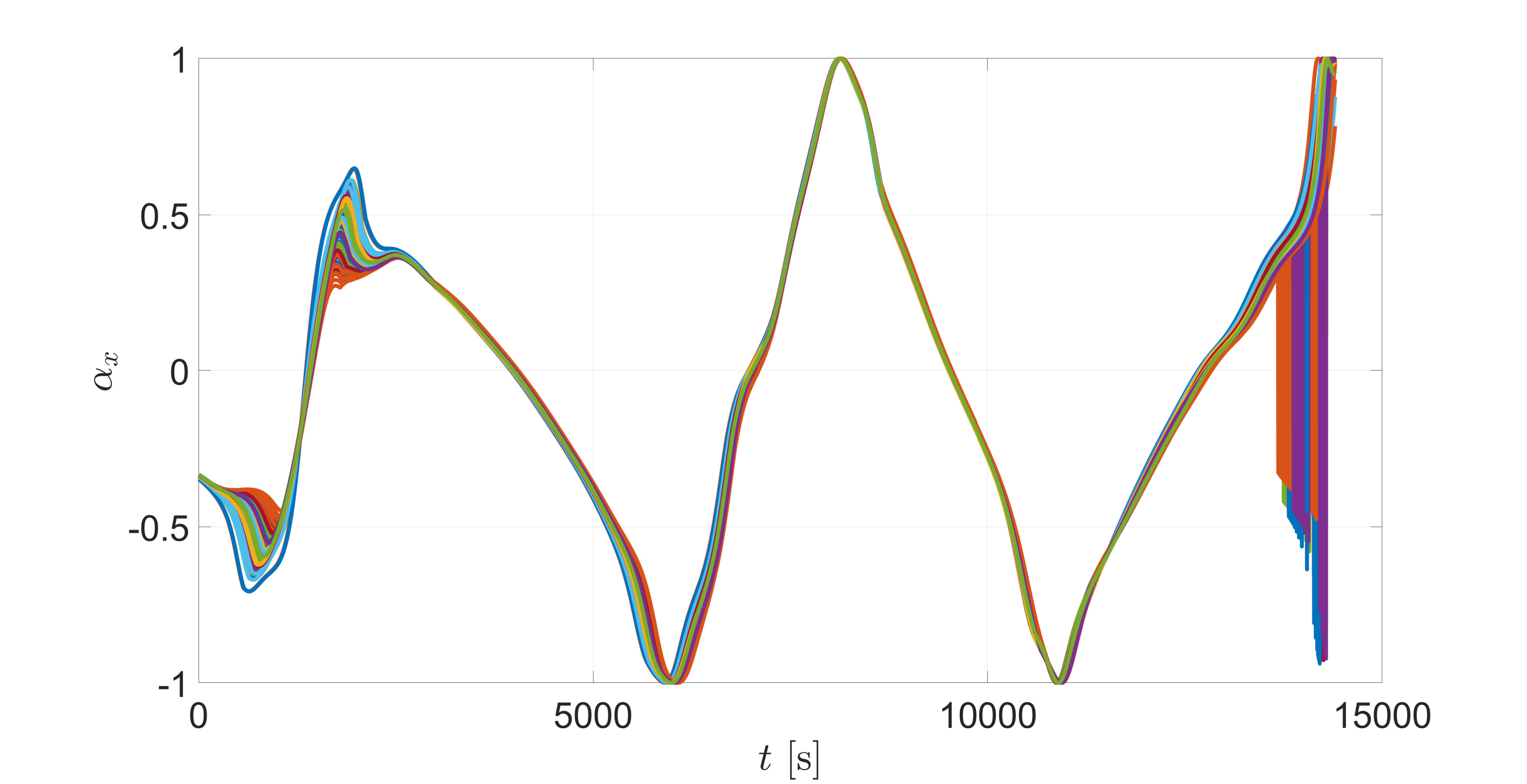}
    \caption{Thrust direction component $\alpha_x$}\label{Fig:CaseB_many__fuel_alpha}
    \end{subfigure}
    \caption{Guidance policy profiles under initial condition perturbations.}
  \label{Fig:CaseA_optimal_many_controlpolicy}
  \vspace{-0.5cm}
    \end{figure}

Similarly, we consider 200 initial conditions by introducing perturbations, as defined in Eq.~(\ref{initial_condition_definition}), to the initial condition of \(\boldsymbol{x}_0 = [550~\text{m}, -550~\text{m}, 1~\text{m/s}, -1~\text{m/s}]^T\). For simplicity, the rendezvous time remains fixed at \(14,400~\text{s}\). 
The convergence region remains consistent with that defined in Eq.~(\ref{ball_definition}). Using the proposed method, a total of 121 cases successfully achieved rendezvous.  
For cases where the states fail to converge to the region specified in Eq.~(\ref{ball_definition}), the maximum position and velocity errors are $50$ m and $0.05$ m/s, respectively.  
Figures~\ref{Fig:CaseA_optimal_many_fuel} and ~\ref{Fig:CaseA_optimal_many_controlpolicy} illustrate the relative position and velocity components, as well as the corresponding guidance policy profiles for the successful cases.

The success rate is expected to be lower compared to the time-optimal problem for the following reasons: (1)  The fuel-optimal problem with a fixed final time is inherently more sensitive to perturbations than the time-optimal problem; (2) For the time-optimal problem, only the thrust direction needs to be replicated, as reflected in the empirical loss function defined in Eq.~(\ref{EQ:existing_method_paper}). In contrast, the fuel-optimal problem requires replicating both the thrust direction and the throttle, as shown in Eq.~(\ref{EQ:existing_method_paper_fuel}).  
Consequently, the discrepancy between the results obtained by the proposed method and the open-loop optimal solution is larger for the fuel-optimal problem than for the time-optimal problem. This difference is also evident in the comparison of thrust directions, as illustrated in Figs.~\ref{Fig:CaseA_direction} and \ref{Fig:CaseA_control_profile_fuel_comparison}.

\begin{figure}[!htp]
  \begin{center}
  \includegraphics[scale=0.2]{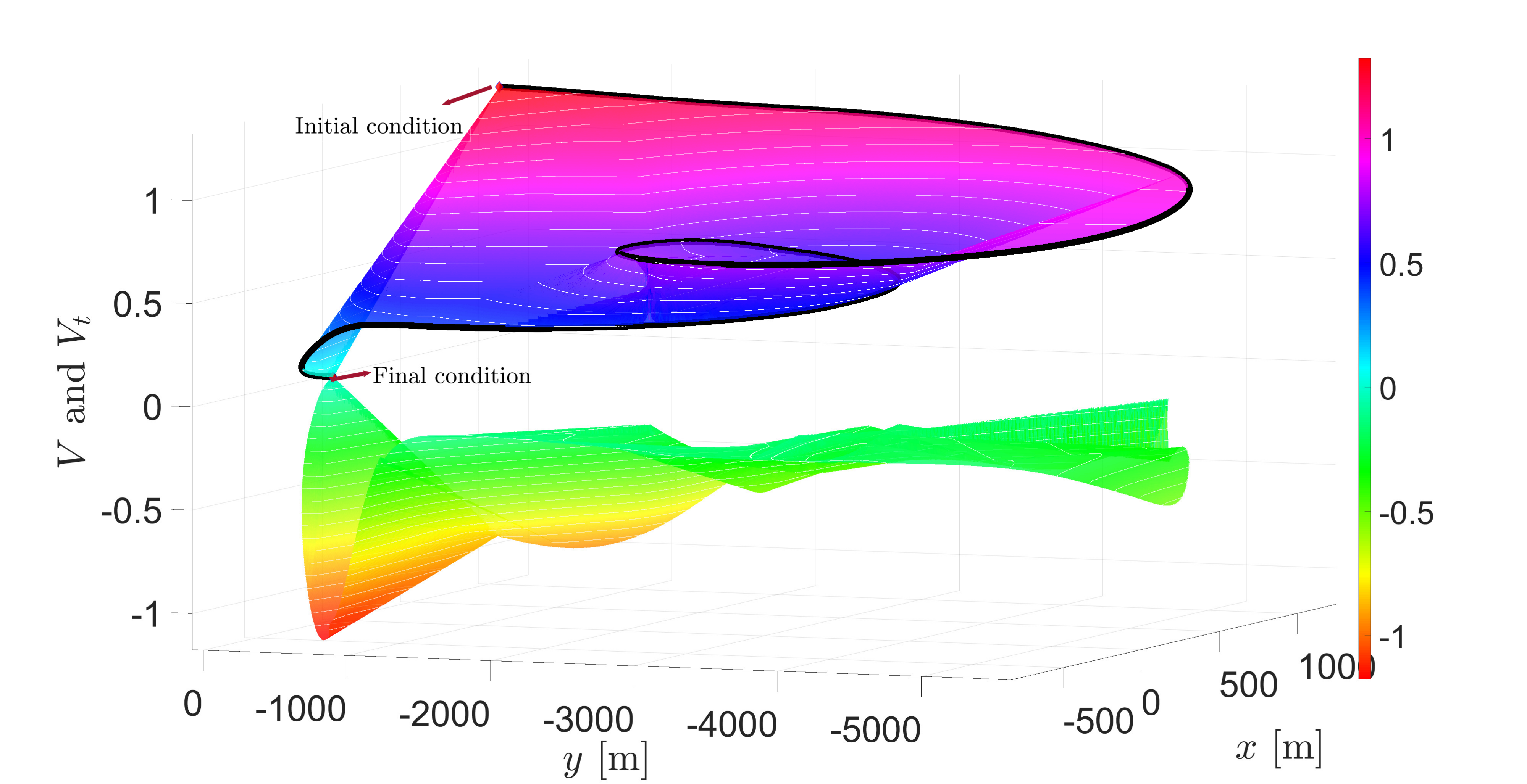}
  \caption{Lyapunov function and its time derivative w.r.t. the relative position components.}\label{Fig:fuel_optimal_many_vel_lya_position}
  \end{center}
  \vspace{-0.4cm}
\end{figure}

\begin{figure}[!htp]
  \begin{center}
  \includegraphics[scale=0.2]{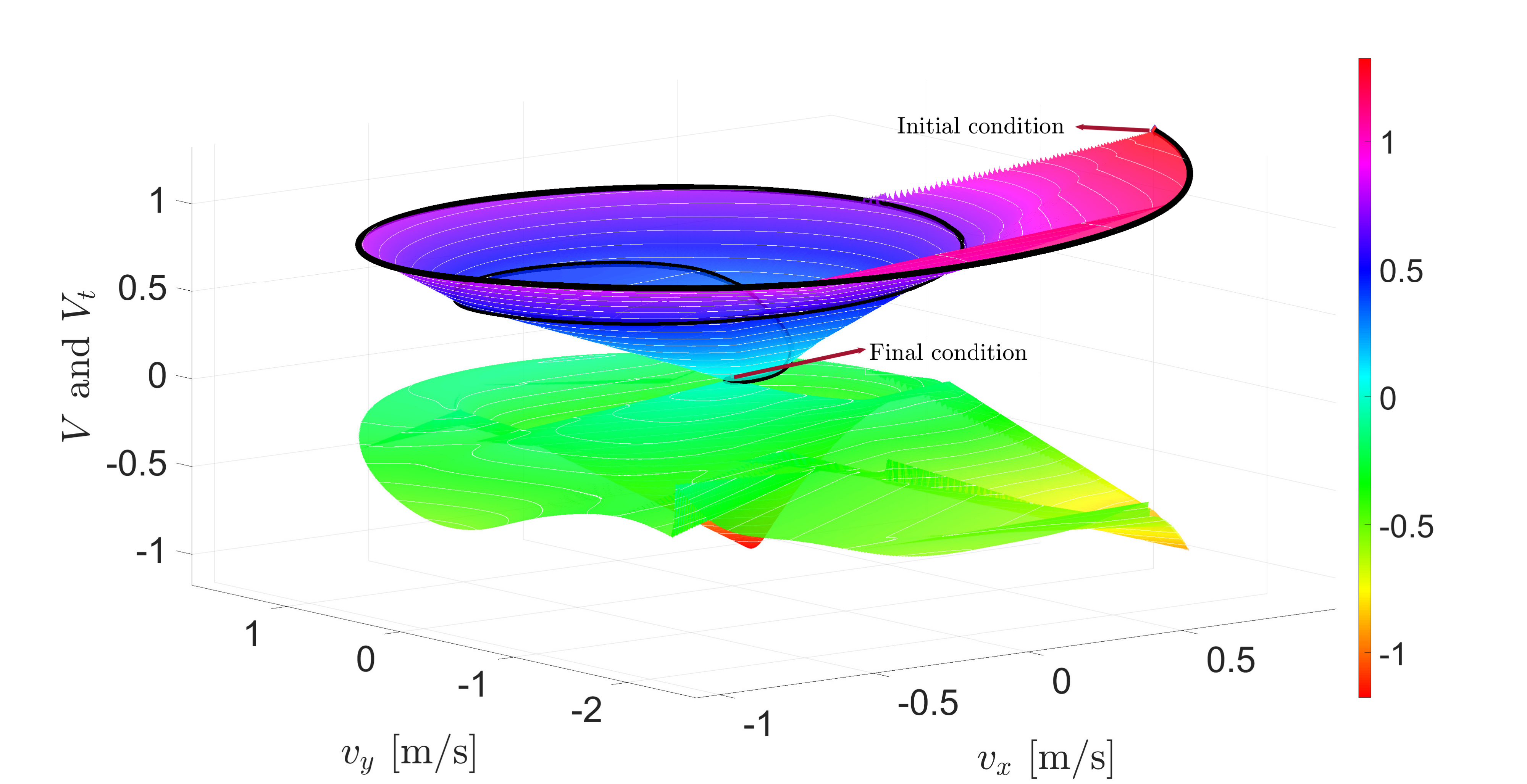}
  \caption{Lyapunov function and its time derivative w.r.t. the relative velocity components.}\label{Fig:fuel_optimal_many_vel_lya}
  \end{center}
  \vspace{-0.4cm}
\end{figure}

For a clearer illustration, Figs.~\ref{Fig:fuel_optimal_many_vel_lya_position} and \ref{Fig:fuel_optimal_many_vel_lya} present the Lyapunov function and its time derivative with respect to the relative position and velocity components, respectively. The upper surface represents the Lyapunov function \( V \), which gradually decreases from the initial to the final condition, reflecting the system's convergence towards the equilibrium point. The lower surface shows the Lyapunov function's time derivative \( V_t \). Both the initial and final conditions are marked, showing that \( V \) starts at a higher value and approaches zero as the state reaches the final condition. Despite the linearity of the system dynamics, the surfaces of both \( V \) and \( V_t \) exhibit considerable complexity, highlighting the intricacy of reaching the equilibrium point stably within a fixed time. The results confirm that the proposed method effectively guides the chaser spacecraft towards a final condition near the equilibrium point while fulfilling both optimality and stability requirements. This is evident from the gradual decrease in the Lyapunov function and its derivative as they approach zero near the final condition. 

\begin{figure}[!htp]
  \begin{center}
  \includegraphics[scale=0.18]{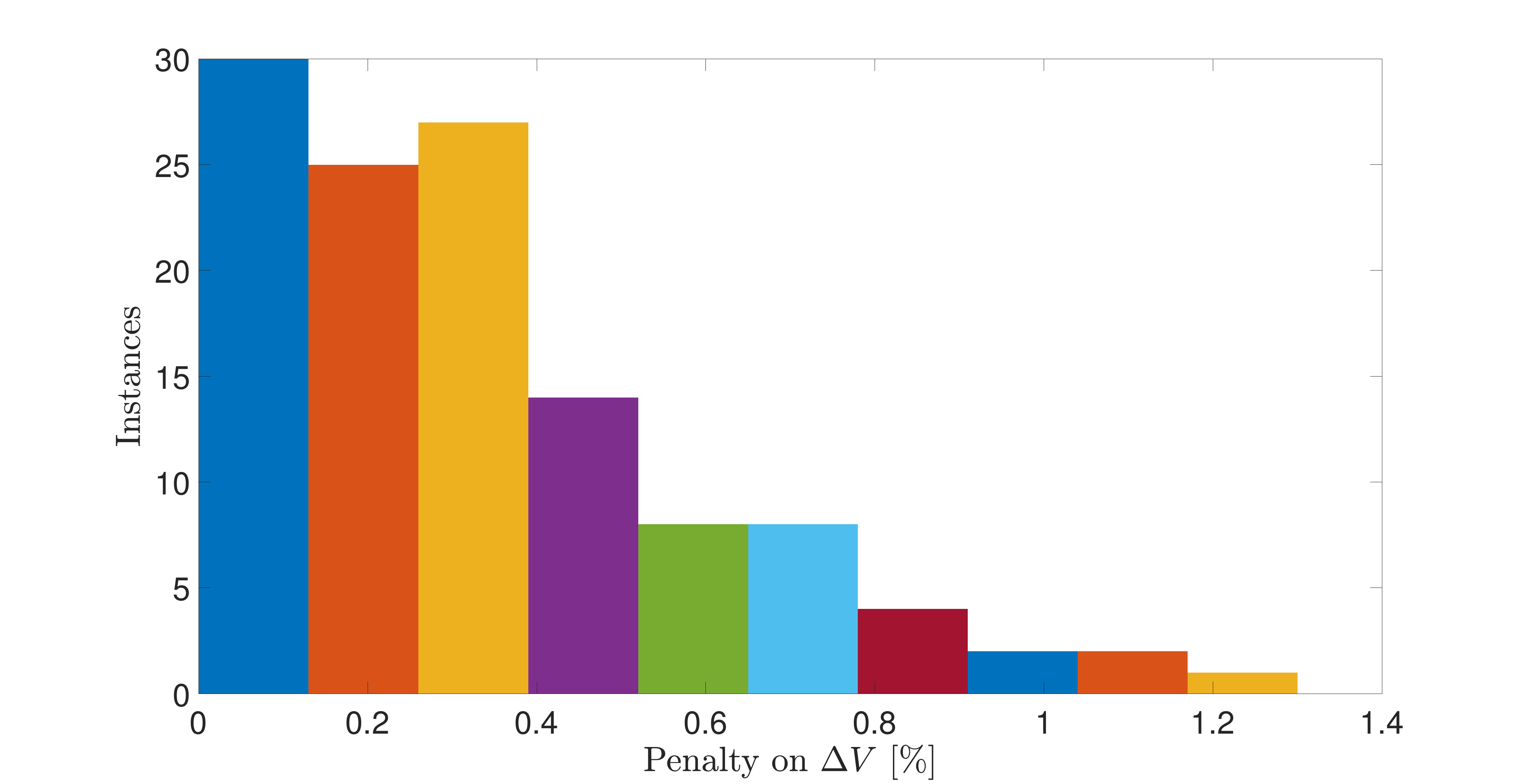}
  \caption{Histogram of the penalty on $\Delta V$.}\label{Fig:fuel_optimal_many_deltaV}
  \end{center}
  \vspace{-0.5cm}
\end{figure}

To assess the penalty on $\Delta V$ caused by the proposed method, we apply the indirect shooting method to solve the shooting function in Eq.~(\ref{EQ:magnitudeUcha6_smooth}) so as to find the optimal $\Delta V$. The solution to the nominal initial condition is used as the initial guess.  The histogram of the penalty on $\Delta V$ for these 121 cases is presented in Fig.~\ref{Fig:fuel_optimal_many_deltaV}, from which we can see that even the largest penalty is less than $1.4 \%$. 
\subsection{Computational Cost}
To implement the proposed method, computing the guidance command first requires obtaining the Lyapunov function \( V \) and its decay rate \( \gamma \) through simple matrix addition and multiplication. Next, the gradient of \( V \) with respect to the state \( \boldsymbol{x} \) is used to determine the thrust direction \( \boldsymbol{\alpha} \) in the time-optimal problem, and both the thrust direction and the minimal required throttle  \( \underline{u} \) for the fuel-optimal problem. A total of 1,000 cases with different states are tested, yielding an average computational time of 0.36 ms for the time-optimal problem and 0.81 ms for the fuel-optimal problem. If the proposed method was executed on a state-of-the-art flight processor with a clock speed of 200 MHz \cite{dueri2017customized}, the computational costs would be 5.76 ms and 12.96 ms, respectively. These results demonstrate the great potential of the proposed method for onboard implementation. 

\section{Conclusions}\label{SE:conclusions}  
This work proposed a learning-based approach to generate time- and fuel-optimal guidance commands with certified stability for rendezvous maneuvers governed by Clohessy–Wiltshire dynamics. An efficient polynomial mapping method was employed to rapidly generate a large dataset of optimal trajectories. A novel neural candidate Lyapunov function was constructed, inherently satisfying the positive definiteness property of control Lyapunov functions. For the control policy, the thrust direction vector was designed to minimize the  Lyapunov function's time derivative. The minimal required throttle was incorporated to replace conventional loss terms related to the Lyapunov function to enforce the decay condition of the control Lyapunov function. A simple yet effective loss function was then proposed to jointly supervise the Lyapunov function and its corresponding control policy using the collected optimal trajectories. This unified approach successfully addressed both stability and optimality. Furthermore, by treating the decay rate in the control Lyapunov function as a state-dependent parameter rather than a constant, we found that it improved the training process and resulted in slightly improved final state errors for the time-optimal problem. Notably, in the fuel-optimal problem, fixing the decay rate as a constant significantly hindered training convergence, underscoring its critical role. Finally, the efficacy of the proposed method in generating certified stable and optimal guidance commands was validated through numerical simulations.  Partially due to the inherent sensitivity of the fixed-final-time fuel-optimal problem to initial state perturbations, the guidance performance of the proposed method is not as effective as that of the time-optimal problem. Future work will focus on enhancing the guidance performance for the fuel-optimal problem.
\section*{Acknowledgement}
This research was supported by the National Natural Science Foundation of China under grant No. 62088101.

\bibliography{main}

\begin{thebibliography}{49}
\newcommand{\enquote}[1]{``#1''}
\providecommand{\natexlab}[1]{#1}
\providecommand{\url}[1]{\texttt{#1}}
\providecommand{\urlprefix}{URL }
\expandafter\ifx\csname urlstyle\endcsname\relax
  \providecommand{\doi}[1]{\discretionary{}{}{}https://doi.org/#1}\else
  \providecommand{\doi}[1]{\discretionary{}{}{}\urlstyle{rm}\url{https://doi.org/#1}}\fi

\bibitem[{Flores-Abad et~al.(2014)Flores-Abad, Ma, Pham, and Ulrich}]{flores2014review}
Flores-Abad, A., Ma, O., Pham, K., and Ulrich, S., \enquote{A review of space robotics technologies for on-orbit servicing,} \emph{Progress in Aerospace Sciences}, Vol.~68, 2014, pp. 1--26.
\newblock \doi{10.1016/j.paerosci.2014.03.002}.

\bibitem[{Polites(1999)}]{polites1999technology}
Polites, M.~E., \enquote{Technology of automated rendezvous and capture in space,} \emph{Journal of Spacecraft and Rockets}, Vol.~36, No.~2, 1999, pp. 280--291.
\newblock \doi{10.2514/2.3443}.

\bibitem[{Dittmar(2003)}]{dittmar2003commercial}
Dittmar, M., \enquote{Commercial avenues for space utilization,} \emph{AIAA Space 2003 Conference \& Exposition}, 2003, p. 6234.

\bibitem[{Betts(1998)}]{betts1998survey}
Betts, J.~T., \enquote{Survey of numerical methods for trajectory optimization,} \emph{Journal of Guidance, Control, and Dynamics}, Vol.~21, No.~2, 1998, pp. 193--207.
\newblock \doi{10.2514/2.4231}.

\bibitem[{Lu(2017)}]{lu2017introducing}
Lu, P., \enquote{Introducing computational guidance and control,} \emph{Journal of Guidance, Control, and Dynamics}, Vol.~40, No.~2, 2017, pp. 193--193.
\newblock \doi{10.2514/1.G002745}.

\bibitem[{Bodin et~al.(2011)Bodin, Noteborn, Larsson, and Chasset}]{bodin2011system}
Bodin, P., Noteborn, R., Larsson, R., and Chasset, C., \enquote{System test results from the {GNC} experiments on the {PRISMA} in-orbit test bed,} \emph{Acta Astronautica}, Vol.~68, No. 7-8, 2011, pp. 862--872.
\newblock \doi{10.1016/j.actaastro.2010.08.021}.

\bibitem[{Di~Cairano et~al.(2012)Di~Cairano, Park, and Kolmanovsky}]{di2012model}
Di~Cairano, S., Park, H., and Kolmanovsky, I., \enquote{Model predictive control approach for guidance of spacecraft rendezvous and proximity maneuvering,} \emph{International Journal of Robust and Nonlinear Control}, Vol.~22, No.~12, 2012, pp. 1398--1427.
\newblock \doi{10.1002/rnc.2827}.

\bibitem[{Weiss et~al.(2015)Weiss, Baldwin, Erwin, and Kolmanovsky}]{weiss2015model}
Weiss, A., Baldwin, M., Erwin, R.~S., and Kolmanovsky, I., \enquote{Model predictive control for spacecraft rendezvous and docking: Strategies for handling constraints and case studies,} \emph{IEEE Transactions on Control Systems Technology}, Vol.~23, No.~4, 2015, pp. 1638--1647.
\newblock \doi{10.1109/TCST.2014.2379639}.

\bibitem[{Hartley(2015)}]{hartley2015tutorial}
Hartley, E.~N., \enquote{A tutorial on model predictive control for spacecraft rendezvous,} \emph{2015 European Control Conference (ECC)}, IEEE, 2015, pp. 1355--1361.
\newblock \doi{10.1109/ECC.2015.7330727}.

\bibitem[{Pagone et~al.(2021)Pagone, Boggio, Novara, and Vidano}]{pagone2021pontryagin}
Pagone, M., Boggio, M., Novara, C., and Vidano, S., \enquote{A {P}ontryagin-based {NMPC} approach for autonomous rendezvous proximity operations,} \emph{2021 IEEE Aerospace Conference (50100)}, IEEE, 2021, pp. 1--9.
\newblock \doi{10.1109/AERO50100.2021.9438226}.

\bibitem[{Bashnick and Ulrich(2023)}]{bashnick2023fast}
Bashnick, C., and Ulrich, S., \enquote{Fast model predictive control for spacecraft rendezvous and docking with obstacle avoidance,} \emph{Journal of Guidance, Control, and Dynamics}, Vol.~46, No.~5, 2023, pp. 998--1007.
\newblock \doi{10.2514/1.G007314}.

\bibitem[{Bayer et~al.(2016)Bayer, Brunner, Lazar, Wijnand, and Allg{\"o}wer}]{bayer2016tube}
Bayer, F.~A., Brunner, F.~D., Lazar, M., Wijnand, M., and Allg{\"o}wer, F., \enquote{A tube-based approach to nonlinear explicit {MPC},} \emph{2016 IEEE 55th Conference on Decision and Control (CDC)}, IEEE, 2016, pp. 4059--4064.
\newblock \doi{10.1109/CDC.2016.7798884}.

\bibitem[{Lu and Liu(2013)}]{lu2013autonomous}
Lu, P., and Liu, X., \enquote{Autonomous trajectory planning for rendezvous and proximity operations by conic optimization,} \emph{Journal of Guidance, Control, and Dynamics}, Vol.~36, No.~2, 2013, pp. 375--389.
\newblock \doi{10.2514/1.58436}.

\bibitem[{Ortolano et~al.(2021)Ortolano, Geller, and Avery}]{ortolano2021autonomous}
Ortolano, N., Geller, D.~K., and Avery, A., \enquote{Autonomous optimal trajectory planning for orbital rendezvous, satellite inspection, and final approach based on convex optimization,} \emph{The Journal of the Astronautical Sciences}, Vol.~68, No.~2, 2021, pp. 444--479.
\newblock \doi{10.1007/s40295-021-00260-5}.

\bibitem[{Malyuta and Acikmese(2023)}]{malyuta2023fast}
Malyuta, D., and Acikmese, B., \enquote{Fast homotopy for spacecraft rendezvous trajectory optimization with discrete logic,} \emph{Journal of Guidance, Control, and Dynamics}, Vol.~46, No.~7, 2023, pp. 1262--1279.
\newblock \doi{10.2514/1.G006295}.

\bibitem[{Izzo et~al.(2020)Izzo, Tailor, and Vasileiou}]{izzo2020stability}
Izzo, D., Tailor, D., and Vasileiou, T., \enquote{On the stability analysis of deep neural network representations of an optimal state feedback,} \emph{IEEE Transactions on Aerospace and Electronic Systems}, Vol.~57, No.~1, 2020, pp. 145--154.
\newblock \doi{10.1109/TAES.2020.3010670}.

\bibitem[{Wang et~al.(2024{\natexlab{a}})Wang, Chen, and Li}]{WANG2024}
Wang, K., Chen, Z., and Li, J., \enquote{Fuel-optimal powered descent guidance for lunar pinpoint landing using neural networks,} \emph{Advances in Space Research}, Vol.~74, No.~10, 2024{\natexlab{a}}, pp. 5006--5022.
\newblock \doi{10.1016/j.asr.2024.07.019}.

\bibitem[{Mulekar et~al.(2024)Mulekar, Cho, and Bevilacqua}]{mulekar2024stable}
Mulekar, O.~S., Cho, H., and Bevilacqua, R., \enquote{Stable optimal feedback control for landers based on machine learning,} \emph{AIAA Journal}, Vol.~62, No.~5, 2024, pp. 1932--1945.
\newblock \doi{10.2514/1.J063682}.

\bibitem[{Izzo and {\"O}zt{\"u}rk(2021)}]{izzo2021real}
Izzo, D., and {\"O}zt{\"u}rk, E., \enquote{Real-time guidance for low-thrust transfers using deep neural networks,} \emph{Journal of Guidance, Control, and Dynamics}, Vol.~44, No.~2, 2021, pp. 315--327.
\newblock \doi{10.2514/1.G005254}.

\bibitem[{Wang et~al.(2024{\natexlab{b}})Wang, Lu, Chen, and Li}]{wang2024real}
Wang, K., Lu, F., Chen, Z., and Li, J., \enquote{Real-time optimal control for attitude-constrained solar sailcrafts via neural networks,} \emph{Acta Astronautica}, Vol. 216, 2024{\natexlab{b}}, pp. 446--458.
\newblock \doi{10.1016/j.actaastro.2024.01.026}.

\bibitem[{Federici et~al.(2021)Federici, Benedikter, and Zavoli}]{federici2021deep}
Federici, L., Benedikter, B., and Zavoli, A., \enquote{Deep learning techniques for autonomous spacecraft guidance during proximity operations,} \emph{Journal of Spacecraft and Rockets}, Vol.~58, No.~6, 2021, pp. 1774--1785.
\newblock \doi{10.2514/1.A35076}.

\bibitem[{Liu et~al.(2024)Liu, Wang, Lee, and Tóth}]{Yuhan2024}
Liu, Y., Wang, P., Lee, C.-H., and Tóth, R., \enquote{Attitude takeover control for noncooperative space targets based on Gaussian processes with online model learning,} \emph{IEEE Transactions on Aerospace and Electronic Systems}, Vol.~60, No.~3, 2024, pp. 3050--3066.
\newblock \doi{10.1109/TAES.2024.3361886}.

\bibitem[{Broida and Linares(2019)}]{broida2019spacecraft}
Broida, J., and Linares, R., \enquote{Spacecraft rendezvous guidance in cluttered environments via reinforcement learning,} \emph{29th AAS/AIAA Space Flight Mechanics Meeting}, American Astronautical Society, 2019, pp. 1--15.

\bibitem[{Qu et~al.(2022)Qu, Liu, Wang, and L{\"u}}]{qu2022spacecraft}
Qu, Q., Liu, K., Wang, W., and L{\"u}, J., \enquote{Spacecraft proximity maneuvering and rendezvous with collision avoidance based on reinforcement learning,} \emph{IEEE Transactions on Aerospace and Electronic Systems}, Vol.~58, No.~6, 2022, pp. 5823--5834.
\newblock \doi{10.1109/TAES.2022.3180271}.

\bibitem[{Federici et~al.(2022)Federici, Scorsoglio, Zavoli, and Furfaro}]{federici2022meta}
Federici, L., Scorsoglio, A., Zavoli, A., and Furfaro, R., \enquote{Meta-reinforcement learning for adaptive spacecraft guidance during finite-thrust rendezvous missions,} \emph{Acta Astronautica}, Vol. 201, 2022, pp. 129--141.
\newblock \doi{10.1016/j.actaastro.2022.08.047}.

\bibitem[{Tipaldi et~al.(2022)Tipaldi, Iervolino, and Massenio}]{tipaldi2022reinforcement}
Tipaldi, M., Iervolino, R., and Massenio, P.~R., \enquote{Reinforcement learning in spacecraft control applications: Advances, prospects, and challenges,} \emph{Annual Reviews in Control}, Vol.~54, 2022, pp. 1--23.
\newblock \doi{10.1016/j.arcontrol.2022.07.004}.

\bibitem[{Fereoli et~al.(2024)Fereoli, Schaub, and Di~Lizia}]{fereoli2024meta}
Fereoli, G., Schaub, H., and Di~Lizia, P., \enquote{Meta-reinforcement learning for spacecraft proximity operations guidance and control in cislunar space,} \emph{Journal of Spacecraft and Rockets}, 2024, pp. 1--13.
\newblock \doi{10.2514/1.A36100}.

\bibitem[{Zhou and Lam(2017)}]{zhou2017global}
Zhou, B., and Lam, J., \enquote{Global stabilization of linearized spacecraft rendezvous system by saturated linear feedback,} \emph{IEEE Transactions on Control Systems Technology}, Vol.~25, No.~6, 2017, pp. 2185--2193.
\newblock \doi{10.1109/TCST.2016.2632529}.

\bibitem[{Shirobokov et~al.(2021)Shirobokov, Trofimov, and Ovchinnikov}]{shirobokov2021survey}
Shirobokov, M., Trofimov, S., and Ovchinnikov, M., \enquote{Survey of machine learning techniques in spacecraft control design,} \emph{Acta Astronautica}, Vol. 186, 2021, pp. 87--97.
\newblock \doi{10.1016/j.actaastro.2021.05.018}.

\bibitem[{Nakamura-Zimmerer et~al.(2022)Nakamura-Zimmerer, Gong, and Kang}]{nakamura2022neural}
Nakamura-Zimmerer, T., Gong, Q., and Kang, W., \enquote{Neural network optimal feedback control with guaranteed local stability,} \emph{IEEE Open Journal of Control Systems}, Vol.~1, 2022, pp. 210--222.
\newblock \doi{10.1109/OJCSYS.2022.3205863}.

\bibitem[{Dalin et~al.(2015)Dalin, Bo, and Youtao}]{dalin2015optimal}
Dalin, Y., Bo, X., and Youtao, G., \enquote{Optimal strategy for low-thrust spiral trajectories using Lyapunov-based guidance,} \emph{Advances in Space Research}, Vol.~56, No.~5, 2015, pp. 865--878.
\newblock \doi{10.1016/j.asr.2015.05.030}.

\bibitem[{Holt et~al.(2021)Holt, Armellin, Baresi, Hashida, Turconi, Scorsoglio, and Furfaro}]{holt2021optimal}
Holt, H., Armellin, R., Baresi, N., Hashida, Y., Turconi, A., Scorsoglio, A., and Furfaro, R., \enquote{Optimal {Q}-laws via reinforcement learning with guaranteed stability,} \emph{Acta Astronautica}, Vol. 187, 2021, pp. 511--528.
\newblock \doi{10.1016/j.actaastro.2021.07.010}.

\bibitem[{Holt and Armellin(2024)}]{holtISSFD}
Holt, H., and Armellin, R., \enquote{Reinforcement learning enhanced {LQR} and control Lyapunov functions for proximity operations,} \emph{International Symposium on Space Flight Dynamics (29th: 2024: Darmstadt, Germany)}, 2024, pp. 1--12.

\bibitem[{Dawson et~al.(2023)Dawson, Gao, and Fan}]{dawson2023safe}
Dawson, C., Gao, S., and Fan, C., \enquote{Safe control with learned certificates: A survey of neural {L}yapunov, barrier, and contraction methods for robotics and control,} \emph{IEEE Transactions on Robotics}, Vol.~39, No.~3, 2023, pp. 1749--1767.
\newblock \doi{10.1109/TRO.2022.3232542}.

\bibitem[{Richards et~al.(2018)Richards, Berkenkamp, and Krause}]{richards2018lyapunov}
Richards, S.~M., Berkenkamp, F., and Krause, A., \enquote{The {L}yapunov neural network: Adaptive stability certification for safe learning of dynamical systems,} \emph{Conference on Robot Learning}, PMLR, 2018, pp. 466--476.

\bibitem[{Abate et~al.(2020)Abate, Ahmed, Giacobbe, and Peruffo}]{abate2020formal}
Abate, A., Ahmed, D., Giacobbe, M., and Peruffo, A., \enquote{Formal synthesis of Lyapunov neural networks,} \emph{IEEE Control Systems Letters}, Vol.~5, No.~3, 2020, pp. 773--778.
\newblock \doi{10.1109/LCSYS.2020.3005328}.

\bibitem[{Yin et~al.(2021)Yin, Seiler, and Arcak}]{yin2021stability}
Yin, H., Seiler, P., and Arcak, M., \enquote{Stability analysis using quadratic constraints for systems with neural network controllers,} \emph{IEEE Transactions on Automatic Control}, Vol.~67, No.~4, 2021, pp. 1980--1987.
\newblock \doi{10.1109/TAC.2021.3069388}.

\bibitem[{Sastry(2013)}]{sastry2013nonlinear}
Sastry, S., \emph{Nonlinear systems: analysis, stability, and control}, Vol.~10, Springer Science \& Business Media, 2013.

\bibitem[{Chang et~al.(2019)Chang, Roohi, and Gao}]{chang2019neural}
Chang, Y.-C., Roohi, N., and Gao, S., \enquote{Neural {L}yapunov control,} \emph{Advances in Neural Information Processing Systems}, Vol.~32, 2019.

\bibitem[{Clohessy and Wiltshire(1960)}]{clohessy1960terminal}
Clohessy, W., and Wiltshire, R., \enquote{Terminal guidance system for satellite rendezvous,} \emph{Journal of the Aerospace Sciences}, Vol.~27, No.~9, 1960, pp. 653--658.
\newblock \doi{10.2514/8.8704}.

\bibitem[{Wang et~al.(2022)Wang, Chen, Wang, Li, and Shao}]{wang2022nonlinear}
Wang, K., Chen, Z., Wang, H., Li, J., and Shao, X., \enquote{Nonlinear optimal guidance for intercepting stationary targets with impact-time constraints,} \emph{Journal of Guidance, Control, and Dynamics}, Vol.~45, No.~9, 2022, pp. 1614--1626.
\newblock \doi{10.2514/1.G006666}.

\bibitem[{Pontryagin(2018)}]{Pontryagin}
Pontryagin, L.~S., \emph{Mathematical theory of optimal processes}, Routledge, 2018.
\newblock \doi{10.1201/9780203749319}.

\bibitem[{Evans et~al.(2024)Evans, Armellin, Pirovano, and Baresi}]{evans2024high}
Evans, A., Armellin, R., Pirovano, L., and Baresi, N., \enquote{High-order guidance for time-optimal low-thrust trajectories with accuracy control,} \emph{Journal of Guidance, Control, and Dynamics}, Vol.~47, No.~2, 2024, pp. 279--290.
\newblock \doi{10.2514/1.G007540}.

\bibitem[{Evans et~al.(2025)Evans, Armellin, Holt, and Pirovano}]{EVANS202517}
Evans, A., Armellin, R., Holt, H., and Pirovano, L., \enquote{Fuel-optimal guidance using costate supervised learning with local refinement,} \emph{Acta Astronautica}, Vol. 228, 2025, pp. 17--29.
\newblock \doi{10.1016/j.actaastro.2024.11.031}.

\bibitem[{Di~Lizia et~al.(2014)Di~Lizia, Armellin, Morselli, and Bernelli-Zazzera}]{di2014high}
Di~Lizia, P., Armellin, R., Morselli, A., and Bernelli-Zazzera, F., \enquote{High order optimal feedback control of space trajectories with bounded control,} \emph{Acta Astronautica}, Vol.~94, No.~1, 2014, pp. 383--394.
\newblock \doi{10.1016/j.actaastro.2013.02.011}.

\bibitem[{Majumdar et~al.(2013)Majumdar, Ahmadi, and Tedrake}]{majumdar2013control}
Majumdar, A., Ahmadi, A.~A., and Tedrake, R., \enquote{Control design along trajectories with sums of squares programming,} \emph{2013 IEEE International Conference on Robotics and Automation}, IEEE, 2013, pp. 4054--4061.
\newblock \doi{10.1109/ICRA.2013.6631149}.

\bibitem[{Gurfil(2023)}]{gurfil2023spacecraft}
Gurfil, P., \enquote{Spacecraft rendezvous using constant-magnitude low thrust,} \emph{Journal of Guidance, Control, and Dynamics}, Vol.~46, No.~11, 2023, pp. 2183--2191.
\newblock \doi{10.2514/1.G007472}.

\bibitem[{Origer et~al.(2023)Origer, De~Wagter, Ferede, de~Croon, and Izzo}]{origer2023guidance}
Origer, S., De~Wagter, C., Ferede, R., de~Croon, G.~C., and Izzo, D., \enquote{Guidance and control networks for time-optimal quadcopter flight,} \emph{arXiv}, 2023.
\newblock \doi{10.48550/arXiv.2305.02705}.

\bibitem[{Dueri et~al.(2017)Dueri, Acikmese, Scharf, and Harris}]{dueri2017customized}
Dueri, D., Acikmese, B., Scharf, D.~P., and Harris, M.~W., \enquote{Customized real-time interior-point methods for onboard powered-descent guidance,} \emph{Journal of Guidance, Control, and Dynamics}, Vol.~40, No.~2, 2017, pp. 197--212.
\newblock \doi{10.2514/1.G001480}.

\end{thebibliography}

\end{document}